\documentclass[apj,twocolumn,twocolappendix,numberedappendix]{openjournal}

\usepackage[dvipsnames]{xcolor}
\usepackage{hyperref}
\hypersetup{colorlinks=true,citecolor=blue,linkcolor=blue,urlcolor=blue,pdftitle={Identifying and Distinguishing Quenching Galaxies with Spatially Resolved Star Formation in the Hubble Frontier Fields},pdfauthor={Cameron Lawlor-Forsyth; Michael L. Balogh; Sean L. McGee; Gregory H. Rudnick},pdfsubject={}}
\usepackage{orcidlink}
\usepackage[hang]{footmisc}
\definecolor{doiPink}{RGB}{255, 0, 255}

\usepackage{enumitem}

\usepackage[largesc]{newtx}
\usepackage{couriers}
\DeclareSymbolFont{CMop}{OT1}{cmr}{m}{n}
\SetSymbolFont{CMop}{bold}{OT1}{cmr}{bx}{n}
\DeclareSymbolFont{CMlet}{OML}{cmm}{m}{it}
\SetSymbolFont{CMlet}{bold}{OML}{cmm}{b}{it}
\DeclareSymbolFont{CMSy}{OMS}{cmsy}{m}{n}
\SetSymbolFont{CMSy}{bold}{OMS}{cmsy}{b}{n}
\DeclareSymbolFont{AMSa}{U}{msa}{m}{n}
\SetSymbolFont{AMSa}{bold}{U}{msa}{m}{n}

\AtBeginDocument{%
    \DeclareMathSymbol{/}{\mathord}{CMop}{"2F}%
    \DeclareMathSymbol{+}{\mathbin}{CMop}{"2B}%
    \DeclareMathSymbol{-}{\mathbin}{CMSy}{"00}%
    \DeclareMathSymbol{=}{\mathrel}{CMop}{"3D}%
    \DeclareMathSymbol{<}{\mathrel}{CMlet}{"3C}%
    \DeclareMathSymbol{>}{\mathrel}{CMlet}{"3E}%
    \DeclareMathSymbol{\leqslant}{\mathrel}{AMSa}{"36}%
    \DeclareMathSymbol{\geqslant}{\mathrel}{AMSa}{"3E}%
    \DeclareMathSymbol{\lesssim}{\mathrel}{AMSa}{"2E}%
    \DeclareMathSymbol{\gtrsim}{\mathrel}{AMSa}{"26}%
    \DeclareMathSymbol{\sim}{\mathrel}{CMSy}{"18}%
    \DeclareMathSymbol{\pm}{\mathrel}{CMSy}{"06}%
    \DeclareMathSymbol{\approx}{\mathrel}{CMSy}{"19}%
    \DeclareMathSymbol{\times}{\mathbin}{CMSy}{"02}%
    \DeclareMathSymbol{\Delta}{\mathalpha}{CMop}{1}%
    \DeclareMathSymbol{\Omega}{\mathalpha}{CMop}{10}%
    \DeclareMathSymbol{\alpha}{\mathalpha}{CMlet}{11}%
    \DeclareMathSymbol{\beta}{\mathalpha}{CMlet}{12}%
    \DeclareMathSymbol{\delta}{\mathalpha}{CMlet}{14}%
    \DeclareMathSymbol{\mu}{\mathalpha}{CMlet}{22}%
    \DeclareMathSymbol{\sigma}{\mathalpha}{CMlet}{27}%
}

\makeatletter
    \patchcmd{\NAT@citex} 
        {\@citea\NAT@hyper@{%
            \NAT@nmfmt{\NAT@nm}%
            \hyper@natlinkbreak{\NAT@aysep\NAT@spacechar}{\@citeb\@extra@b@citeb}%
            \NAT@date}}
        {\@citea
        \NAT@nmfmt{\NAT@nm}%
        \NAT@aysep\NAT@spacechar
        \NAT@hyper@{\NAT@date}}%
        {}{}
    \patchcmd{\NAT@citex} 
        {\@citea\NAT@hyper@{%
            \NAT@nmfmt{\NAT@nm}%
            \hyper@natlinkbreak{\NAT@spacechar\NAT@@open\if*#1*\else#1\NAT@spacechar\fi}%
            {\@citeb\@extra@b@citeb}%
            \NAT@date}}
        {\@citea
        \NAT@nmfmt{\NAT@nm}%
        \NAT@spacechar\NAT@@open\if*#1*\else#1\NAT@spacechar\fi
        \NAT@hyper@{\NAT@date}}%
        {}{}
\makeatother

\newcommand{\CC}{C\nolinebreak\hspace{-.05em}\raisebox{.4ex}{\tiny\bf +}\nolinebreak\hspace{-.10em}\raisebox{.4ex}{\tiny\bf +}}
\def\CC{{C\nolinebreak[4]\hspace{-.05em}\raisebox{.4ex}{\tiny\bf ++}}}

\newlength{\licenseiconwidth}
\newlength{\licenseicongap}
\newlength{\licenseindent}
\newcommand{\licensebox}{%
  \begin{figure}[!b]
    \footnotesize\linespread{1.2}\selectfont
    \setlength{\parindent}{0pt}
    \begingroup
      \parshape=3
        \licenseindent \dimexpr\columnwidth-\licenseindent\relax
        \licenseindent \dimexpr\columnwidth-\licenseindent\relax
        0pt \columnwidth
      \noindent
      \makebox[0pt][l]{%
        \hspace*{-\licenseindent}%
        \smash{%
          \raisebox{-1.1\baselineskip}[0pt][0pt]{%
            \includegraphics[width=\licenseiconwidth]{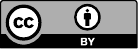}%
          }%
        }%
      }%
      Original content from this work may be used under the terms of the \href{https://creativecommons.org/licenses/by/4.0/}{Creative Commons Attribution 4.0 licence}. Any further distribution of this work must maintain attribution to the author(s) and the title of the work, journal citation and DOI.\par
    \endgroup
  \end{figure}
}

\journalinfo{The Open Journal of Astrophysics}
\historyline{Received 2026 July 23}
\shortauthors{Lawlor-Forsyth et al.}
\defcitealias{lawlor2026a}{Paper~I}
\defcitealias{lawlor2026b}{Paper~II}

\newcommand{\rotacp}{(ROTAC et al.~\citeyear{rotac2025})}
\newcommand{\rotact}{ROTAC et al.~\citeyear{rotac2025}}

\begin{document}

\title{Identifying and Distinguishing Quenching Galaxies with Spatially Resolved Star Formation in the Hubble Frontier Fields\vspace{-1.5cm}}

\author{Cameron~Lawlor-Forsyth$^{\text{\color{blue}1,2}}$\orcidlink{0000-0002-2958-0593}}
\author{Michael~L.~Balogh$^{\text{\color{blue}1,2}}$\orcidlink{0000-0003-4849-9536}}
\author{Sean~L.~McGee$^{\text{\color{blue}3}}$\orcidlink{0000-0003-3255-3139}}
\author{Gregory~H.~Rudnick$^{\text{\color{blue}4}}$\orcidlink{0000-0001-5851-1856}\vspace{0.1cm}}

\affiliation{$^{\text{1}}$Department of Physics and Astronomy, University of Waterloo, Waterloo, ON N2L 3G1, Canada; \href{mailto:clawlorforsyth@uwaterloo.ca}{clawlorforsyth@uwaterloo.ca}}
\affiliation{$^{\text{2}}$Waterloo Centre for Astrophysics, University of Waterloo, Waterloo, ON N2L 3G1, Canada}
\affiliation{$^{\text{3}}$School of Physics and Astronomy, University of Birmingham, Birmingham B15 2TT, UK}
\affiliation{$^{\text{4}}$Department of Physics and Astronomy, University of Kansas, Lawrence, KS 66045, USA}

\begin{abstract}
We investigate the nature and prevalence of different quenching signatures for 1437 galaxies with $M_{*} \geqslant 10^{8}~\mathrm{M}_{\odot}$ and $\text{SFR} \geqslant 10^{-3}~M_{\odot}~\text{yr}^{-1}$ in the Hubble Frontier Fields through spatially resolved spectral energy distribution fitting with \texttt{FAST++}. We use the morphological metrics previously presented in our series of investigations to quantify the distribution of star formation, and use a $k$-nearest neighbors algorithm to classify quenching galaxies into different quenching pathways, including an inside-out pathway and an outside-in pathway. We find 129 galaxies have morphologies consistent with an inside-out quenching pathway, and 70 are consistent with an outside-in pathway. Inside-out quenching galaxies are $0.8^{+0.2}_{-0.1}~\text{dex}$ more massive compared to outside-in quenching galaxies, where both populations are more massive in clusters than the field, by $0.8^{+0.3}_{-0.1}~\text{dex}$. Inside-out quenching galaxies are found more often in clusters (106/129), compared to outside-in quenching galaxies (25/70). In clusters, the fraction of inside-out quenching galaxies strongly increases with mass, representing ${\sim}$30\% of the non-quenched galaxy population at high masses. A milder evolution is seen in the field. The fraction of outside-in quenching galaxies is independent of mass, representing ${\lesssim}$10\% of the non-quenched population, in both the cluster and field. We find no strong dependence for the fraction of any population with estimated infall time, except massive inside-out quenching galaxies, which increase in fraction with increasing infall time.
\end{abstract}

\keywords{\href{http://astrothesaurus.org/uat/584}{Galaxy clusters (584)}; \href{http://astrothesaurus.org/uat/594}{Galaxy evolution (594)}; \href{http://astrothesaurus.org/uat/2040}{Galaxy quenching (2040)}; \href{http://astrothesaurus.org/uat/597}{Galaxy groups (597)}; \href{http://astrothesaurus.org/uat/621}{Galaxy stellar content (621)}; \href{http://astrothesaurus.org/uat/2016}{Quenched galaxies (2016)}}

\maketitle

\section{Introduction}
\setcounter{footnote}{4}

\licensebox

Galaxy quenching is a central unsolved problem in the field of astrophysics \citep[e.g.,][]{tumlinson2017}, given the complex physical mechanisms and associated length- and timescales involved. Understanding how galaxies quench, or have their star formation suppressed, is key to understanding galaxy evolution, as quenching is the primary driver for galaxies moving from the star forming blue cloud to the red sequence \citep{bell2004,faber2007,martin2007}. Much progress has been made in our understanding of quenching, including considerations related to the roles of mass and environment for low-mass and high-mass galaxies \citep[e.g.,][]{peng2010,zhang2021}, for central and satellite galaxies \citep[e.g.,][]{peng2012,davies2019,bluck2020b}, as well as at low- \citep[e.g.,][]{gomez2003,peng2010,wetzel2012} and high-redshift \citep[e.g.,][]{muzzin2012,balogh2016,nantais2017,vanderburg2020}. However, there remain results that are currently unexplained by formation and evolutionary paradigms, including the large body of evidence of massive quenched galaxies at early cosmic times \citep[$ z > 3$; e.g.,][]{carnall2023a,carnall2024,alberts2024,bluck2024,nanayakkara2024} as discovered with the James Webb Space Telescope \citep[JWST;][]{gardner2006,gardner2023}, for example. Other poorly constrained phenomena include timescales associated with quenching \citep[e.g.,][]{schawinski2014,smethurst2015,bremer2018,phillipps2019}, as well as the role of mergers \citep[e.g.,][]{birnboim2007,peng2014,ellison2022} and the circumgalactic medium \citep[e.g.,][]{davies2020}. Though large cosmological simulations have recently proved integral in better understanding and testing different quenching mechanisms \citep[e.g.,][]{donnari2019,donnari2021a,donnari2021b,wright2019,nelson2021,wang2023b,lawlor2026a}, large observational samples of quenched galaxies remain crucial as input for subsequent generations of simulations \citep[e.g.,][]{vogelsberger2020,crain2023,feldmann2026}, as well as for studying quenching in the real Universe.

Quenching mechanisms in the literature are commonly divided into internal and external regimes \citep[e.g.,][]{peng2010,peng2012}, where the internal mechanisms are driven by the galaxy itself, while the external mechanisms consider the environment and surroundings in which a galaxy is embedded. Internal mechanisms include the effects of active galactic nucleus feedback \citep[e.g.,][]{dimatteo2005,bower2006,croton2006,mcnamara2007,mcnamara2012,fabian2012}, stellar feedback \citep[e.g.,][]{cole2000,su2019}, and secular evolution from disk and bar instabilities \citep[e.g.,][]{martig2009,gensior2020}. External quenching mechanisms driven by the environment include ram pressure stripping \citep[e.g.,][]{gunn1972,abadi1999,poggianti2017,boselli2022} and starvation$/$strangulation \citep[e.g.,][]{larson1980,kawata2008,peng2015}. Gravitational harassment is also a potential driver for environmental quenching \citep[e.g.,][]{moore1996,moore1998,moore1999}, in addition to mergers \citep[e.g.,][]{ellison2022,ellison2024}, though the extent to which mergers are important for galaxy quenching remains debated \citep[e.g.,][]{birnboim2007,moustakas2013,weigel2017,quai2021}.

As well, it has been suggested that quenching can partially be described using a model where high-mass central galaxies quench first in the center, before this quenching proceeds outward, producing an ``inside-out'' signature \citep[e.g.,][]{tacchella2015,tacchella2016,tacchella2018,gonzalez2016,abdurrouf2018,belfiore2018,rowlands2018b,guo2019,lin2019,bluck2020b,nelson2021,mcdonough2023,mcdonough2025,lawlor2026a}. Conversely, for low-mass satellite galaxies, quenching first occurs on the outskirts of the galaxy and proceeds inward, producing an ``outside-in'' signature \citep[e.g.,][]{bluck2020b,wang2023b,mcdonough2025,lawlor2026a}. These inside-out and outside-in explanations chiefly consider feedback from an active galactic nucleus for the former, and environmental effects such as ram pressure stripping and starvation$/$strangulation for the latter \citep[e.g.,][]{bluck2020b,nelson2021,mcdonough2025}. However, such a model does not account for other objects such as high-mass satellites, though it has been suggested that these objects are predominantly quenching inside-out, with minor outside-in effects \citep[e.g.,][]{bluck2020b,mcdonough2025}.

In addition, many of the physical mechanisms that quench galaxies act over a range of length- \citep[e.g.,][]{peng2010,bluck2020a} and timescales \citep[e.g.,][]{wetzel2013,schawinski2014,wright2019}, highlighting the need for detailed observations that can spatially resolve the young stellar populations that are suppressed through the act of quenching. Naturally, integral field unit observations have therefore been used to study quenching phenomena \citep[e.g.,][]{gonzalez2015,tacchella2015,goddard2017,schaefer2017,sanchez2018,lin2019}. As well, studies using H$\alpha$ to directly probe the extent of star formation over short timescales have been completed \citep[e.g.,][]{koopmann2004a,koopmann2004b,gavazzi2013,matharu2021,matharu2022}. Beyond these investigations, studies leveraging high-resolution photometry with spectral energy distribution fitting have been completed \citep[e.g.,][]{abdurrouf2018,morselli2019,nelson2021,abdurrouf2022a,abdurrouf2023,tan2022}. These spatially resolved photometric studies are compelling, given their scalability and the number of galaxies that will be observed in large-scale surveys in the coming years \citep[e.g.,][]{lsst2009,euclid2022,rotac2025}.

Thus, to better understand the nature of quenching, the various physical mechanisms, evolutionary pathways, and associated length- and timescales, samples of spatially resolved galaxies are critical. We then require a way to efficiently identify different quenching modes in such observations, to first facilitate direct investigations, and secondly to highlight cases for future follow-up. In \citet[][hereafter \citetalias{lawlor2026a}]{lawlor2026a}, we used the highest resolution run of the IllustrisTNG simulation \citep{nelson2018,pillepich2018b,springel2018}, TNG50 \citep{nelson2019b,pillepich2019}, to develop observationally motivated metrics that are based on the spatial distribution and morphology of star formation in the simulated galaxies. We showed that these metrics can be used to distinguish simulated quenching galaxies from simulated star forming galaxies on the basis of morphology alone. Further, we showed that simulated quenching galaxies can be additionally separated into different quenching evolutionary signatures, including an inside-out morphological signature, and an outside-in signature. We found that the inside-out population was often higher mass field galaxies, while the outside-in population was commonly lower mass satellite galaxies. These populations additionally experienced different quenching timescales, where the outside-in population took ${\sim}~1.5~\text{Gyr}$ to quench, while the inside-out class took ${\sim} 2.5~\text{Gyr}$. 

In \citetalias{lawlor2026a}, we limited our analysis to simulated galaxies, leaving an application to real galaxies for future work. However, it was still necessary to evaluate how these metrics, as developed for simulated galaxies, would perform on real data. As an intermediate step, in \citet[][hereafter \citetalias{lawlor2026b}]{lawlor2026b}, we created mock observations of the simulated galaxies from \citetalias{lawlor2026a}. These mock observations were designed to mimic data from upcoming galaxy surveys \citep{cote2025,rotac2025}, but are comparable in quality and wavelength coverage to those available with the Hubble Space Telescope (HST). By performing spatially resolved spectral energy distribution fitting, we found that we were able to recover key parameters like stellar mass and star formation rate. Using these physical quantities, we were then able to recover the morphological metrics as determined in \citetalias{lawlor2026a} to high fidelity, suggesting that this methodology was mature for application to samples of real galaxies. Therefore, we now look to apply our procedure to galaxies in the Hubble Frontier Fields \citep{lotz2017} to identify and investigate different quenching signatures.

{\renewcommand{\arraystretch}{1.5}
\begin{deluxetable*}{lcccccccccc}
    \tablecaption{The Hubble Frontier Fields.\label{tab:HFF}}[t]
    \tablehead{\colhead{Cluster} & \colhead{RA\tablenotemark{a}} & \colhead{Dec\tablenotemark{a}} & \colhead{$z_{\text{clus}}$} & \colhead{$\sigma$} & \colhead{$R_{200}$} & \colhead{Parallel RA\tablenotemark{a}} & \colhead{Parallel Dec\tablenotemark{a}} & \colhead{Reference} \\[-0.1cm]
    \colhead{} & \colhead{(h m s)} & \colhead{(d m s)} & \colhead{} & \colhead{($\text{km s}^{-1}$)} & \colhead{(Mpc)} & \colhead{(h m s)} & \colhead{(d m s)}}
    \startdata
    Abell~2744                   & 00:14:21.2 & $-$30:23:50.1 & 0.308 & 1497 & 2.35 & 00:13:53.6 & $-$30:22:54.3 & $z$ [1], $\sigma$ [2], $R_{200}$ [3] \\
    Abell~S1063\tablenotemark{b} & 22:48:44.4 & $-$44:31:48.5 & 0.348 & 1840 & 2.38 & 22:49:17.7 & $-$44:32:43.8 & $z$ [4, 5], $\sigma$ [6], $R_{200}$ [7] \\
    Abell~370                    & 02:39:52.9 & $-$01:34:36.5 & 0.375 & 1170 & 2.66 & 02:40:13.4 & $-$01:37:32.8 & $z$ [8, 9], $\sigma$ [10], $R_{200}$ [11] \\
    MACS~J0416.1$-$2403          & 04:16:08.9 & $-$24:04:28.7 & 0.396 &  955 & 1.89 & 04:16:33.1 & $-$24:06:48.7 & $z$ [12, 13], $\sigma$ [14], $R_{200}$ [7] \\
    MACS~J1149.5$+$2223          & 11:49:36.3 & $+$22:23:58.1 & 0.543 & 1840 & 2.35 & 11:49:40.5 & $+$22:18:02.3 & $z$ [15], $\sigma$ [15], $R_{200}$ [7] \\
    MACS~J0717.5$+$3745          & 07:17:34.0 & $+$37:44:49.0 & 0.545 & 1660 & 2.36 & 07:17:17.0 & $+$37:49:47.3 & $z$ [15, 16], $\sigma$ [15], $R_{200}$ [7] \vspace{0.1cm}
    \enddata
    \tablenotetext{a}{Cluster and parallel positions taken from \citet{lotz2017}.}
    \tablenotetext{b}{Also known as RXC~J2248.7$-$4431.}
    \tablerefs{1. \citet{couch1984}; 2. \citet{owers2011}; 3. \citet{medezinski2016}; 4. \citet{bohringer2004}; 5. \citet{gomez2012}; 6. \citet{gomez2012}; 7. \citet{umetsu2016}; 8. \citet{kristian1978}; 9. \citet{struble1991}; 10. \citet{dressler1999}; 11. \citet{umetsu2011}; 12. \citet{mann2012}; 13. \citet{ebeling2014}; 14. \citet{jauzac2014}; 15. \citet{ebeling2007}; 16. \citet{edge2003}}
\end{deluxetable*}}

This paper is structured as follows: in Section~\ref{sec:data}, we describe the available high-resolution processed data, and data preparation and reduction steps. In Section~\ref{sec:methods}, we describe our analysis procedure for extracting physical measurements, as well as the morphological metrics as introduced in \citetalias{lawlor2026a} and \citetalias{lawlor2026b} for classifying our sample into different quenching evolutionary signatures. In Section~\ref{sec:results}, we present our results, in particular focusing on the differences between the quenching populations when considering stellar mass and environmental influences like location in phase space and phase space-inferred infall time. In Section~\ref{sec:discussion}, we discuss our results, the implications of our findings for upcoming large-scale galaxy surveys, and next steps. Finally, in Section~\ref{sec:summary}, we present our summary and conclusion.

Throughout this paper we adopt a flat $\Lambda$CDM cosmology that is consistent with the TNG simulation \citep{weinberger2017,pillepich2018a}, based on the Planck intermediate results \citep{planck2016}: $H_{0} = 67.74~\text{km s}^{-1}~\text{Mpc}^{-1}$, $\Omega_{\text{m}} = 0.3089$, $\Omega_{\Lambda} = 0.6911$, and $\Omega_{\text{b}} = 0.0486$.

\newpage
\section{Data}\label{sec:data}

\subsection{Hubble Frontier Fields}\label{subsec:HFF}

The Hubble Frontier Fields\footnote{\href{https://frontierfields.org}{https:$//$frontierfields.org}}\textsuperscript{,}\footnote{\href{https://archive.stsci.edu/prepds/frontier}{https:$//$archive.stsci.edu$/$prepds$/$frontier}} \citep{lotz2017} are a sample of six massive lensing clusters drawn from the Abell catalog \citep{abell1958,abell1989} and the Massive Cluster Survey \citep{ebeling2001,ebeling2007}: Abell~2744, Abell~S1063 (also known as RXC~J2248.7$-$4431), Abell~370, MACS~J0416.1$-$2403 (hereafter MACS~J0416), MACS~J0717.5$+$3745 (hereafter MACS~J0717), and MACS~J1149.5$+$2223 (hereafter MACS~J1149). The survey includes imaging of each cluster along with adjacent (or parallel) fields offset from the clusters \citep{lotz2017}. Properties of the clusters and their parallel fields, including the cluster redshifts, velocity dispersions, and radii within which the mean density of the cluster halo is 200 times the critical density of the Universe at the time of observation, $R_{200}$, are tabulated in Table~\ref{tab:HFF}.

The Frontier Fields program builds on the success of the Cluster Lensing And Supernova Survey with Hubble \citep[CLASH;][]{postman2012}, that included four of the six selected clusters. At the time of their completion, the Frontier Fields represented some of the deepest observations ever conducted with HST, totaling more than 840 orbits across the six clusters and their parallel fields, reaching $5 \sigma$ point source depths of ${\sim} 29~\text{mag}_{\text{AB}}$ in the parallel fields \citep{lotz2017}. Over small portions of the cluster pointings, the imaging reaches depths of ${\sim}30$--$33~\text{mag}_{\text{AB}}$ given the lensing nature of the clusters \citep{lotz2017}. As part of the original survey design, each of the clusters and their parallel fields were imaged with seven filters using the Advanced Camera for Surveys Wide Field Channel \citep[ACS$/$WFC; e.g.,][]{stark2025} and the Wide Field Camera 3 near-infrared channel \citep[WFC3$/$IR; e.g.,][]{marinelli2025}. In addition to this, imaging in up to ten other filters\footnote{We use \textit{UV}, \textit{UV'}, \textit{u}, \textit{C}, \textit{B}, \textit{g}, \textit{V}, \textit{V}$^{\text{W}}$, \textit{r}, \textit{i}, \textit{I}, \textit{z}, \textit{Y}, \textit{Y}$^{\text{W}}$, \textit{J}, \textit{JH}, and \textit{H} to refer to the HST F225W, F275W, F336W, F390W, F435W, F475W, F555W, F606W, F625W, F775W, F814W, F850LP, F105W, F110W, F125W, F140W, and F160W filters, respectively, consistent with \citet{marinelli2025} and \citet{stark2025}.}  exists from the Wide Field Camera 3 ultraviolet channel \citep[WFC3$/$UVIS; e.g.,][]{postman2012} and other ACS$/$WFC and WFC3$/$IR observations \citep[e.g.,][]{shipley2018}. Filter transmission curves for the complete filter set are shown in Figure~\ref{fig:filters}. In Appendix~\ref{app:filters} we tabulate the filter set available per cluster$/$parallel field.

\begin{figure*}
    \centering
    \includegraphics[width=\textwidth]{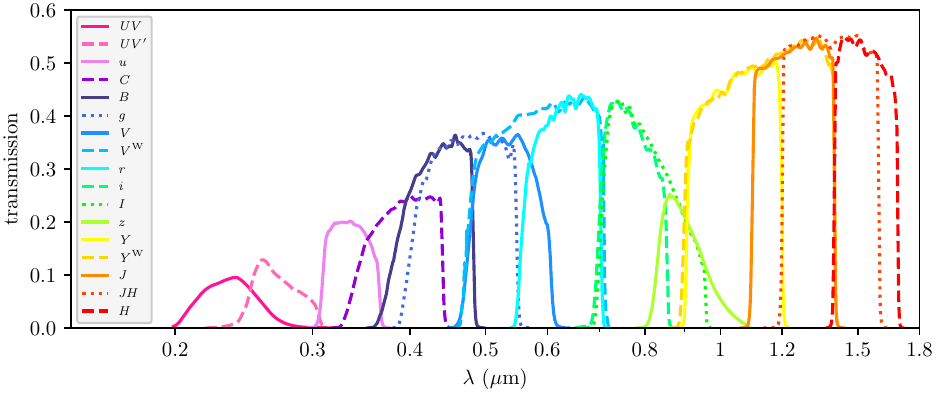}
    \caption{HST filter transmission curves for the Frontier Fields bands across all clusters and parallel fields. Not every band is available for every pointing (i.e. cluster or parallel field). See Appendix~\ref{app:filters} for full details regarding which filters are available per pointing.}
    \label{fig:filters}
\end{figure*}

Following the work completed by \citet{lotz2017}, several groups including the AstroDeep team\footnote{\href{http://www.astrodeep.eu/frontier-fields}{http:$//$www.astrodeep.eu$/$frontier-fields}}\textsuperscript{,}\footnote{\href{http://www.astrodeep.eu/ff34/}{http:$//$www.astrodeep.eu$/$ff34}}\textsuperscript{,}\footnote{\href{https://archive.stsci.edu/hlsp/hffcatalogs}{https:$//$archive.stsci.edu$/$hlsp$/$hffcatalogs}} \citep{castellano2016a,merlin2016,dicriscienzo2017,bradac2019}, the DeepSpace team \citep{shipley2018,nedkova2021}, and those involved in the Beyond Ultra-deep Frontier Fields and Legacy Observations\footnote{\href{https://archive.stsci.edu/hlsp/buffalo}{https:$//$archive.stsci.edu$/$hlsp$/$buffalo}} \citep[BUFFALO;][]{steinhardt2020,niemiec2023,pagul2024} have produced extended photometric and redshift catalogs for the Frontier Fields. More recently, several groups have completed large programs targeting the Frontier Fields with JWST \citep[e.g.,][]{treu2022,willott2022,windhorst2023,bezanson2024,suess2024,atek2025,sarrouh2026}; we discuss these additional data in Section~\ref{sec:discussion}.

\subsubsection{DeepSpace Catalogs and Data}

The DeepSpace team\footnote{\href{https://cosmos.phy.tufts.edu/~danilo/HFF/Home.html}{https:$//$cosmos.phy.tufts.edu$/{\sim}$danilo$/$HFF$/$Home.html}} \citep{shipley2018,nedkova2021} have produced science-ready mosaics for the Frontier Fields, including the clusters and their parallel fields, and associated 
multi-wavelength photometric, redshift, and stellar population synthesis catalogs. This dataset is described fully in \citet{shipley2018}. These catalogs include photometric fluxes and uncertainties, photometric redshifts determined with EAzY \citep{brammer2008}, spectroscopic redshifts from the literature when available, flux radii that describe circular apertures that enclose half the total flux \citep[determined with \textsc{sextractor};][]{bertin1996}, and various quality flags. These catalogs also include physical parameters like integrated stellar mass and star formation rate (SFR) derived through spectral energy distribution (SED) fitting using FAST \citep{kriek2009}. In their SED fitting, \citet{shipley2018} employed a simple delayed exponential (or ``tau'') model for their star formation history along with a fixed solar metallicity ($Z = 0.02$).

\begin{figure*}
    \centering
    \includegraphics[width=\textwidth]{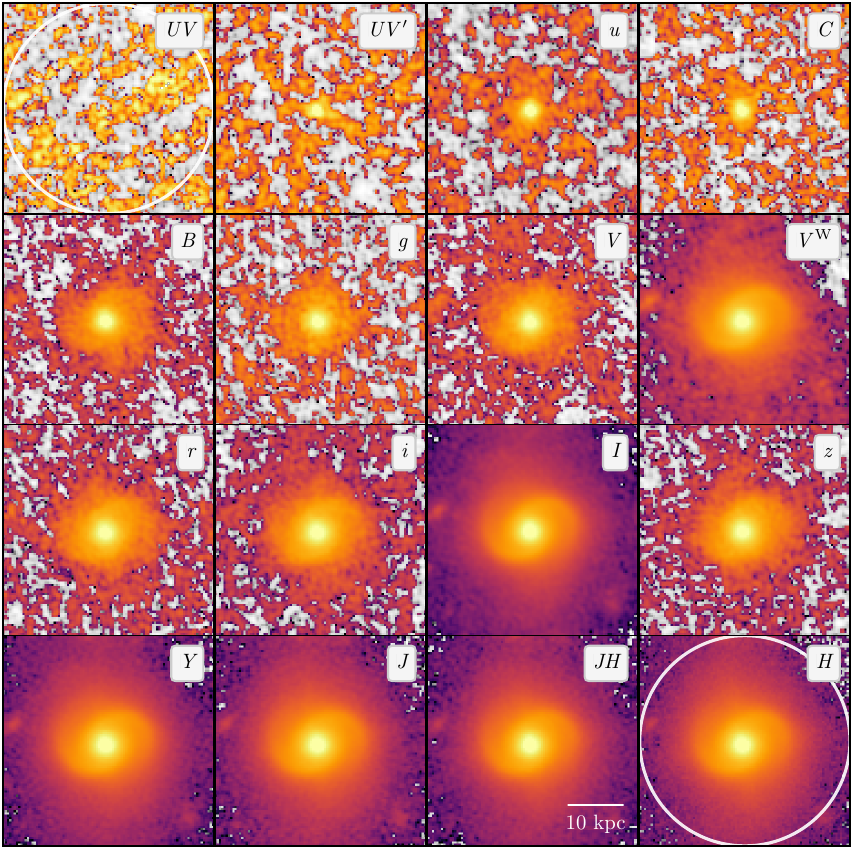}
    \caption{Example cutouts for galaxy ID~5809 in MACS~J0717 ($z_{\text{spec}} = 0.546$, $M_{*} = 10^{10.44}~M_{\odot}$, $\text{SFR} = 0.36~M_{\odot}~\text{yr}^{-1}$). In each cutout, pixels with negative values are shown in a log-grayscale, while positive pixels are shown in color with a log scale. Each cutout uses a per-cutout scaling, meaning a universal scaling for all cutouts has not been applied. Cutouts are arranged according to increasing wavelength, from left to right and top to bottom. Each cutout is labeled in the top right corner with the corresponding filter. Inscribed within the \textit{UV}- and \textit{H}-band cutouts are circles denoting the extent of $5~R_{\text{e}}$ (see Section~\ref{subsec:annular_phot}), for visual reference. A scale bar is additionally shown in the \textit{JH}-band image. This galaxy does not have any quality flag issues per band (Section~\ref{subsec:annular_phot}).}
    \label{fig:cutout_example}
\end{figure*}

The imaging available from \citet{shipley2018} includes science-ready images from the rest-frame ultraviolet to the near-infrared in up to 17 filters, as noted above, where bright cluster galaxies have been modeled and removed. These bright cluster galaxies (bCGs) are distinct from the traditional use of the term ``brightest cluster galaxy'' in the literature to describe the brightest galaxy \citep[commonly a massive elliptical; e.g.,][]{delucia2007,vonderlinden2007,lauer2014} within a cluster. In practice, these bright cluster galaxies all have extended profiles that dominate the majority of light and that will contaminate other nearby galaxies \citep{shipley2018}. The bCG-subtracted science-ready imaging has been point spread function (PSF) matched to the \textit{H}-band imaging, as the \textit{H}-band photometry is the longest wavelength imaging available in the dataset and has the largest PSF (full width at half maximum ${\sim} 0 \farcs 147$\footnote{\href{https://hst-docs.stsci.edu/wfc3ihb/chapter-7-ir-imaging-with-wfc3/7-6-ir-optical-performance}{https:$//$hst-docs.stsci.edu$/$wfc3ihb$/$chapter-7-ir-imaging-with-wfc3$/\\$7-6-ir-optical-performance}}) of the available filters. We show cutouts of the bCG-subtracted PSF-matched science-ready imaging for an example galaxy in Figure~\ref{fig:cutout_example}. In addition, weight maps, bCG models, point spread functions, and segmentation maps are also available \citep{shipley2018}. However, though the bCG-subtracted images are science-ready for non-bCG galaxies, science-ready images including the bCGs are not available, and so must be prepared separately. As well, the weight maps and bCG model images are not PSF-matched. Further, the segmentation maps were similarly produced after the bCGs had been removed \citep{shipley2018}, and so bCG-only segmentation maps must also be generated.

\subsection{Data Preparation}

In order to prepare the data for analysis, we begin by PSF-matching the weight (inverse variance) maps and model images through convolution with the available point spread function images, as described earlier \citep[i.e.,][]{shipley2018}. As above, we PSF-match to the \textit{H}-band image per pointing. We additionally determine dimensionless correction$/$calibration factors for subsequent use when determining the per-pixel noise with the variance maps. We determine these correction factors by calculating the empirically-measured root mean square (RMS) value of blank sky regions in the science images and the predicted RMS from the PSF-matched variance maps for the same pixels. The ratio of the empirical RMS to the predicted RMS provides the global correction factor that will be used subsequently to rescale the variance maps on a per filter, per pointing basis.

Given that the imaging produced by \citet{shipley2018} includes bCG-subtracted science-ready images for the non-bCGs, we must account for the bCGs themselves, as they are not present in the existing segmentation maps. We therefore generate segmentation maps for the bCGs from the model fits to these galaxies, in order to maintain consistency with how the photometry of fainter galaxies is defined. Since detection and segmentation algorithms require a noise threshold, we add random Gaussian noise based on the median RMS from the inverse weight map to the model images.

With these noisy model images, we next create co-added bCG-only detection images for segmentation mapping. We follow \citet{shipley2018} and use the \textit{I}-, \textit{Y}-, \textit{J}-, \textit{JH}-, and \textit{H}-bands to create the bCG-only detection image by first combining the model images into a mean mosaic, weighted by the corresponding weight images. We additionally create a total weight image by combining the weight images \citep{shipley2018}. We produce a final bCG model detection image by dividing the weighted mean mosaic with the total error image, which is the square root of the total weight image, consistent with \citet{shipley2018}. Using these detection images we then produce bCG-only segmentation maps with \texttt{Photutils} \citep{bradley2025}.

When creating photometric cutouts, we start with the PSF-matched bCG-subtracted science imaging for non-bCGs, and add the PSF-matched noiseless bCG models for bCGs. We create noise cutouts from the science images and PSF-matched variance images according to
\begin{equation*}
    \text{noise} = \sqrt{\text{var} \times \text{corr}^2 + \text{sci}/t_{\text{exp}}},
\end{equation*}
where ``var'' is the variance map, ``corr'' is the correction$/$calibration as determined above, ``sci'' is the science image, and $t_{\text{exp}}$ is the exposure time, in seconds. We additionally create segmentation map cutouts for all galaxies and bCG-segmentation map cutouts for the bCGs.

\subsection{Sample Selection}\label{subsec:sample}

We next determine a galaxy sample, including bCGs, based on the existing catalogs from \citet{shipley2018}. We begin by creating a single catalog that includes every source in the \citet{shipley2018} catalogs, numbering 81,315 sources. We additionally adopt select attributes from the catalogs of \citet{pagul2024}, who produced photometric and redshift catalogs for the Frontier Fields through the BUFFALO survey \citep{steinhardt2020}. We adopt the spectroscopic redshifts as tabulated by \citet{pagul2024}, collected from the literature \citep{lagattuta2019,treu2022} and the NASA$/$IPAC Extragalactic Database\footnote{\href{https://ned.ipac.caltech.edu}{https:$//$ned.ipac.caltech.edu}}.

We then form an initial sample based on redshift, integrated stellar mass, and quality flags, as described further below. Given our analysis requirements, forming a final galaxy sample requires multiple steps that will be outlined below. Where available, we use spectroscopic redshifts in place of photometric redshifts, but use photometric redshifts otherwise. We select galaxies with $M_{*} \geqslant 10^{8}~M_{\odot}$, $z \geqslant 0.14$, and $z \leqslant 0.67$ from the \citet{shipley2018} catalogs. The redshift limits are chosen to span the range of redshifts for cluster members, as defined in Section~\ref{subsubsec:z_limits}. We additionally require $\delta z < 0.2$ for the sources that only have photometric redshifts. Further, we require that all sources have \texttt{use\_phot = 1} \citep[as introduced by][]{skelton2014}, that selects ``OK'' sources in a consistent fashion \citep{shipley2018}. We also require that the sources are extended (\texttt{star\_flag $\neq$ 1}) and that there are no problematic \textit{H}-band flags (\texttt{flag\_F160W $\geqslant$ 0}), that ensures that the \textit{H}-band photometry is not aligned with a region where the weight map is negative \citep{shipley2018}. These combined cuts produce a catalog of 3643 sources.

Next, we impose an \textit{H}-band signal-to-noise ratio (SNR) floor for the sources. Our analysis is based on photometry in 20 circular annuli, as described in Section~\ref{subsec:annular_phot}. We require \textit{H}-band $\text{SNR} \geqslant 3$ for all annuli to ensure that reliable stellar mass estimates can be determined when completing SED fitting (see below), resulting in 1987 sources. Furthermore, we conservatively exclude another 48 sources that are partially masked \citep{shipley2018} or have other visually identified artifacts (as described in Section~\ref{subsec:annular_phot}), resulting in a sample of 1939 galaxies. We also exclude 502 galaxies with low integrated SFR as determined from SED fits to the total photometry (see Section~\ref{subsec:fitting}), and thus have a sample of 1437 galaxies. We consider these 1437 galaxies to be our non-quenched ``primary'' sample, while the 502 galaxies with low global SFR forms our quenched extended sample (see below).

\subsubsection{Redshift Limits}\label{subsubsec:z_limits}

We seek global limits on redshift, informed by redshift limits for each cluster. In their work, \citet{shipley2018} compared the scatter ($\sigma$) of photometric redshifts with known spectroscopic redshifts on a per pointing (i.e. cluster$/$parallel field) basis, producing normalized median absolute deviations \citep[$\sigma_{\text{NMAD}}$;][]{beers1990} per cluster. Motivated by \citet{morishita2017}, we follow \citet{nedkova2021} and adopt the normalized median absolute deviations for the Frontier Fields, and use these as redshift limits to determine the global upper and lower limits for the sample. These limits are defined as \citep{nedkova2021}:
\begin{equation*}
    3 \times \sigma_{\text{NMAD}} \geqslant \frac{z \pm z_{\text{clus}}}{1 + z_{\text{clus}}},
\end{equation*}
where $z$ represents the global upper$/$lower redshifts of the sample, depending on the sign in the numerator. Using the lowest redshift cluster and its associated normalized median absolute deviation (Abell~2744; $z_{\text{clus}} = 0.308$, $\sigma_{\text{NMAD}} = 0.043$), we find $z_{\text{min}} = 0.14$. Following a similar procedure for the highest redshift cluster$/$normalized median absolute deviation combination (MACS~J1149; $z_{\text{clus}} = 0.543$, $\sigma_{\text{NMAD}} = 0.027$), we find $z_{\text{max}} = 0.67$. We use these redshift limits to select our sample as described above. For galaxies beyond the extent of each respective cluster (see Section~\ref{subsubsec:phase_space}), these limits additionally define a foreground and background field sample.

\subsubsection{Phase Space-inferred Infall Time}\label{subsubsec:phase_space}

Recently, \citet{dou2025} completed an investigation evaluating the performance of caustics \citep{regos1989,diaferio1997,diaferio1999} and relations from the literature \citep[e.g.,][]{rhee2017,pasquali2019}, in order to accurately estimate the infall times for simulated galaxies into clusters at $z = 0$. They used the TNG300 simulation \citep{nelson2019a} of the IllustrisTNG simulation suite \citep[.e.g,][]{nelson2018,pillepich2018b,springel2018}, and created phase space diagrams \citep[e.g.,][]{diaferio1999,mahajan2011} in order to determine the infall time \citep[e.g.,][]{oman2013,oman2016,rhee2017,pasquali2019} for cluster galaxies based on location in phase space by considering their orbital histories \citep{dou2025}. We adopt the method of \citet{dou2025}, but extend their analysis to all snapshots with redshifts similar to the Frontier Fields ($0.14 < z < 0.67$), and additionally use the TNG-Cluster simulation \citep{nelson2024} in combination with the TNG300 simulation. We therefore adopt the method of \citet{lawlor2026c}, to determine infall times for our Frontier Fields cluster galaxies, based on redshift and location in phase space.

With these considerations, we separate our Frontier Fields galaxies into a cluster and field sample based on position in phase space. We define cluster galaxies to be those with $R_{\text{2D}} \leqslant 0.5 R_{200}$ and $|V_{\text{LoS}}| \leqslant 3.5 \sigma_{\text{LoS}}$. Of the 1437 galaxies that comprise the non-quenched sample (those with $\text{SFR}_{\text{total}} \geqslant 10^{-3}~M_{\odot}~\text{yr}^{-1}$), we find that 723 galaxies reside in clusters, while the remaining 714 reside in the field. Of the 502 galaxies that make up the quenched extended sample, 338 reside in clusters and 164 reside in the field.

\section{Methodology}\label{sec:methods}

\subsection{Annular Photometry and Visual Inspection}\label{subsec:annular_phot}

Following \citetalias{lawlor2026b}, for all galaxies in the sample, we bin the photometry into 20 equally spaced (linearly in radius) concentric circular annuli out to five times the effective radius (also known as the half flux radius), $R_{\text{e}}$ \citep[as tabulated by][]{shipley2018}, for each galaxy. When determining the flux and noise for each annulus, we use the segmentation maps as a mask, where only pixels associated with the galaxy of interest or the background are included in the flux$/$noise determination. For bCGs, the bCG-only segmentation maps are similarly used as a mask. Total noise per annulus is determined by considering the individual noise contributions per pixel, added in quadrature.

We inspect every cutout and compare with the tabulated quality flags from the \citet{shipley2018} catalogs to ensure that our annular bins do not overlap with masked regions and that there are no remaining artifacts. The masked regions in the science-ready images come directly from \citet{shipley2018}, where they masked regions associated with bright stars causing halos and diffraction spikes, bCG model residuals, satellite trails, cosmic rays, and pixels with negative weights \citep{shipley2018}. Where our cutout photometry does overlap with a masked region, we record flags for the affected filters in such cases, to ensure that these bands are not used subsequently when performing SED fitting. After visually inspecting every cutout and recording ``use'' flags for the sources, we find that 48 sources have partially masked regions on their \textit{H}-band cutouts, such that performing our analysis is not possible for these sources. We remove these and subsequently follow the procedure as described above to determine the annular photometry across the 20 radial bins for each source. As stated in Section~\ref{subsec:sample}, our final sample satisfying all our requirements is 1939 galaxies.

\subsection{Spectral Energy Distribution Fitting}\label{subsec:fitting}

We are ultimately interested in applying the morphological metrics introduced in \citetalias{lawlor2026a} and refined in \citetalias{lawlor2026b} to our sample of galaxies. We will discuss our morphological metrics in further detail below (Section~\ref{subsec:metrics}), but for now, this requirement means that our sample galaxies must have non-negligible levels of star formation. To determine this, we perform SED fitting on the integrated photometry. Further, we require SED-fitted physical parameters for the annuli in order to calculate our morphological metrics. We now describe our SED fitting procedure.

We perform spectral energy distribution (SED) fitting using the SED fitting code \texttt{FAST++}\footnote{\href{https://github.com/cschreib/fastpp}{https:$//$github.com$/$cschreib$/$fastpp}} \citep[e.g.,][]{schreiber2018a,schreiber2018b}. \texttt{FAST++} is an updated {\CC} version of the original SED fitting code \texttt{FAST}, as introduced in \citet{kriek2009}. Each parameter involved in describing the star formation history for a target galaxy, from which the spectral energy distribution is derived, is defined on a fixed grid \citep{schreiber2018a}. Adopting the core characteristics of the SED fitting setup from \citetalias{lawlor2026b}, we use the standard and preferred \textsc{galaxev} simple stellar population synthesis model \citep{bruzual2003}, that employ the ``Padova 1994 library'' evolutionary tracks \citep{alongi1993,bressan1993,fagotto1994a,fagotto1994b,fagotto1994c,girardi1996}. We use the updated 2016\footnote{\href{https://www.bruzual.org/bc03/Updated_version_2016}{https:$//$www.bruzual.org$/$bc03$/$Updated\_version\_2016}} version of the \textsc{galaxev} code to achieve better wavelength and metallicity coverage compared to earlier releases: 221 unequally spaced ages from $0$ to $20~\text{Gyr}$; 2023 unequally spaced wavelengths from $91~\Angstrom$ to 36~mm; and a total of seven unequally spaced metallicities between $10^{-4}$ and $0.1$ \citep{bruzual2003}. We additionally use a \citet{chabrier2003} initial mass function along with the dust attenuation law from \citet{calzetti2000}.

\begin{figure*}[t]
    \centering
    \includegraphics[width=\textwidth]{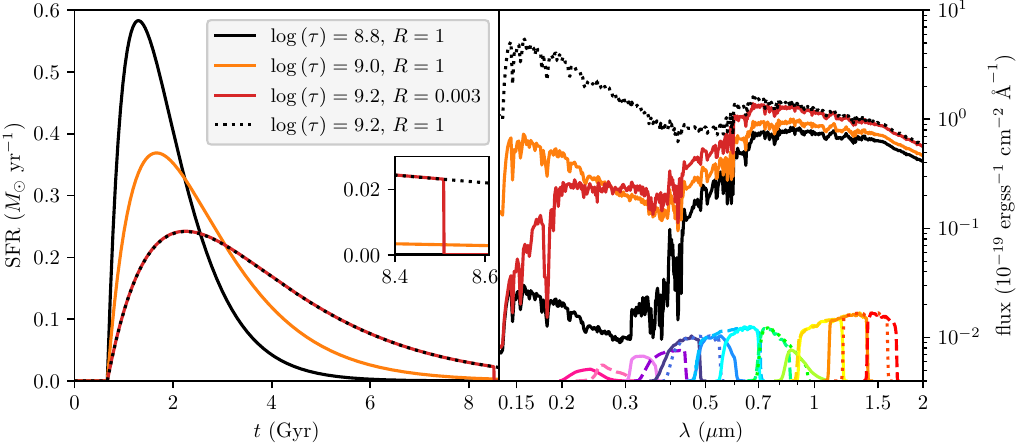}
    \caption{Left: example star formation histories drawn from the parameter space investigated during the integrated fits. These star formation histories show different combinations of the multiplicative factor $R$ (that controls the overall scaling of the star formation history in the final $100~\text{Myr}$) and the $e$-folding time $\tau$. The inset shows a zoom into the last $200~\text{Myr}$ to better highlight the effect of $R$. Right: the resulting spectral energy distributions for the star formation histories on the left, using the same color scheme, without the effect of any foreground dust (i.e. $A_{\text{V}} = 0$). We also show the filter transmission curves for the filter set available in the Frontier Fields.}
    \label{fig:SFH_and_SED}
\end{figure*}

We adopt the star formation history as described in \citetalias{lawlor2026b}: a flexible hybrid star formation history that builds on a delayed-tau model \citep[e.g.,][]{moustakas2013,speagle2014,ciesla2017,carnall2019,johnson2021} until the final $100~\text{Myr}$. At this point, a multiplicative factor, $R$, is introduced, that controls the overall scaling of the star formation history over those final $100~\text{Myr}$ \citep[e.g.,][]{fumagalli2011,ciesla2016,ciesla2018,ciesla2021,schreiber2018b}. This star formation history has the form \citepalias{lawlor2026b}
\begin{equation}\label{eq:sfh}
    \text{SFR}(t) \propto \begin{cases}
        R \times (t - t_{\text{f}}) e^{-(t - t_{\text{f}})/\tau}, & t_{\text{lb}} \leqslant \Delta t_{\text{SF}}, \\
        (t - t_{\text{f}}) e^{-(t - t_{\text{f}})/\tau}, & t_{\text{lb}} > \Delta t_{\text{SF}} \cap t \geqslant t_{\text{f}}, \\
        0, & t < t_{\text{f}},
    \end{cases}
\end{equation}
where $t$ is time, $t_{\text{f}}$ is the formation time of the first stars, $\tau$ is the $e$-folding time of the stellar population, $t_{\text{lb}}$ is the lookback time, and $\Delta t_{\text{SF}} = 100~\text{Myr}$. The various times included in the star formation history are related: $t + t_{\text{lb}} = t_{\text{age}}$, where $t_{\text{age}}$ is the age of the stellar population. Further, $t_{\text{f}} + t_{\text{age}} = t_{0}$, where $t_{0}$ is the current age of the Universe \citepalias{lawlor2026b}. As was described in \citetalias{lawlor2026b}, this star formation history can capture broad evolution of old stellar populations over long timescales while also being capable of describing rapid truncations or bursts of star formation over the defined $100~\text{Myr}$ timescale. We list the parameter grid for all components of the SED fitting procedure in Table~\ref{tab:fitting}, noting that the entire parameter space encompasses 59,535,360 possible combinations. We additionally show example star formation histories and their resulting SEDs for specific values available in our fitting prescription in Figure~\ref{fig:SFH_and_SED}.

{\renewcommand{\arraystretch}{1.5}
\begin{deluxetable}{ccccc}
    \tablecaption{Spectral energy distribution fitting model parameters.\label{tab:fitting}}
    \tablehead{\colhead{Parameter} & \colhead{Minimum} & \colhead{Maximum} & \colhead{Step} & \colhead{$N_{\text{steps}}$}}
    \startdata
    $\log{(t_{\text{age}}/\text{yr})}$ & 8.5    & 10.0 & 0.1  & 16 \\
    $\log{(\tau/\text{yr})}$           & 8.0    & 9.9  & 0.1  & 20 \\
    $\log{(R)}$                         & $-$3.0 & 2.0  & 0.1  & 51 \\
    $A_{\text{V}}$ (mag)                & 0.0    & 1.8  & 0.1  & 19 \\
    $Z$                                 & 0.008  & 0.05 & irr. & 3 \\
    $z$                                 & 0.1    & 0.73 & 0.01 & 64 \\
    \hline
    stellar populations & \multicolumn{4}{c}{\citet{bruzual2003}} \\
    initial mass function & \multicolumn{4}{c}{\citet{chabrier2003}} \\
    dust attenuation & \multicolumn{4}{c}{\citet{calzetti2000}} \vspace{0.1cm}
    \enddata
\end{deluxetable}}

We then perform SED fitting on the integrated photometry for the non-quenched+quenched sample of 1939 galaxies, ensuring that we use the results of our visual inspection and the corresponding ``use'' flags to properly mask any requisite filters from the SED fitting analysis. Consistently with \citetalias{lawlor2026a} and \citetalias{lawlor2026b}, we choose to define SFR as the star formation rate averaged over the final $100~\text{Myr}$, as it is the timescale that is well traced by ultraviolet and infrared observations \citep{kennicutt1998a,madau2014}, and is the duration associated with ``recent'' galaxy-wide star formation \citep{calzetti2013}.

After completing the integrated SED fitting, we find that 502 of the 1939 galaxies have integrated $\text{SFR}_{\text{total}} < 10^{-3}~M_{\odot}~\text{yr}^{-1}$. Based on findings from \citetalias{lawlor2026b}, integrated star formation rates below this level become unreliable for subsequent annular analysis, and so we set a threshold requirement at $10^{-3}~M_{\odot}~\text{yr}^{-1}$ for our galaxies. Removing these 502 galaxies, we arrive at a sample of 1437 galaxies with $M_{*} \geqslant 10^{8}~M_{\odot}$ and $\text{SFR} \geqslant 10^{-3}~M_{\odot}~\text{yr}^{-1}$ that we apply our morphological metrics to (Section~\ref{subsec:metrics}). We additionally find that we recover well the stellar mass estimates from the integrated fitting of \citet{shipley2018}, where the median difference and scatter between their integrated stellar masses and our estimates are $0.01~\text{dex}$ and $0.16~\text{dex}$, respectively.

We next complete SED fitting on the annular photometry (from Section~\ref{subsec:annular_phot}), producing physical properties for the 20 radial bins per galaxy, specifically stellar mass and SFR radial profiles. We show an example set of these profiles in Figure~\ref{fig:radial_profiles}. As well, we sum across our stellar mass and SFR profiles on a per galaxy basis, and compare with the integrated fits determined above. With the summed annular properties, we recover well the integrated stellar masses and SFRs. The median difference and scatter between the integrated stellar masses and the annular sums are $-0.01~\text{dex}$ and $0.10~\text{dex}$, respectively, while the difference and scatter for SFR are $-0.05~\text{dex}$ and $0.30~\text{dex}$, respectively.

\subsection{Star Forming Main Sequence Determination}\label{subsec:SFMS}

We determine each galaxy's location relative to the star forming main sequence \citep[SFMS; e.g.,][]{brinchmann2004,peng2010,speagle2014}. Given our sample size, as well as how the Frontier Fields target clusters that host galaxies with lower specific star formation rates \citep[e.g.,][]{balogh1997,balogh1998,balogh1999,balogh2000,lewis2002,gomez2003,kauffmann2004}, in order to better sample and define the SFMS for normal star forming galaxies at these redshifts, we use the results from the recent data release of the Cosmic Evolution Survey \citep[COSMOS;][]{scoville2007}, that includes JWST data: COSMOS-Web\footnote{\href{https://cosmos2025.iap.fr}{https:$//$cosmos2025.iap.fr}} \citep{casey2023,shuntov2025}. We describe the data and SED fitting procedure employed by the COSMOS-Web team \citep[][]{arangotoro2025,shuntov2025} in Appendix~\ref{app:COSMOS_Web}.

As in \citetalias{lawlor2026a}, we adopt the ``ridge line method'' \citep[e.g.,][]{pillepich2019,donnari2021a,donnari2021b,bottrell2024} introduced in \citet{donnari2019} to determine the location of the SFMS. In this method, the COSMOS-Web galaxies are binned by stellar mass and the median (log) SFR is then calculated for each bin \citep{donnari2019}. We use a bin width of $0.2~\text{dex}$, and limit our analysis to galaxies with $M_{*} \leqslant 10^{10.1}~M_{\odot}$, the mass scale where, roughly, the star forming main sequence deviates from linearity \citep[e.g.,][]{whitaker2012,whitaker2014}. Based on their star formation rates, COSMOS-Web galaxies that are ${>} 1~\text{dex}$ away from the current median SFR are removed from the bin and from further analysis, and the median is then recalculated. This iterative procedure is then complete when the median SFR converges, and these median SFRs (across all bins) are used, along with the bin centers, to perform a linear fit \citep{donnari2019}:
\begin{equation*}
    \langle \log{(\text{SFR}/M_{\odot}~\text{yr}^{-1})} \rangle = \alpha \log{(M_{*}/M_{\odot})} + \beta,
\end{equation*}
where $\langle \log(\text{SFR}) \rangle$ are the median (log) SFRs, and $\alpha$ and $\beta$ are the slope and intercept, respectively. For higher masses, $M_{*} > 10^{10.1}~M_{\odot}$, the star forming main sequence is linearly extrapolated using the fitted $\alpha$ and $\beta$, consistent with \citet{donnari2021b}. We then use the fitted relation from the COSMOS-Web galaxies to classify our Frontier Fields galaxies as being part of the star forming main sequence or the green valley. Motivated by \citet{pillepich2019}, we define SFMS galaxies as those with $\Delta \log(\text{SFR}) \geqslant -0.5$, consistent with our procedure in \citetalias{lawlor2026a}. Galaxies below this threshold we define as green valley galaxies, as we have already excluded galaxies with very low global star formation rates ($\text{SFR}_{\text{total}} < 10^{-3}~M_{\odot}~\text{yr}^{-1}$, as above). With the fit and accompanying threshold (as described above) applied to our galaxies, we find 681 star forming main sequence galaxies, along with 756 green valley galaxies.

\newpage
\subsection{Quenching Process Classification}\label{subsec:metrics}

\begin{figure*}
    \centering
    \includegraphics[width=\textwidth]{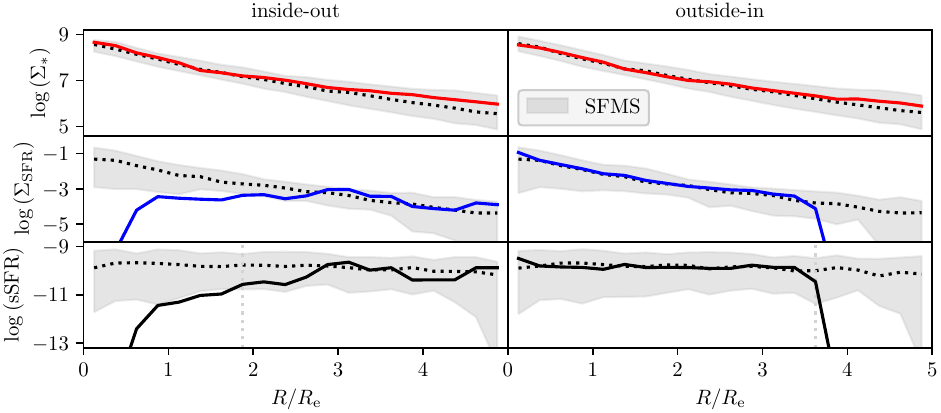}
    \caption{Example radial profiles for different types of galaxies using the classification scheme from Section~\ref{subsec:metrics}. An inside-out quenching galaxy (left) is shown along with an outside-in quenching galaxy (right). Moving from top to bottom, these panels show the stellar mass radial profile (red, units of $M_{\odot}~\text{kpc}^{-2}$), the SFR radial profile (blue, units of $M_{\odot}~\text{yr}^{-1}~\text{kpc}^{-2}$), and finally the sSFR radial profile (black, units of yr$^{-1}$). In all panels we show the median (black dotted line) and ${\pm} 1 \sigma$ spread (gray shaded region) for the radial profiles, based on using comparison SFMS galaxies (Section~\ref{subsec:SFMS}) that are close in stellar mass: $\Delta \log{(M_{*}/M_{\odot})} \leqslant 0.1~\text{dex}$. In the bottom panels, we denote the location of $R_{\text{inner}}$ (left) and $R_{\text{outer}}$ (right) with a vertical light gray dotted line, respectively.}
    \label{fig:radial_profiles}
\end{figure*}

In \citetalias{lawlor2026a}, using a machine learning classification algorithm, we showed that observationally motivated morphological metrics can first separate simulated quenching galaxies from star forming galaxies, and further, that quenching galaxies can be additionally separated into two evolutionary signatures: a pathway where the star formation is initially quenched in the interior regions of a galaxy, and subsequently proceeds outward, creating an ``inside-out'' signature. The second pathway evolves in the opposite fashion, where star formation is first suppressed in the outer regions of a galaxy, before proceeding inward, defining an ``outside-in'' signature. We now apply the same scheme to our Frontier Fields galaxies, in order to classify them as quenching inside out or outside in.

We adopt the observationally motivated morphological metrics first introduced in \citetalias{lawlor2026a}, and further refined in \citetalias{lawlor2026b}, that are based on the spatial distribution of star formation. The morphological metrics consider the concentration of star formation, the size of the star forming disk relative to the stellar disk, and radii that describe abrupt changes or truncations of star formation \citepalias{lawlor2026a}:
\begin{align}
    C_{\text{SF}} &= \text{SFR}_{< 1~\text{kpc}}/\text{SFR}_{\text{total}}, \\
    R_{\text{SF}} &= \log{(R_{\text{e, SF}}/R_{\text{e}})}, \\
    R_{\text{inner}} &= \log[\text{sSFR}(r)] \leqslant -10.5 \cap \text{d} (\text{sSFR})/\text{d}r \geqslant 1,\label{eq:Rinner} \\
    R_{\text{outer}} &= \log[\text{sSFR}(r)] \leqslant -10.5 \cap \text{d} (\text{sSFR})/\text{d}r \leqslant -1,\label{eq:Router}
\end{align}
where $C_{\text{SF}}$ is the concentration of star formation, measured via the SFR in the inner $1~\text{kpc}$ of the galaxy relative to the total SFR. The size of the star forming disk, $R_{\text{e, SF}}$ is next compared to the stellar disk, $R_{\text{e}}$, to define $R_{\text{SF}}$. $R_{\text{inner}}$ and $R_{\text{outer}}$ are then the inner and outer truncation radii, respectively, and are determined using the specific star formation radial profile, $\text{sSFR}(r)$, along with its first derivative, $\text{d} (\text{sSFR})/\text{d}r$.

Though these metrics were initially designed for use when considering simulated galaxies from IllustrisTNG \citep{nelson2018,pillepich2018b,springel2018} as was done in \citetalias{lawlor2026a}, in \citetalias{lawlor2026b} we adapted the metrics for use with mock observations of the same simulated galaxies. We determined that 20 radial bins (extending out to $5~R_{\text{e}}$) were sufficient to properly recover the stellar mass and SFR radial profiles after SED fitting, and with these SED fits we could likewise recover the morphological metrics with high precision, using data of quality$/$coverage comparable to the Frontier Fields data. For this reason we adopt the morphological metrics from \citetalias{lawlor2026b}, using the same adaptations as described there: we interpolate the SFR profile to find the value at $1~\text{kpc}$ (projected) and compare with the integrated profile for $C_{\text{SF}}$. For $R_{\text{SF}}$, we likewise interpolate the SFR and stellar mass profiles to find the radius containing fifty percent of the profile, and subsequently take the logarithm of the ratio. Both truncation radii require no observational adaptations \citepalias{lawlor2026b}. We calculate these quantities for all 1437 galaxies in our non-quenched primary sample with $\text{SFR}_{\text{total}} \geqslant 10^{-3}~M_{\odot}~\text{yr}^{-1}.$

We next employ a \textit{k}-nearest neighbor algorithm \citep[\textit{k}NN; e.g.,][]{fix1951,fix1989,cover1967}, from the \texttt{scikit-learn} package \citep{pedregosa2011}, in order to classify our galaxies as quenching from the inside out or the outside in. The \textit{k}NN algorithm uses the \textit{k}-nearest neighbors to a point in a given parameter space as training samples, to infer the class or label of the point under consideration, with optional weighting based on distance \citep{pedregosa2011}. For the training set, we use the simulated galaxies (drawn from IllustrisTNG, from \citetalias{lawlor2026a}), but employ the observational proxies of the four morphological metrics (from \citetalias{lawlor2026b}), along with the distance from the SFMS, $\Delta \text{SFMS}$ (measured in dex), that compares the expected (log) SFR for a galaxy based on its (log) stellar mass if it was on the SFMS, to its actual (log) SFR. We additionally use the true population labels (from \citetalias{lawlor2026a}; i.e. star forming, inside-out, and outside-in). For the observed target sample, we use the morphological metrics and $\Delta \text{SFMS}$ for the 1437 galaxies in the non-quenched primary sample.

\begin{figure*}[t]
    \centering
    \includegraphics[width=\textwidth]{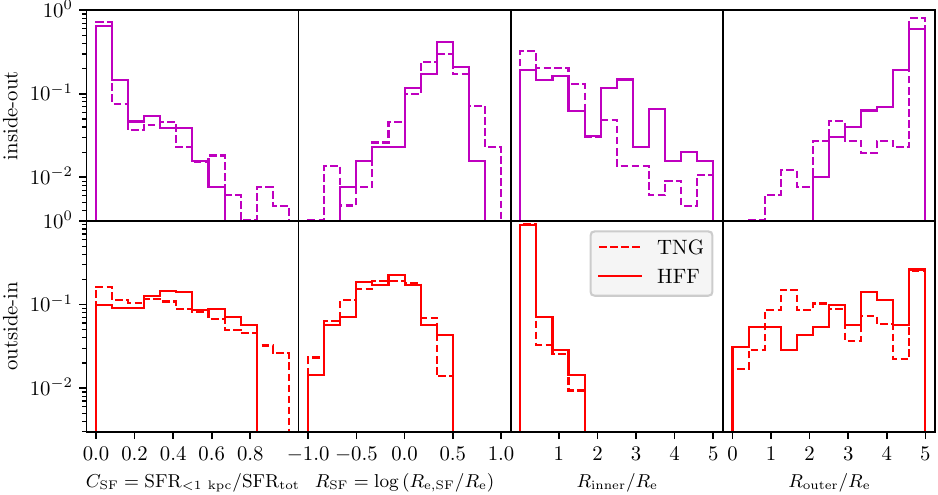}
    \caption{Distributions for each morphological metric (columns), separating inside-out quenching galaxies (top, magenta), from outside-in quenching galaxies (bottom, red), after classification (solid lines). Comparison distributions for the corresponding population of simulated quenching galaxies from \citetalias{lawlor2026a} are shown with dashed lines, respectively. These simulated galaxies formed the training set during classification (see Section~\ref{subsec:metrics}). Moving from left to right, the columns show $C_{\text{SF}}$, $R_{\text{SF}}$, $R_{\text{inner}}$, and $R_{\text{outer}}$, respectively. Each distribution has been normalized such that the $y$-axis can simply be read as the frequency of a given value occurring in the sample.}
    \label{fig:MMs}
\end{figure*}

For both the training set and target sample, we standardize$/$feature-scale the respective morphological metrics prior to performing the classification. This ensures that metrics with large dynamic range are not dominant when classifying \citep{pedregosa2011}. Similarly to \citetalias{lawlor2026a} and \citetalias{lawlor2026b}, we use the five nearest neighbors, and weight the contribution of the neighbors using the inverse of their distance. We use the complete set of galaxies available from \citetalias{lawlor2026a} and \citetalias{lawlor2026b}, some 69,493 unique galaxies with which we have morphological metrics, to complete the classification. After applying the \textit{k}NN classification, we find 129 inside-out quenching galaxies, along with 70 outside-in quenching galaxies. In Figure~\ref{fig:radial_profiles}, we show example radial profiles for an inside-out quenching galaxy on the left, and those for an outside-in quenching galaxy on the right. In all panels we compare with galaxies close in stellar mass, $\Delta \log{(M_{*}/M_{\odot})} \leqslant 0.1~\text{dex}$, that lie on the SFMS, where the  ${\pm} 1 \sigma$ region is shown with a gray band, while the the median is shown with a dotted black line. In the bottom row, we denote the location of $R_{\text{inner}}$ for the inside-out quenching galaxy (left), and $R_{\text{outer}}$ for the outside-in quenching galaxy (right), with a vertical light gray dotted line, respectively.

\section{Results}\label{sec:results}

In Figure~\ref{fig:MMs}, we show the distributions of the morphological metrics for the classified galaxies, compared to similarly classified simulated galaxies, where the metrics are taken from \citetalias{lawlor2026a}. In the top row, we show inside-out quenching galaxies (magenta), while in the bottom row we show outside-in quenching galaxies (red). The distributions for the simulated galaxies from \citetalias{lawlor2026a} are shown with dashed lines, while the Frontier Fields galaxies investigated here are shown with solid lines. In each panel, we normalize the distributions so that the $y$-axis can be read as frequency. We show similar plots for star forming galaxies in Appendix~\ref{app:SF_MM_dists}, where the distributions shown there are consistent between the simulated galaxies and the Frontier Fields galaxies.

We first consider the inside-out quenching population in Figure~\ref{fig:MMs}, and compare the distributions between the simulated galaxies (that form the training set,) and the Frontier Fields galaxies (that form the observed target sample) using a two-sample Kolmogorov-Smirnov (KS) test for goodness of fit \citep[e.g.,][]{hodges1958}. We find good agreement, where for each of the morphological metrics, the distribution for the sample galaxies is consistent with being drawn from the same distribution as the simulated galaxies. The \textit{p}-values are $0.54$, $0.87$, $0.10$, and $0.54$ for $C_{\text{SF}}$, $R_{\text{SF}}$, $R_{\text{inner}}$, and $R_{\text{outer}}$, respectively. Notably, the distributions for $C_{\text{SF}}$ and $R_{\text{SF}}$ are quite similar, while $R_{\text{inner}}$ shows a higher fraction of galaxies with large values when compared with the simulation. For $R_{\text{outer}}$, we again find very good agreement, though Frontier Fields galaxies generally take larger values of $R_{\text{outer}}$ compared to their simulated counterparts from \citetalias{lawlor2026a}.

\begin{figure*}
    \centering
    \includegraphics[width=\textwidth]{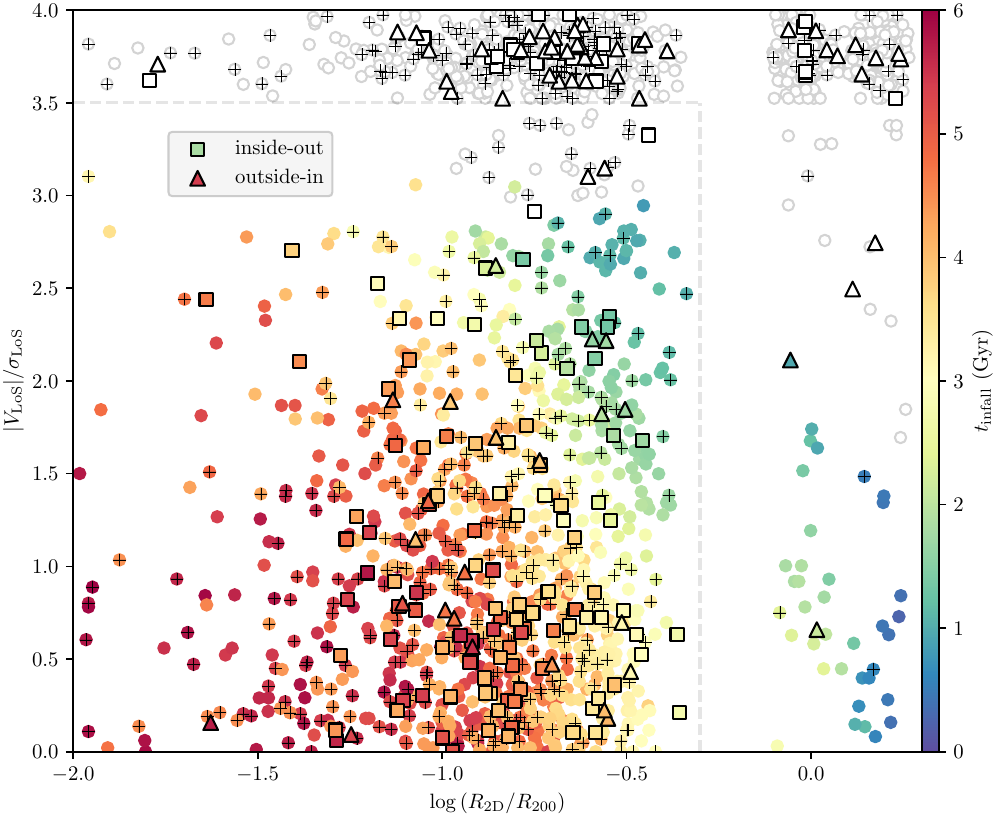}
    \caption{Phase space locations for the non-quenched+quenched sample of 1939 galaxies. Here we plot galaxies that do not have infall time estimates with white, while inside-out and outside-in quenching galaxies are plotted with squares and triangles, respectively. We plot galaxies that are part of the non-quenched sample but not classified as inside-out or outside-in with circles, and color-code these galaxies, along with the inside-out and outside-in populations, by infall time. Galaxies that are part of the quenched extended sample are additionally color-coded by infall time when possible, and are highlighted with small crosses. We denote the boundary that separates the field from the cluster ($R_{\text{2D}}/R_{200} < 0.5 \cap |V_{\text{LoS}}|/\sigma_{\text{LoS}} < 3.5$) with a dashed light gray line. Field galaxies with large line-of-sight velocities beyond $|V_{\text{LoS}}|/\sigma_{\text{LoS}} = 3.5$ are plotted between this value and $|V_{\text{LoS}}|/\sigma_{\text{LoS}} = 4$ for visualization purposes.}
    \label{fig:phase_space}
\end{figure*}

For the outside-in quenching galaxies in the bottom row of Figure~\ref{fig:MMs}, we likewise find very good agreement between our sample galaxies and the simulated galaxies. Employing a two-sample KS test as before, for each metric for the outside-in quenching Frontier Fields galaxies, we find no significant differences when compared with the simulated galaxies. The \textit{p}-values are $0.99$, $0.99$, $0.99$, and $0.54$, for $C_{\text{SF}}$, $R_{\text{SF}}$, $R_{\text{inner}}$, and $R_{\text{outer}}$, respectively. In particular, the distributions for $R_{\text{SF}}$, and $R_{\text{inner}}$ show excellent agreement. The distributions for $C_{\text{SF}}$ show that our outside-in quenching sample galaxies are less likely to take very high concentration values (${\sim} 1$) when compared with the simulated galaxies, and that they similarly show a relatively uniform distribution in $R_{\text{outer}}$. Of course, by construction these distributions should be similar, as in Section~\ref{subsec:metrics} we used the simulated galaxies as the training set, in order to classify the sample galaxies according to their morphological metrics. Given this, the $k$NN classification should only classify galaxies that have morphological metrics similar to simulated galaxies as inside-out or outside-in, for example. Nonetheless, it is encouraging to find broad similarities between the simulated and observed galaxies when considering the suite of morphological metrics simultaneously, for a given population.

We next consider the phase space distribution of our non-quenched+quenched samples in Figure~\ref{fig:phase_space}, following the analysis completed in Section~\ref{subsubsec:phase_space}. The cluster-normalized line-of-sight velocity is plotted as a function of projected 2D distance from the respective cluster center, where we take the logarithm of the projected 2D distance for visual clarity. Galaxies without phase space-inferred infall times are plotted in white, otherwise galaxies are color-coded by their infall time: galaxies with recent infall times ($t_{\text{infall}} \lesssim 1~\text{Gyr}$) are plotted with blue, while galaxies with ancient infall times ($t_{\text{infall}} \sim 6~\text{Gyr}$) are plotted with red, for example. Inside-out quenching galaxies are plotted with squares, while outside-in quenching galaxies are plotted with triangles. We additionally denote the boundary (light gray line) that separates the field from the cluster, where the boundary is $R_{\text{2D}} < 0.5 R_{200} \cap |V_{\text{LoS}}| < 3.5 \sigma_{\text{LoS}}$. Galaxies that make up the quenched extended sample ($\text{SFR}_{\text{total}} < 10^{-3}~M_{\odot}~\text{yr}^{-1}$, numbering 502), and galaxies from the non-quenched primary sample that were not classified as either inside-out or outside-in quenching (numbering 682, excluding 556 $k$NN-classified star forming galaxies), are plotted with circles and are color-coded according to infall time. Both of these latter subpopulations (quenched+unclassified) are predominantly found within the cluster. Quenched extended sample galaxies are further denoted with small crosses. Field galaxies with large line-of-sight velocities beyond the plot boundaries are plotted at random $y$-values between $|V_{\text{LoS}}|/\sigma_{\text{LoS}} = 3.5$ and $|V_{\text{LoS}}|/\sigma_{\text{LoS}} = 4$ for visualization purposes.

We first consider the cluster galaxies in Figure~\ref{fig:phase_space}, and will discuss the field population below. Of the 129 inside-out quenching galaxies we classified, 106 are in clusters. For the outside-in population, 25 of the 70 galaxies reside in clusters. Figure~\ref{fig:phase_space} shows that both populations of cluster quenching galaxies reside in similar regions of the phase space, commonly with $-1.3 < \log{(R_{\text{2D}}/R_{200})} < -0.5$ and $0 < |V_{\text{LoS}}|/\sigma_{\text{LoS}} < 2.5$. Further, the inside-out and outside-in populations have a similar distribution to the unclassified galaxies (colored circles) along the line-of-sight velocity, but are somewhat more centrally concentrated radially. Given these locations in phase space, these galaxies tend to have old infall times. The median infall time (and spread) for the inside-out quenching population is $4.0^{+1.1}_{-1.0}~\text{Gyr}$, while it is $4.0^{+0.9}_{-2.1}~\text{Gyr}$ for the outside-in quenching population.

Beyond the two main quenching populations, there exist many galaxies at small distances from the cluster core with correspondingly ancient infall times ($t_{\text{infall}} \sim 6~\text{Gyr}$). A large proportion of these galaxies populate the quenched extended sample (denoted with small crosses), the 502 galaxies with global $\text{SFR} < 10^{-3}~M_{\odot}~\text{yr}^{-1}$ that we were unable to determine morphological metrics for, consistent with observations of clusters with high fractions of quenched galaxies \citep[e.g.,][]{wetzel2013,davies2019,hewitt2025}.

\begin{figure*}[t]
    \centering
    \includegraphics[width=\textwidth]{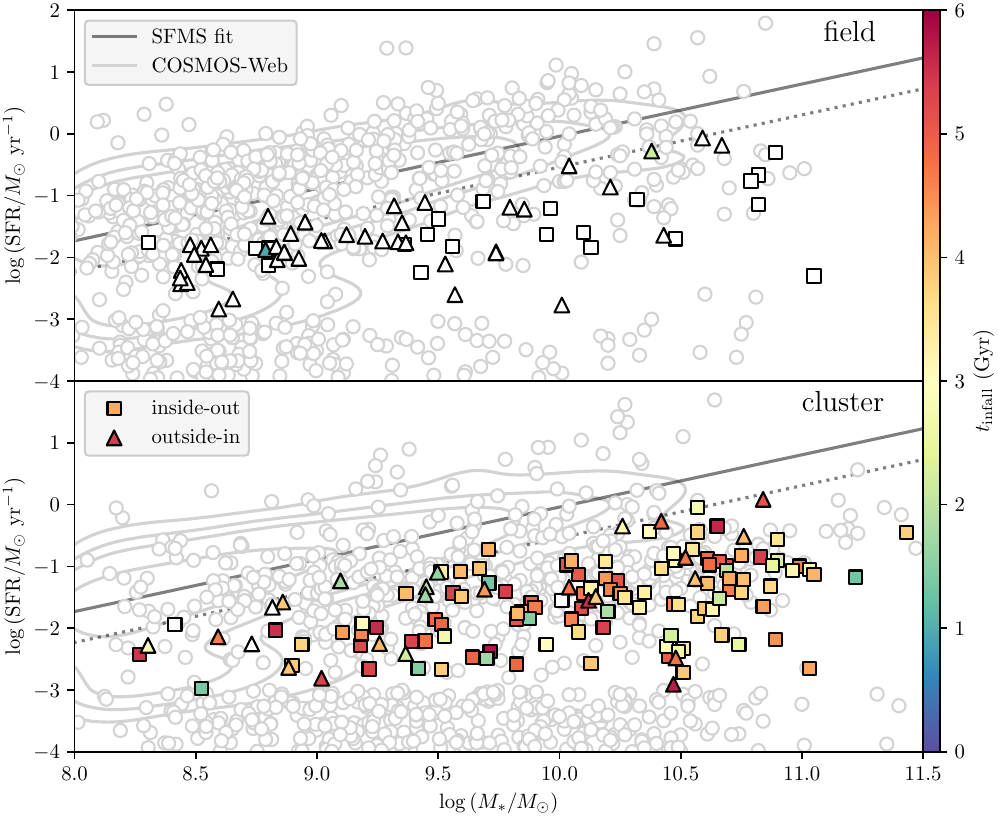}
    \caption{Global star formation rate as a function of stellar mass for the non-quenched+quenched sample of 1939 galaxies. In the background we show unfilled contours (light gray) that denote the location and density of galaxies drawn from COSMOS-Web \citep{casey2023} from which we determine the star forming main sequence (dark gray solid line). Galaxies ${>} 0.5~\text{dex}$ below the fit for the star forming main sequence (dark gray dotted line) are classified as green valley galaxies, while those above this line are classified as star forming main sequence galaxies. Galaxies with very low levels of star formation from the quenched extended sample, that we cannot apply our morphological metrics to, are plotted at small, random nonzero star formation rates from $10^{-3}~M_{\odot}~\text{yr}^{-1}$ to $10^{-4}~M_{\odot}~\text{yr}^{-1}$ for visualization purposes. Inside-out and outside-in quenching galaxies are plotted with squares and triangles, respectively, and are color-coded according to infall time. Galaxies that are part of the non-quenched sample not classified as inside-out or outside-in quenching are plotted with white dots, as are galaxies from the quenched extended sample. We separate galaxies that reside in the field (top) from those present in the cluster (bottom).}
    \label{fig:SFMS}
\end{figure*}

In Figure~\ref{fig:SFMS}, we show the total star formation rates for our non-quenched+quenched sample as a function of stellar mass. In the background we show contours (light gray) that describe the location and density of the COSMOS-Web \citep{casey2023} galaxies we used to determine the SFMS. Our best fit to the SFMS is shown with a solid dark gray line, while the $\text{SFMS} -0.5~\text{dex}$ boundary for classification as green valley is shown with a dotted gray line. In the foreground we show all galaxies from our non-quenched+quenched samples as colored symbols and white dots. The 502 galaxies with $\text{SFR} < 10^{-3}~M_{\odot}~\text{yr}^{-1}$ that make up the quenched extended sample are plotted with white dots at random SFRs between $10^{-3}~M_{\odot}~\text{yr}^{-1}$ and $10^{-4}~M_{\odot}~\text{yr}^{-1}$ for visualization purposes. Galaxies that are part of the non-quenched primary sample that were not classified as either inside-out or outside-in quenching, numbering 682 galaxies and excluding classified star forming galaxies (556), are likewise plotted with white dots. The inside-out quenching galaxies are plotted with squares, and are color-coded by infall time, based on the phase space analysis completed in Section~\ref{subsubsec:phase_space}. Outside-in quenching galaxies are plotted with triangles, and are similarly color-coded. We separate galaxies that reside in the field (top), from those that live in clusters (bottom).

\begin{figure*}[t]
    \centering
    \includegraphics[width=\textwidth]{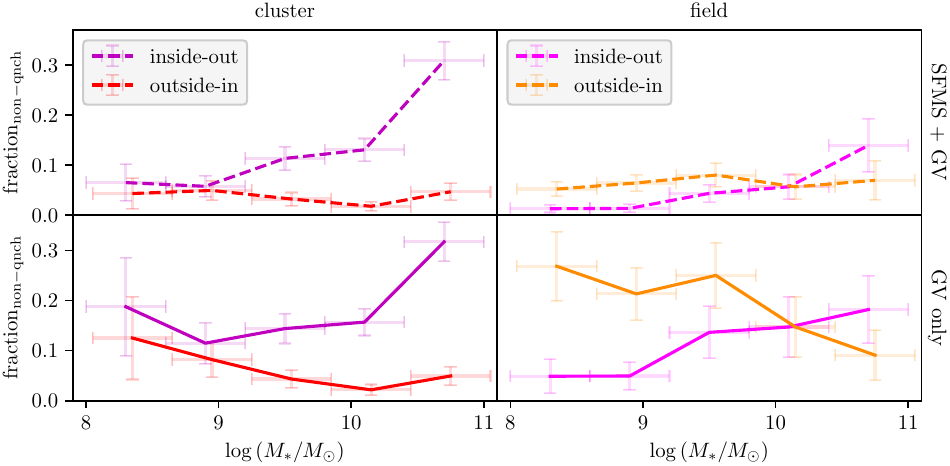}
    \caption{Fraction of non-quenched galaxies with $\text{SFR} > 10^{-3}~M_{\odot}~\text{yr}^{-1}$ experiencing either inside-out (pink--magenta) or outside-in (orange--red) quenching, as a function of stellar mass. We bin galaxies by environment, separating cluster galaxies (left) from those in the field (right). We additionally separate all galaxies in the non-quenched primary sample (top, dashed lines), from green valley-only galaxies in the non-quenched sample (bottom, solid lines). In all panels, we show binomial errors for each fraction ($y$-axis error bars), and use $x$-axis error bars to illustrate the considered mass range for each mass bin. The results for the outside-in population are slightly offset horizontally from the inside-out populations for visual clarity.}
    \label{fig:fractions_cl_fd}
\end{figure*}

In both the field and cluster, inside-out and outside-in quenching galaxies are almost exclusively found in the green valley (also see Appendix~\ref{app:UNCOVER_CANUCS}), which is likely a consequence of including the distance from the SFMS, $\Delta \text{SFMS}$, in the \textit{k}NN classification above. However, we investigated the distributions of $\Delta \text{SFMS}$ for the star forming, inside-out, and outside-in populations drawn from the simulation, and found that there was significant overlap between the quenching populations ($\mu = -0.27~\text{dex}$, $\sigma = 0.30~\text{dex}$, when combined) and the star forming population ($\mu = 0.01~\text{dex}$, $\sigma = 0.20~\text{dex}$). Further, we additionally used a random forest classifier \citep[from \texttt{scikit-learn};][]{pedregosa2011}, which uses a collection of 100 randomized decision trees to determine the relative importance \citep[or mean decrease in impurity, or Gini impurity;][]{pedregosa2011} of each morphological metric and $\Delta \text{SFMS}$ when training. We find that $\Delta \text{SFMS}$ has an importance of $0.25$, which is less than, or comparable to, the importance values for $C_{\text{SF}}$ ($0.24$) and $R_{\text{SF}}$ ($0.31$), but more than the importance values for $R_{\text{inner}}$ ($0.07$) and $R_{\text{outer}}$ ($0.13$), consistent with findings from \citetalias{lawlor2026a}. This indicates that while $\Delta \text{SFMS}$ is helpful when classifying, ${\sim}$75\% of the classification is driven by the morphological metrics alone.

Considering the field and cluster together, inside-out quenching galaxies are more massive, with a median stellar mass of $\log{(M_{*}/M_{\odot})} = 10.1^{+0.6}_{-0.7}$, compared with $\log{(M_{*}/M_{\odot})} = 9.3^{+0.9}_{-0.7}$ for outside-in quenching galaxies, representing a higher median stellar mass of $\Delta \log{(M_{*}/M_{\odot})} = 0.8^{+0.2}_{-0.1}~\text{dex}$, where the 68\% confidence interval was found through bootstrap resampling. The inside-out population is also found more commonly in clusters (106/129), compared to the outside-in population (25/70), which is most often found in the field. As shown above, the two populations have similar median infall times ($4.0^{+1.1}_{-1.0}~\text{Gyr}$ for inside-out, and $4.0^{+0.9}_{-2.1}~\text{Gyr}$ for outside-in) when considering the field and cluster populations simultaneously. We find some evidence that the field population of inside-out+outside-in quenching galaxies has higher specific star formation rates, with a median sSFR of $\log{(\text{sSFR}/\text{yr}^{-1})} = -10.9^{+0.3}_{-0.7}$, compared with $\log{(\text{sSFR}/\text{yr}^{-1})} = -11.6^{+0.6}_{-0.6}$ for the cluster population of these quenching galaxies. However, these differences are comparable to the spread for each population per environment.

In the top panel of Figure~\ref{fig:SFMS}, we see that there are 23 inside-out and 45 outside-in quenching galaxies present in the field. The inside-out quenching galaxies span the range of stellar masses under investigation, with a median stellar mass of $\log{(M_{*}/M_{\odot})} = 9.7^{+1.1}_{-0.9}$. Considering the outside-in quenching galaxies in the field, we find that they are generally less massive than the inside-out quenching galaxies, with a median stellar mass of $\log{(M_{*}/M_{\odot})} = 9.0^{+0.8}_{-0.5}$. In the bottom (cluster) panel, considering first the inside-out population, we find that these galaxies (106) are generally more massive than their field counterparts, with a median stellar mass of $\log{(M_{*}/M_{\odot})} = 10.2^{+0.5}_{-0.7}$. In a similar fashion, the outside-in quenching galaxies in the cluster (25) are more massive than outside-in quenching galaxies in the field, with a median stellar mass of $\log{(M_{*}/M_{\odot})} = 9.5^{+0.9}_{-0.6}$. Taken together, the inside-out+outside-in population in the cluster is $\Delta \log{(M_{*}/M_{\odot})} = 0.8^{+0.3}_{-0.1}~\text{dex}$ more massive than in the field, with 68\% bootstrapped confidence intervals.

\begin{figure*}[t]
    \centering
    \includegraphics[width=\textwidth]{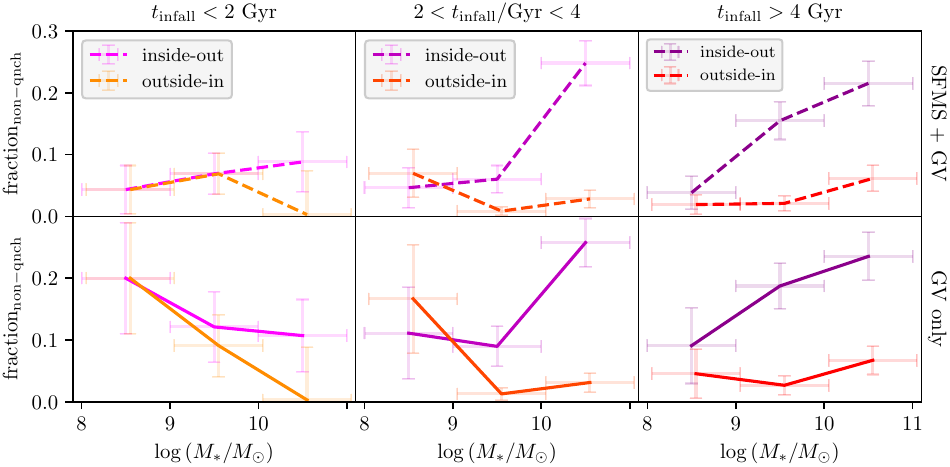}
    \caption{Fraction of non-quenched galaxies with $\text{SFR} > 10^{-3}~M_{\odot}~\text{yr}^{-1}$ experiencing either inside-out (pink--dark magenta) or outside-in (orange--red) quenching, as a function of stellar mass. We bin galaxies by phase space-based estimated infall time, $t_{\text{infall}}$, separating recent infall galaxies (left, $t_{\text{infall}} < 2~\text{Gyr}$), from intermediate infall galaxies (middle, $2 < t_{\text{infall}}/\text{Gyr} < 4$), and ancient infall galaxies (right, $t_{\text{infall}} > 4~\text{Gyr}$). We additionally separate all galaxies in the non-quenched primary sample (top, dashed lines), from green valley-only galaxies in the non-quenched sample (bottom, solid lines). As in Figure~\ref{fig:fractions_cl_fd}, we use binomial errors to describe uncertainties on the fractions ($y$-axis error bars), while the range for each mass bin is given by the $x$-axis error bars. Fractions for the outside-in population are slightly offset horizontally for clarity.}
    \label{fig:fractions_tinfall}
\end{figure*}

We next look to understand the prevalence of these different quenching processes as a function of stellar mass, binned by environment, and additionally separated by star forming+green valley versus green valley only. The results of this investigation are shown in Figure~\ref{fig:fractions_cl_fd}. Galaxies present in the cluster (left) are separated from those in the field (right), while green valley galaxies in the non-quenched primary sample (bottom, solid lines) are also considered distinct from all galaxies in the non-quenched sample (top, dashed lines; i.e. star forming+green valley). In all panels, fractions are relative to the non-quenched sample: those galaxies with $\text{SFR}_{\text{total}} \geqslant 10^{-3}~M_{\odot}~\text{yr}^{-1}$ that we were able to compute morphological metrics for. Binomial errors are used to convey the uncertainty for each fraction, while the range of each mass bin is illustrated with $x$-axis error bars.

In Figure~\ref{fig:fractions_cl_fd}, by first considering the evolution for all galaxies (top row), we see that the fraction of cluster inside-out quenching galaxies (dashed magenta line) increases with increasing stellar mass, where at high masses ($M_{*} \gtrsim 10^{10.5}~M_{\odot}$), the inside-out population makes up ${\sim}$30\% of the galaxy population with $\text{SFR} > 10^{-3}~M_{\odot}~\text{yr}^{-1}$. By fitting a linear function, we find a slope of $m = 0.08 \pm 0.02$. Conversely, the fraction of outside-in galaxies (dashed red line) is relatively independent of stellar mass ($m = -0.01 \pm 0.01$). Moving next to the field while still considering all galaxies with $\text{SFR} > 10^{-3}~M_{\odot}~\text{yr}^{-1}$, we find a similar situation for the inside-out galaxies but less pronounced: the fraction of inside-out galaxies (dashed pink line) increases with stellar mass ($m = 0.02 \pm 0.01$), amounting to ${\sim}$15\% of the galaxy population at high stellar mass ($M_{*} \gtrsim 10^{10.5}~M_{\odot}$). By comparing the trends for the inside-out population in the cluster and the field, we find that the difference in slopes is statistically significant, with $p = 0.024$ from a \textit{t}-test. In the field, the fraction of outside-in quenching galaxies is again relatively uniform with stellar mass ($m = 0.01 \pm 0.01$), as was found in the cluster. Given that there is no statistically significant difference between the linear relations (slopes or intercepts) for the outside-in populations in the cluster and field, the rate of finding an outside-in quenching galaxy in the field is the same as in the cluster, when considering all galaxies.

We expect, and find (see above), quenching galaxies to be primarily located in the green valley. Therefore it may be more appropriate to consider the green valley as the parent population, though this of course depends on the relative timescales of morphologically-identifiable transition signatures with green valley lifetime. We consider only the green valley population, numbering 756 galaxies, in the bottom row of Figure~\ref{fig:fractions_cl_fd}. By simultaneously considering both quenching populations, we find that the fraction of all green valley galaxies in the cluster (543 galaxies) is made up of ${\sim}$30\% inside-out+outside-in quenching galaxies for $M_{*} \lesssim 10^{9}~M_{\odot}$, where this fraction decreases at intermediate masses ($M_{*} \sim 10^{9.5}~M_{\odot}$) to ${\sim}$20\%, before increasing at high stellar mass ($M_{*} \sim 10^{10.5}~M_{\odot}$), reaching ${\sim}$40\%. For the green valley galaxies in the field (213 galaxies), by again considering both quenching populations simultaneously, we find that the fraction of inside-out+outside-in quenching galaxies is ${\sim}$30\%, where this fraction is relatively constant with stellar mass, showing a modest increase at intermediate stellar masses to ${\sim}$40\%. The similarity between the cluster and field proportions is then worth considering, given that the green valley itself represents a much larger fraction of the total population in clusters (${\sim}$72\%). Conversely, when considering all galaxies in the field (top right panel; 714 galaxies), these mostly reside on the SFMS (501 galaxies), as shown above. Therefore, the fraction of field inside-out+outside-in quenching galaxies in the green valley is much higher than the fraction for all galaxies, which is not the case for cluster galaxies.

Given the large population of galaxies that we find in clusters (723 galaxies), we next explore the fraction of inside-out and outside-in quenching galaxies as a function of stellar mass, but now binned by phase space-inferred infall time. We consider three infall time bins: recent infall galaxies ($t_{\text{infall}} < 2~\text{Gyr}$), intermediate infall galaxies ($2 < t_{\text{infall}}/\text{Gyr} < 4$), and ancient infall galaxies ($t_{\text{infall}} > 4~\text{Gyr}$). We show the results of this investigation in Figure~\ref{fig:fractions_tinfall}, where error bars are as in Figure~\ref{fig:fractions_cl_fd}.

In Figure~\ref{fig:fractions_tinfall}, we see that the trends noticed in Figure~\ref{fig:fractions_cl_fd} for the general cluster population are largely recovered across the three infall time bins as well. The strong mass dependence of the inside-out quenching population noted in Figure~\ref{fig:fractions_cl_fd} is fairly robust, seen in almost all panels. As above, these panels show that the fraction of inside-out quenching galaxies increases as a function of stellar mass. The exception is the recent infall panel, $t_{\text{infall}} < 2~\text{Gyr}$, for green valley galaxies, which has large uncertainties. As well, the outside-in quenching population is likewise mass independent given the uncertainties. This mimics the situation for this population noted above in Figure~\ref{fig:fractions_cl_fd}. Given the large uncertainties, especially in the green valley only panels (bottom row), there is no strong dependence on infall time for any population, except perhaps the most massive ($M_{*} \gtrsim 10^{10.5}~M_{\odot}$) inside-out population. At these masses, the fraction of inside-out quenching galaxies increases with increasing infall time. This is seen when considering all galaxies (top row), as well as green valley-only galaxies (bottom row). It is somewhat surprising, in particular, that the outside-in quenching population shows no dependence on infall time. However, this is possibly a consequence of the small sample size, where only 25 outside-in quenching galaxies reside in the cluster.

\section{Discussion}\label{sec:discussion}

Given the results presented above, we have found that using the observationally motivated morphological metrics introduced in \citetalias{lawlor2026a} and refined in \citetalias{lawlor2026b}, that trace the distribution of star formation within galaxies, we can identify galaxies in the Hubble Frontier Fields that have morphological metrics consistent with those predicted from the TNG simulation to have resulted from different quenching pathways. For these quenching pathways, the first is consistent with a scenario in which star formation is suppressed in the interior of a galaxy, and proceeds outward. The other pathway has star formation that is quenched initially on the outskirts, and proceeds inward. Considering their mass and environmental dependence, the inside-out quenching population in the Frontier Fields is more massive than the outside-in quenching population, and is commonly found in cluster galaxies. Outside-in quenching galaxies are commonly found in the field. This suggests that the high-mass galaxies, regardless of environment and central$/$satellite status, are quenching inside-out, while the low mass satellites are quenching outside-in.

When viewed together, these results suggest that the nature of galaxy quenching, as inferred here, is more complicated than our previous investigations would suggest. In particular, we find a high rate of inside-out quenching galaxies present in the cluster, and indeed that have resided in the cluster over extended periods. In \citetalias{lawlor2026a}, and as followed-up in \citetalias{lawlor2026b}, when we investigated simulated galaxies, we found inside-out quenching galaxies to largely reside in the field. Further, in \citetalias{lawlor2026a}, we did find that inside-out galaxies have quenching that is extended over several gigayears, on the order of the infall times we found above, so perhaps some environmental effect is promoting the funneling of material down onto the active galactic nuclei (AGN) in these systems, thereby producing the expected inside-out signature. This physical scenario seems plausible, but will require additional work (see below).

Concerning the outside-in population, we find fewer galaxies, especially in the cluster, when viewed in the context of \citetalias{lawlor2026a}. In that work, we found that outside-in galaxies make up the majority of the quenching population and are dominant in clusters, while above we showed that the fraction of outside-in quenching galaxies in the Frontier Fields sample is higher in the field (${\sim} 6.3$\%), compared with the cluster (${\sim} 3.5$\%). As well, in \citetalias{lawlor2026a}, we found that simulated outside-in quenching galaxies in TNG had a quenching timescale of the order of ${\sim} 1.5~\text{Gyr}$, while above we found outside-in galaxies residing in the cluster with infall time estimates of ${>} 3~\text{Gyr}$. This situation is puzzling, as we might expect quenching driven by environmental processes to act in a more efficient manner \citep[e.g.,][]{balogh2000,quilis2000}. However, we did find that the outside-in quenching galaxies have a mild preference towards more recent infall times. As above, this scenario requires additional work, as described below.

From these considerations, we find that the quenching of simulated galaxies as investigated in \citetalias{lawlor2026a} is somewhat simplified compared to what we have inferred about quenching using observations of real galaxies. Specifically, in \citetalias{lawlor2026a}, we could summarize the quenching pathways as thus: high-mass central galaxies in the field quench inside out, likely driven by AGN feedback, while low-mass satellites quench outside in, as a consequence of environment, consistent with expectations for ram pressure stripping. Here, for the Frontier Fields, we find that high-mass satellites additionally quench inside out, in a similar manner to their high-mass central counterparts. This result agrees well with the findings of \citet{bluck2020b} and \citet{mcdonough2025}, and suggests that the quenching prescription in the TNG simulations does not completely capture the various quenching pathways experienced by real galaxies.

\subsection{Comparison with Literature Results}

In the context of studies in the literature, we are not currently aware of any such work that has completed a directly similar analysis as presented above. However, there have been a number of recent investigations using spatially resolved SED fitting to investigate quenching in different environments \citep[e.g.,][]{nelson2021,abdurrouf2022a,abdurrouf2023,olsen2026}, and other works have investigated the fraction of quenching galaxies at various redshifts \citep[e.g.,][]{schawinski2014,mcnab2021}.

Using imaging and slitless spectroscopy from the 3D-HST survey \citep{brammer2012,skelton2014,momcheva2016}, \citet{nelson2021} completed spatially resolved spectral energy distribution fitting for 3200 star forming main sequence galaxies at $0.7 < z < 1.5$ with $10^{9} < M_{*}/M_{\odot} < 10^{11}$. In their work, \citet{nelson2021} used spatially resolved near-infrared spectra to create H$\alpha$ emission line maps, which trace the distribution of star formation for short timescales \citep[${\sim} 10~\text{Myr}$; e.g.,][]{kennicutt1998b,kennicutt2012}, and \textit{JH}-band imaging to trace the distribution of stellar mass. \citet{nelson2021} additionally complete spatially resolved SED fitting using existing rest frame ultraviolet to near-infrared imaging in eight bands from the Cosmic Assembly Near-infrared Deep Extragalactic Legacy Survey \citep[CANDELS;][]{grogin2011,koekemoer2011} to correct their SFR maps for dust. They then bin galaxies by stellar mass, and create sSFR radial profiles for galaxies below, on, and above the star forming main sequence by stacking galaxies with a mass bin. \citet{nelson2021} find that at high stellar masses ($10^{10.5} < M_{*}/M_{\odot} < 10^{11}$), galaxies below the main sequence are strongly centrally suppressed, indicative of inside-out quenching. Indeed, they even find evidence of inside-out quenching for massive 3D-HST galaxies that reside on the main sequence \citep{nelson2021}. For less massive galaxies ($M_{*} < 10^{10.5}~M_{\odot}$), \citet{nelson2021} find no evidence for inside-out or outside-in quenching for any galaxies across the main sequence, finding flat sSFR profiles on average. In our work, we found evidence of inside-out quenching for massive green valley galaxies in the field, consistent with the findings of \citet{nelson2021} at $M_{*} > 10^{10.5}~M_{\odot}$. However, \citet{nelson2021} do not find any evidence for outside-in quenching within their sample of 3D-HST galaxies, while we find a smaller population of outside-in quenching galaxies in the Frontier Fields.

\citet{abdurrouf2022a} completed spatially resolved panchromatic SED fitting for 10 nearby (${<} 20~\text{Mpc}$) spiral galaxies with $10^{9.7} < M_{*}/M_{\odot} < 10^{11.2}$ on a pixel-by-pixel basis, using 24 broad band filters covering the rest frame far-ultraviolet to the far-infrared. Using maps of physical properties including SFR and stellar mass, along with archival high-resolution maps of atomic and molecular gas, they investigate radial variations, and find central SFR suppressions for many of their galaxies, which correlate with reduced molecular gas fractions \citep{abdurrouf2022a}. They additionally find that with increasing distance from the star forming main sequence, the central suppressions seen in their sSFR radial profiles become more pronounced \citep{abdurrouf2022a}. They conclude that these features are consistent with inside-out quenching, where this quenching started a few gigayears prior to observation \citep{abdurrouf2022a}. We find good agreement with the results of \citet{abdurrouf2022a} when considering the intermediate-to-high mass population in the field, where we found inside-out quenching. In particular, the finding of \citet{abdurrouf2022a}, for an increasing central suppression of SFR with increasing distance from the SFMS, is complementary to our results above. We found a higher fraction of inside-out quenching galaxies (at high masses) in the green valley, when compared with the combined SFMS+green valley population. In a similar fashion to \citet{nelson2021}, \citet{abdurrouf2022a} do not comment on outside-in quenching in their sample, though their radial sSFR profiles for galaxies away from the SFMS (${\gtrsim} 0.5~\text{dex}$) do show evidence for a negative sSFR profile at $R > 1.5~R_{\text{e}}$, especially at $\Delta \text{SFMS} < -1$ (their Figure~11).

As well, \citet{abdurrouf2023} investigated quenching in 444 galaxies with $10^{7.5} < M_{*}/M_{\odot} \lesssim 10^{11.8}$ at $0.3 < z < 6.0$ present in two intermediate redshift clusters, along with a corresponding field sample. \citet{abdurrouf2023} complete spatially resolved SED fitting using high-resolution HST and JWST imaging covering the rest frame blue-optical to the near-infrared. As above, they create maps of physical properties, from which they form radial profiles, including sSFR radial profiles \citep{abdurrouf2023}. They bin their galaxies by distance from the SFMS and by stellar mass, from which they produce average radial profiles \citep{abdurrouf2023}. Here, we focus on their results for $0.3 < z < 0.8$ galaxies, which is the redshift range most comparable with our analysis of the Frontier Fields. They find that $M_{*} < 10^{11}~M_{\odot}$ green valley ($-1 < \Delta \text{SFMS} < -0.4$) galaxies have suppressed central star formation with active star formation in their disks, while quiescent galaxies ($\Delta \text{SFMS} < -1$) show similar profiles for $M_{*} > 10^{9.5}~M_{\odot}$ \citep[their Figure~8;][]{abdurrouf2023}. \citet{abdurrouf2023} conclude that their results point toward inside-out quenching. We find that our findings are consistent with those of \citet{abdurrouf2023}, where we have likewise found inside-out quenching across similar mass scales as those investigated by \citet{abdurrouf2023}, for galaxies in clusters. Indeed, we found that inside-out quenching galaxies are more often found in the cluster, compared to other quenching modes.

\begin{figure*}[t]
    \centering
    \includegraphics[width=\textwidth]{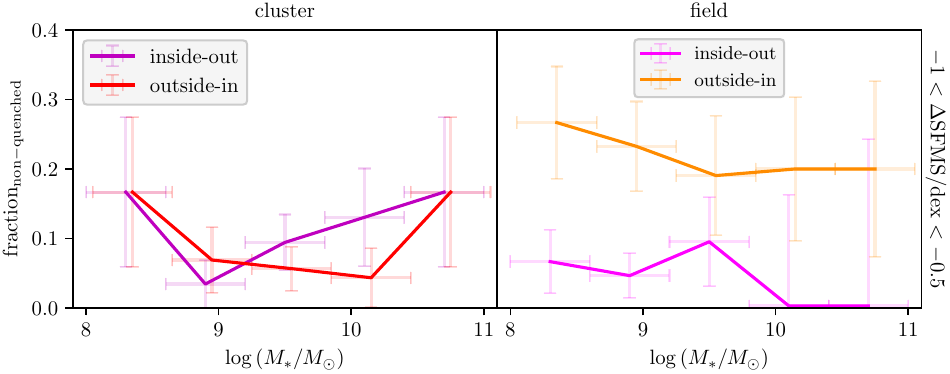}
    \caption{Fraction of non-quenched galaxies, per population and environment, as a function of stellar mass, that reside in a narrower definition of the green valley, similar to \citet{abdurrouf2023}. We separate the cluster (left) from the field (right). Populations are color-coded, and error bars are determined as in Figure~\ref{fig:fractions_cl_fd}, with the outside-in population slightly offset horizontally from the inside-out population for visual clarity.}
    \label{fig:GV_frac}
\end{figure*}

Recently, \citet{olsen2026} completed spatially resolved spectral energy distribution fitting for eight $z < 0.3$ well-resolved star forming galaxies with $10^{9.5} < M_{*}/M_{\odot} \lesssim 10^{11}$ in the UltraViolet Imaging of the Cosmic Assembly Near-infrared Deep Extragalactic Legacy Survey \citep[UVCANDELS;][]{wang2025}, an extension to CANDELS \citep{grogin2011,koekemoer2011}. The sample of \citet{olsen2026} is comprised mostly of face-on galaxies, but is supplemented with edge-on disks, with some showing signatures of tidal interactions. \citet{olsen2026} used ten-band HST observations from the rest-frame ultraviolet to the near-infrared and fit voronoi-binned photometry \citep[e.g.,][]{cappellari2003} to produce maps of stellar mass, star formation rate, dust attenuation, and the time at which half the stellar mass was formed, $t_{50}$. Using these maps, \citet{olsen2026} additionally produce radial profiles of (logarithmic) sSFR at the time of observation and at $1~\text{Gyr}$ lookback time to investigate the evolution of star formation, in the context of galaxy quenching. They find evidence for different quenching pathways in their sample by comparing the slopes of the resolved SFR--$M_{*}$ relation: an inside-out pathway, and an outside-in pathway \citep{olsen2026}. They further find that the sSFR profiles show an overall decrease over the past $1~\text{Gyr}$, where five of the eight galaxies in particular show a larger change in the outskirts than in the centers \citep{olsen2026}. They conclude that the evolution for these five systems indicates a possible early onset of quenching in these galaxies \citep{olsen2026}. The results of \citet{olsen2026} are in very good qualitative agreement with the results we found above over a broader mass range ($10^{8} < M_{*}/M_{\odot} < 10^{11.5}$), specifically the identification of quenching pathways suggestive of an inside-out mechanism, as well as an outside-in mechanism. In our work, we found that the outside-in quenching population makes up a smaller overall fraction of the identified quenching pathways (70 galaxies, compared to 129 for inside-out), while \citet{olsen2026} find that five of their eight galaxies show early signs of outside-in quenching. However, due to the smaller sample size investigated by \citet{olsen2026}, it is difficult to drawn firm conclusions on the prevalence of different quenching mechanisms in their sample.

Regarding the fraction of quenching galaxies, \citet{schawinski2014} used a combined analysis of Sloan Digital Sky Survey \citep{york2000}, Galaxy Evolution Explorer \citep[GALEX;][]{martin2005}, and Galaxy Zoo \citep{lintott2008,lintott2011} data to investigate the green valley for local ($0.02 < z < 0.05$) galaxies with $10^{9} < M_{*}/M_{\odot} < 10^{11.5}$. \citet{schawinski2014} define the green valley based on cuts in the \textit{u}--\textit{r} color--mass diagram, and find ${\sim}$17\% of early- and late-type galaxies populate the green valley. At higher redshifts, \citet{mcnab2021} investigated the fraction of green valley galaxies with $10^{9.5} < M_{*}/M_{\odot} < 10^{11.5}$ in $1.0 < z < 1.4$ clusters. \citet{mcnab2021} define the green valley with respect to cuts in the \textit{NUV}--\textit{V}--\textit{J} color--color diagram, and find that the fraction of their green valley galaxies is relatively independent of stellar mass, representing ${\sim}$13\% of their sample in the cluster, and ${\sim}$14\% in the field, where the field galaxies are photometric members that are ${>} 1.5~\text{Mpc}$ from their cluster centers. The results of \citet{schawinski2014} and \citet{mcnab2021} at first seem inconsistent with the results we described above, where we found ${\sim}$30\% of the cluster and field populations (with $\text{SFR} > 10^{-3}~M_{\odot}~\text{yr}^{-1}$) to consist of our combined inside-out+outside-in quenching populations in the green valley (see Figure~\ref{fig:fractions_cl_fd}). However, above we were investigating the fraction of inside-out and outside-in quenching galaxies relative to the green valley, not the fraction of green valley galaxies itself. As well, we defined the green valley with respect to the star forming main sequence, where green valley galaxies have $\Delta \text{SFMS} < - 0.5$, and we did not enforce a lower threshold, unlike both \citet{schawinski2014} and \citet{mcnab2021}. If we instead defined green valley galaxies in the Frontier Fields to be those with $-1 < \Delta \text{SFMS} < -0.5$, similar to \citet{abdurrouf2023}, we find fractions similar to those of \citet{schawinski2014} for their mixed field sample and \citet{mcnab2021} for their cluster and field samples, of ${\sim} 12$\% and ${\sim} 14$\%, respectively. In Figure~\ref{fig:GV_frac}, we repeat the analysis from above, as shown in Figure~\ref{fig:fractions_cl_fd}, using this narrower definition of the green valley. We find results that are consistent with those presented above, given the large uncertainties.

\subsection{Feasibility with Galaxy Surveys}

With the results presented above in mind, we are confident that the methodology as shown can be adapted to large samples of galaxies, in order to collect samples of quenching galaxies for further analysis. Specifically, current, upcoming, and proposed large-scale galaxy surveys with high spatial resolution will be able to map the distribution of star formation for galaxies across a range of redshifts. As described in \citetalias{lawlor2026b}, the Cosmological Advanced Survey Telescope for Optical and UV Research \citep[CASTOR;][]{cote2025}, a proposed far ultraviolet--blue-optical space telescope, will complete a number of community-driven surveys to image the ultraviolet$/$blue-optical sky. The resulting image quality will be comparable to HST, where the PSF full width at half maximum (FWHM) is $0 \farcs 15$ \citep{cheng2024a,cote2025}. Of the proposed surveys, the Wide, Deep, and Ultradeep surveys will form a tiered ``wedding cake'' approach, where the Wide survey will image $2227~\text{deg}^{2}$ to a $5 \sigma$ point source depth of ${\sim} 27.2~\text{AB mag}$ \citep{cote2025,marshall2025}. The Deep and Ultradeep surveys image smaller areas to greater depths, with the Ultradeep survey in particular imaging $1~\text{deg}^{2}$ to a depth of ${\sim} 30.6~\text{AB mag}$ across the available filters. As shown in \citetalias{lawlor2026b}, the Ultradeep survey alone will provide tens of thousands of intermediate-to-massive galaxies ($M_{*} > 10^{9.5}~M_{\odot}$) at intermediate redshifts, from which our methodology can be applied. Beyond CASTOR, the Ultraviolet Explorer \citep[UVEX;][]{kulkarni2021} will additionally provide imaging of the ultraviolet sky, though at lower resolution.

Aside from the critical ultraviolet observations for probing star formation \citep[e.g.,][]{vink2020}, upcoming missions like the Nancy Grace Roman Space Telescope \citep[NGRST;][]{spergel2013,spergel2015,akeson2019} will provide the complementary red-optical--near-infrared observations to trace the distribution of stellar mass. The imaging from NGRST will likewise be comparable in quality to HST, with a PSF FWHM less than $0 \farcs 15$ for all filters but the \textit{K}$_{\text{F213}}$-band (FWHM of $0 \farcs 169$; \citealt{rotac2025}, hereafter \rotact). NGRST will complete a High-Latitude Wide-Area Survey (HLWAS; e.g., \citealt{dore2018}; \rotact), which will similarly be comprised of wedding cake-style tiers, where the Medium tier will image $2415~\text{deg}^{2}$ to a point source depth of ${\sim} 26.4~\text{AB mag}$ across the \textit{Y}$_{\text{F106}}$-, \textit{J}$_{\text{F129}}$-, and \textit{H}$_{\text{F158}}$-bands. The Deep and Ultra deep tiers add additional filters and depth, at the expense of area, where the Ultradeep tier will image $5~\text{deg}^{2}$ to a depth of ${\sim} 28.1~\text{AB mag}$ \rotacp. As well, as shown in \citetalias{lawlor2026b}, we expect that the NGRST HLWAS Deep tier will include on the order of tens of massive ($M > 10^{14}~M_{\odot}$) galaxy clusters at intermediate redshifts similar to the Frontier Fields, which will enable a complementary environmental analysis, similar to the one presented here. In addition to NGRST, Euclid \citep{euclid2025a} and its Wide Survey \citep[${>} 14,600~\text{deg}^{2}$ to a depth of $24.4~\text{AB mag}$;][]{euclid2022,euclid2025a} could also provide imaging to determine stellar mass distributions, but care would need to be taken given the larger PSF \citep[${\sim} 0 \farcs 34$;][]{euclid2025a}.

\subsection{Limitations and Future Directions}

When considering the limitations of the current work, we find a few points that merit commentary. The first concerns our required SNR threshold and extent, in order to include prospective galaxies in our sample. Above, we required \textit{H}-band $\text{SNR} \geqslant 3$ for all annuli extending out to $5~R_{\text{e}}$, to ensure that reliable stellar mass estimates can be determined when fitting spectral energy distributions. Prior to this step, we had 3643 potential galaxies, while after we had the aforementioned 1987 galaxies. This represents a loss of ${\sim}$45\% of the initial sample of 3643 galaxies. If instead we could limit our analysis to $3~R_{\text{e}}$, then we would arrive at a sample of 3001 galaxies with \textit{H}-band $\text{SNR} \geqslant 3$ out to $3~R_{\text{e}}$. This potential sample of 3001 galaxies would then represent a loss of only ${\sim}$18\% of the initial sample. While this appears as an attractive option to increase the sample size, we do not seriously consider this, given the methodology presented in \citetalias{lawlor2026a} and refined in \citetalias{lawlor2026b}. In particular, as an initial consideration, we have shown above that the radial profiles for normal star forming galaxies clearly encode information out to $5~R_{\text{e}}$ (see Figure~\ref{fig:radial_profiles}). Further, much of the differing morphological information encoded by the metrics is captured beyond $3~R_{\text{e}}$, as seen in Figure~\ref{fig:MMs}, for the inside-out and outside-in quenching populations with $R_{\text{inner}}$ and $R_{\text{outer}}$. For these reasons, we believe that simply ``trimming down'' to an arbitrarily smaller radial extent is seriously detrimental to the methodology presented above, where it remains unclear what value that adds to the physical investigation, beyond merely increasing the size of the sample. There clearly remain superior methods for increasing the sample size, including using large galaxy surveys as detailed above.

An additional concern is the similarity between the simulated galaxies found in TNG50, that were used to form the training set during classification (see Section~\ref{subsec:metrics}), and the real galaxies found in the Frontier Fields. When comparing observations to simulated galaxies, it has been suggested that the AGN feedback model in TNG is overly strong \citep[e.g.,][]{mitchell2020,terrazas2020,donnari2021a,habouzit2022,ma2022,voit2024,wright2024}. This could result in dramatic inside-out quenching that is not representative of real data. Indeed, in \citetalias{lawlor2026a} we found examples of inside-out quenching galaxies with large central suppressions of star formation, though we did not compare with observations in that work. In addition, previous studies have proposed that the process regulating environmental gas removal may be overly efficient in TNG compared to observations \citep[e.g.,][]{diemer2019,starkenburg2019,stevens2019,chen2024}, though other studies have found good agreement \citep{stevens2021,stevens2023}. This could possibly result in outside-in quenching for TNG galaxies that is pronounced compared to real galaxies.

In the context of the work presented here, we find that while there is strong similarity on a morphological basis between corresponding populations of galaxies in TNG and the Frontier Fields (i.e. star forming, inside-out, and outside-in; see Figure~\ref{fig:MMs}), populations are not directly representative of one another. Therefore, as detailed in Section~\ref{subsec:metrics}, we opted to use the distance from the star forming main sequence, $\Delta \text{SFMS}$, as an additional input to the $\textit{k}$NN algorithm when classifying. We view this addition as an appropriate adaptation in order to leverage the results from the simulation, as presented in \citetalias{lawlor2026a}, for classification purposes. Furthermore, as discussed in Section~\ref{sec:results}, the distributions for $\Delta \text{SFMS}$ in TNG show significant overlap. Given this, in addition to the results of the random forest classification (completed to understand the importance of $\Delta \text{SFMS}$; see Section~\ref{sec:results}), we conclude that by including $\Delta \text{SFMS}$, we are not introducing biases toward expected results. Instead, we are using the available information to best adapt parameters that were designed for use with simulated galaxies, for use with real galaxies.

Finally, an additional limitation we face in the current work is the inability to predict quenching progress values for the quenching populations, based on the morphological metrics. In \citetalias{lawlor2026a}, once again using a \textit{k}NN classifier, we were able to make predictions for the elapsed normalized duration of quenching, based on the overall evolution of the metrics seen there. In the current work, we attempted to apply this analysis to the Frontier Fields quenching populations using a \textit{k}NN regressor, comparing it with the phase space-inferred infall time estimates. We found no strong correlation between the values for either population, which may suggest that there truly exists no correlation. However, performing a \textit{k}NN regression is inherently more difficult than a more straightforward classification, as while the classification only needs broad boundaries, regression requires the fine structure of the mapping, from metrics to quenching progress, to be more robust. This analysis thus requires additional consideration, and is one of the aspects that we will revisit in future work.

Regarding future directions, a logical next step is to apply the methodology outlined here to larger samples of galaxies. As discussed above, the next decade will provide observations of many tens of thousands of galaxies that we can test on, especially when considering CASTOR and NGRST. However, in the immediate future, we can look to existing high spatial resolution datasets including COSMOS \citep{scoville2007} and CANDELS \citep{grogin2011,koekemoer2011} to investigate field environments. We can additionally employ data from the Cluster Lensing And Supernova Survey with Hubble\footnote{\href{https://www.stsci.edu/~postman/CLASH}{https:$//$www.stsci.edu$/{\sim}$postman$/$CLASH}} \citep[CLASH;][]{postman2012} and the Reionization Lensing Cluster Survey\footnote{\href{https://relics.stsci.edu}{https:$//$relics.stsci.edu}} \citep[RELICS;][]{coe2019} to investigate dense cluster environments. Compelling programs which incorporate HST data with new JWST data include the Canadian NIRISS Unbiased Cluster Survey\footnote{\href{https://niriss.github.io}{https:$//$niriss.github.io}} \citep[CANUCS;][]{willott2022,sarrouh2026}, the JWST Advanced Deep Extragalactic Survey\footnote{\href{https://jades-survey.github.io}{https:$//$jades-survey.github.io}} \citep[JADES;][]{rieke2023b,eisenstein2026}, the JWST Extragalactic Medium-band Survey \citep[JEMS;][]{williams2023}, the Prime Extragalactic Areas for Reionization and Lensing Science\footnote{\href{https://sites.google.com/view/jwstpearls}{https:$//$sites.google.com$/$view$/$jwstpearls}} \citep[PEARLS;][]{windhorst2023} program, the Ultradeep NIRSpec and NIRCam Observations before the Epoch of Reionization\footnote{\href{https://jwst-uncover.github.io}{https:$//$jwst-uncover.github.io}} \citep[UNCOVER;][]{bezanson2024,suess2024} program, and the Gravitational Lensing \& NIRCam Imaging to Probe Early Galaxy Formation and Sources of Reionization \citep[GLIMPSE;][]{atek2025} program. As mentioned above, and as found in \citetalias{lawlor2026b}, the SNR requirements to accurately estimate stellar mass radial profiles are somewhat conservative (e.g., \textit{H}-band $\text{SNR} \geqslant 3$), but the limiting factor when using additional datasets is the availability of ultraviolet or blue-optical imaging, as is available in the Frontier Fields.

However, instead of relying on ultraviolet imaging to determine spatially resolved SFRs after SED fitting, one could alternatively probe the distribution of star formation using H$\alpha$ imaging. Such observations, specifically for the Frontier Fields and parallels, have been produced \citep{vulcani2015,vulcani2016,vulcani2017} from grating prism spectroscopy with HST in the Grism Lens-Amplified Survey from Space\footnote{\href{https://archive.stsci.edu/prepds/glass}{https:$//$archive.stsci.edu$/$prepds$/$glass}} \citep[GLASS;][]{treu2015}. Integral field unit observations have also been undertaken for the Frontier Fields clusters \citep{caminha2017,karman2017,lagattuta2017,mahler2018} and field galaxies \citep[e.g.,][]{poggianti2017}, additionally providing spatially resolved H$\alpha$ imaging. Besides the programs mentioned above, other works have completed narrow band H$\alpha$ imaging for intermediate redshift clusters \citep[e.g.,][]{kodama2004,koyama2010,koyama2011,koyama2018}, and the field \citep[e.g.,][]{sobral2013,matharu2022}. 

Finally, future work could additionally apply the morphological metrics to observations of galaxies known to be experiencing inside-out or outside-in quenching. For instance, Type 1 AGN, particularly Type 1 Seyferts \citep[e.g.,][]{seyfert1943,antonucci1993,urry1995}, as well as ram pressure stripped galaxies \citep[e.g.,][]{poggianti2017,poggianti2025,roberts2020}, are attractive classes for application. In addition to providing direct measurements of the metrics on such systems, this would help to further refine the metrics for use with real galaxies.

\section{Summary and Conclusion}\label{sec:summary}

In this paper, we have investigated quenching populations of galaxies present in the Hubble Frontier Fields. Using the high quality data available for the clusters and their parallel fields, we first selected a non-quenched primary sample of 1437 galaxies with $\text{SFR}_{\text{total}} \geqslant 10^{-3}~M_{\odot}~\text{yr}^{-1}$ at $0.14 < z < 0.67$, along with a quenched extended sample of 502 galaxies with lower SFRs from the same redshifts. We then applied the morphological metrics introduced in \citetalias{lawlor2026a} and refined in \citetalias{lawlor2026b} to our sample, where the morphological metrics collectively describe the spatial distribution of star formation. Through comparison with the simulated galaxies from \citetalias{lawlor2026a}, we classified quenching galaxies in the Frontier Fields into two main quenching pathways, which are described with different morphological signatures. The first is consistent with an inside-out quenching mechanism, where star formation is first suppressed in the interior of the galaxy, before proceeding outward. The second evolves in the opposite direction, where star formation is initially suppressed on the outskirts, and is consistent with outside-in quenching. Our main findings are as follows:
\begin{enumerate}[noitemsep]
    \item We find 129 galaxies have quenching consistent with an inside-out pathway, and an additional 70 galaxies have quenching consistent with an outside-in pathway. We find good agreement between the distributions of the metrics for our classified sample galaxies and their simulated counterparts (Figure~\ref{fig:MMs}). Inside-out quenching galaxies are more massive ($\log{[M_{*}/M_{\odot}]} = 10.1^{+0.6}_{-0.7}$) than outside-in quenching galaxies ($\log{[M_{*}/M_{\odot}]} = 9.3^{+0.9}_{-0.7}$) by $\Delta \log{(M_{*}/M_{\odot})} = 0.8^{+0.2}_{-0.1}~\text{dex}$, and both populations almost exclusively populate the green valley.
    
    \item Inside-out quenching galaxies are found more often in clusters, while the outside-in population predominantly reside in the field (Figure~\ref{fig:SFMS}). In the cluster, the two populations reside in similar regions of phase space (Figure~\ref{fig:phase_space}). These populations have the same median infall time, $4.0~\text{Gyr}$, but the inside-out population has a smaller spread in values (${\sim} 1.1~\text{Gyr}$) compared to the outside-in population (${\sim} 1.4~\text{Gyr}$), which may have a mild preference for more recent infall times. Galaxies present in clusters are additionally more massive than those in the field, by ${\sim} 0.5~\text{dex}$ on a per-population basis.
    
    \item Considering the prevalence of finding these systems in the cluster and the field relative to the entire non-quenched primary sample, we find that the fraction of inside-out quenching galaxies is strongly a function of stellar mass in the cluster (Figure~\ref{fig:fractions_cl_fd}). This population represents ${\sim}$30\% of the non-quenched sample at high masses ($M_{*} \gtrsim 10^{10.5}~M_{\odot}$), and also increases with mass in the field, but to a lesser degree. In both environments, the outside-in population shows no mass dependence, representing ${\lesssim}$10\% of the non-quenched sample at all masses. When considering the green valley, the combined quenching populations make up ${\sim}$30\% of the overall cluster population at all stellar masses, with a similar fraction found across all masses in the field.

    \item For cluster galaxies, when binned by infall time, the prevalence of these systems is similar to the overall cluster situation, where inside-out quenching galaxies similarly show a strong mass dependence (Figure~\ref{fig:fractions_tinfall}). The high mass end ($M_{*} \gtrsim 10^{10.5}~M_{\odot}$) shows mild infall time dependence as well, where these galaxies make up a larger fraction of the non-quenched sample at intermediate and ancient infall times, compared to recent infall times. The outside-in population shows no dependence on infall time, and is once again mass independent.
\end{enumerate}

Viewed holistically, our results suggest that, for a subset of green valley galaxies present in the Frontier Fields, the spatial distribution of their star formation shows quenching morphologies consistent with the two pathways identified in \citetalias{lawlor2026a}, where galaxies experience inside-out and outside-in quenching. When comparing our results from the simulation to real data, the inside-out population is similarly more massive than the outside-in population. However, observed inside-out quenching is more common in clusters, and further, observed high-mass satellites experience inside-out quenching in clusters. This mode additionally dominates over outside-in quenching in clusters, and is not what is expected from simple environmental quenching models \citep[e.g.,][]{abadi1999,roediger2005,hester2006}.

These results demonstrate that star formation quenching is more complicated in real life compared to simulations, but that simulations can be readily used to help infer quenching pathways for observed samples of galaxies. Current surveys and existing datasets offer many avenues for follow-up, and with upcoming large-scale galaxy surveys, the number of galaxies that can be investigated using this methodology will number in the millions. We are thus optimistic that our procedure will be a powerful tool for the future of galaxy evolution studies.

\section*{Acknowledgments}

C.L.F. thanks Kristi Webb for help with fitting software in the early stages of this project. The authors would like to acknowledge research grant funding that has enabled this research, including the support of the Natural Sciences and Engineering Research Council of Canada (NSERC) grants RGPIN-2018-03820 and RGPIN-2024-03878 for C.L.F. and M.L.B. S.L.M. acknowledges support from Science and Technology Facilities Council grants ST$/$W000946$/$1 and ST$/$Y000692$/$1. G.H.R. acknowledges the support of National Science Foundation grants AST-2206473 and AST-2308126, and grant 80NSSC21K0641 issued through the NASA Astrophysics Data Analysis Program (ADAP).

This research is based on observations made with the NASA$/$ESA Hubble Space Telescope obtained from the Mikulski Archive for Space Telescopes at the Space Telescope Science Institute, which is operated by the Association of Universities for Research in Astronomy, Inc., under NASA contract NAS 5-26555. These observations are associated with programs 09722, 10420, 10493, 10793, 11103, 11108, 11507, 11582, 11591, 11689, 12068, 12103, 12458, 12459, 13386, 13389, 13459, 13495, 13496, 13498, 13504, 13790, 14037, 14038, 14041, 14209, 14216. This work is additionally based on data and catalog products from HFF-DeepSpace, funded by the National Science Foundation and Space Telescope Science Institute.

The authors acknowledge the use of the Canadian Advanced Network for Astronomy Research (CANFAR) Science Platform operated by the Canadian Astronomy Data Center (CADC) and the Digital Research Alliance of Canada (DRAC), with support from the National Research Council of Canada (NRC), the Canadian Space Agency (CSA), CANARIE, and the Canadian Foundation for Innovation (CFI).

This research has made use of the Astrophysics Data System, funded by NASA under Cooperative Agreement 80NSSC25M7105, as well as \texttt{TOPCAT},\footnote{\href{https://www.star.bris.ac.uk/~mbt/topcat}{https:$//$www.star.bris.ac.uk$/{\sim}$mbt$/$topcat}} an interactive graphical viewer and editor for tabular data \citep{taylor2005}, in addition to \texttt{Astropy},\footnote{\href{https://www.astropy.org}{https:$//$www.astropy.org}} a community-developed core Python package and an ecosystem of tools and resources for astronomy \citep{astropy2013,astropy2018,astropy2022}, and \texttt{Photutils}, an Astropy package for detection and photometry of astronomical sources \citep{bradley2025}.

\vspace{0.3cm}
\textit{Facility:} HST (ACS, WFC3)

\vspace{0.3cm}
\textit{Software:} \texttt{Astropy} \citep{astropy2013,astropy2018,astropy2022}, \texttt{FAST++} \citep{kriek2009,schreiber2018a}, \texttt{h5py} \citep{collette2023}, \texttt{Matplotlib} \citep{hunter2007}, \texttt{NumPy} \citep{harris2020}, \texttt{Photutils} \citep{bradley2025}, \texttt{scikit-learn}\footnote{\href{https://scikit-learn.org}{https:$//$scikit-learn.org}} \citep{pedregosa2011}, \texttt{SciPy} \citep{virtanen2020}, \texttt{TOPCAT} \citep{taylor2005}

\appendix

{\renewcommand{\arraystretch}{1.5}
\begin{deluxetable*}{cll}
    \tablecaption{Available HST filters per pointing (clusters and parallel fields).\label{tab:available_filters}}
    \tablehead{\colhead{Cluster} & \colhead{Cluster Filters} & \colhead{Parallel Filters}}
    \startdata
    Abell~2744  & \textit{UV'}, \textit{u}, \textit{B}, \textit{V}$^{\text{W}}$, \textit{I}, \textit{Y}, \textit{J}, \textit{JH}, \textit{H} & \textit{B}, \textit{V}$^{\text{W}}$, \textit{I}, \textit{Y}, \textit{J}, \textit{JH}, \textit{H} \\[0.1cm]
    Abell~370   & \textit{UV'}, \textit{u}, \textit{B}, \textit{g}, \textit{V}$^{\text{W}}$, \textit{r}, \textit{I}, \textit{Y}, \textit{Y}$^{\text{W}}$, \textit{J}, \textit{JH}, \textit{H} & \textit{B}, \textit{V}$^{\text{W}}$, \textit{I}, \textit{Y}, \textit{J}, \textit{JH}, \textit{H} \\[0.1cm]
    MACS~J0416  & \textit{UV}, \textit{UV'}, \textit{u}, \textit{C}, \textit{B}, \textit{g}, \textit{V}$^{\text{W}}$, \textit{r}, \textit{i}, \textit{I}, \textit{z}, \textit{Y}, \textit{Y}$^{\text{W}}$, \textit{J}, \textit{JH}, \textit{H} & \textit{B}, \textit{V}$^{\text{W}}$, \textit{i}, \textit{I}, \textit{z}, \textit{Y}, \textit{J}, \textit{JH}, \textit{H} \\[0.1cm]
    MACS~J0717  & \textit{UV}, \textit{UV'}, \textit{u}, \textit{C}, \textit{B}, \textit{g}, \textit{V}, \textit{V}$^{\text{W}}$, \textit{r}, \textit{i}, \textit{I}, \textit{z}, \textit{Y}, \textit{Y}$^{\text{W}}$, \textit{J}, \textit{JH}, \textit{H} & \textit{B}, \textit{V}$^{\text{W}}$, \textit{I}, \textit{Y}, \textit{J}, \textit{JH}, \textit{H} \\[0.1cm]
    MACS~J1149  & \textit{UV}, \textit{UV'}, \textit{u}, \textit{C}, \textit{B}, \textit{g}, \textit{V}, \textit{V}$^{\text{W}}$, \textit{r}, \textit{i}, \textit{I}, \textit{z}, \textit{Y}, \textit{Y}$^{\text{W}}$, \textit{J}, \textit{JH}, \textit{H} & \textit{B}, \textit{V}$^{\text{W}}$, \textit{I}, \textit{Y}, \textit{J}, \textit{JH}, \textit{H} \\[0.1cm]
    Abell~S1063 & \textit{UV}, \textit{UV'}, \textit{u}, \textit{C}, \textit{B}, \textit{g}, \textit{V}$^{\text{W}}$, \textit{r}, \textit{i}, \textit{I}, \textit{z}, \textit{Y}, \textit{Y}$^{\text{W}}$, \textit{J}, \textit{JH}, \textit{H} & \textit{B}, \textit{V}$^{\text{W}}$, \textit{I}, \textit{Y}, \textit{J}, \textit{JH}, \textit{H} \vspace{0.1cm}
    \enddata
\end{deluxetable*}}

\newpage
\section{Filters per Cluster$/$Field}\label{app:filters}

Here, we tabulate the available filters per cluster$/$field for the Frontier Fields. In Table~\ref{tab:available_filters}, for each cluster we show the available HST filters, as well as the passbands available for that cluster's parallel field.

\section{SED Fitting Procedure for COSMOS-Web}\label{app:COSMOS_Web}

The COSMOS-Web team has produced catalogs of stellar population parameters, including stellar masses and global star formation rates, for galaxies present in the COSMOS survey footprint \citep{shuntov2025}. We select all secure (\texttt{warn\_flag = 0}) galaxies (\texttt{type = 0}) from the photometric catalog with \texttt{LePhare}-derived \citep{arnouts1999,arnouts2002,ilbert2006} photometric redshifts between $0.25 \leqslant z \leqslant 0.6$, and also require that the photometric redshifts are of high quality: $\sigma_{z, \text{norm}} = (z_{84} - z_{16})/2(1 + z_{50}) \leqslant 0.1$, where $z_{50}$, $z_{84}$, and $z_{16}$ are the tabulated median, upper, and lower $1 \sigma$ confidence values, respectively, computed from the photometric redshift probability distribution functions \citep{shuntov2025}. The COSMOS-Web galaxy properties we adopt are determined using the \texttt{CIGALE} \citep{boquien2019} SED modeling code through a Bayesian-like analysis \citep{arangotoro2025,shuntov2025}, that uses a nonparametric star formation history \citep{ciesla2023,ciesla2024}, \citet{bruzual2003} stellar population models, the \citet{dale2014} dust emission library, and a modified \citet{calzetti2000} attenuation law, as described in \citet{arangotoro2025}. \citet{arangotoro2025,shuntov2025} use 27 filters when completing their SED fitting with \texttt{CIGALE}, with coverage from the ultraviolet to the mid-infrared, including archival HST$/$ACS imaging, as well as JWST$/$Near Infrared Camera \citep[NIRCam;][]{rieke2023a,rigby2023} and JWST$/$Mid-Infrared Instrument \citep[MIRI;][]{wright2023a} data. From the \texttt{CIGALE}-derived catalogs we adopt the inferred stellar masses and star formation rates for the galaxies that meet our quality cuts described above.

\begin{figure*}[t]
    \centering
    \includegraphics[width=\textwidth]{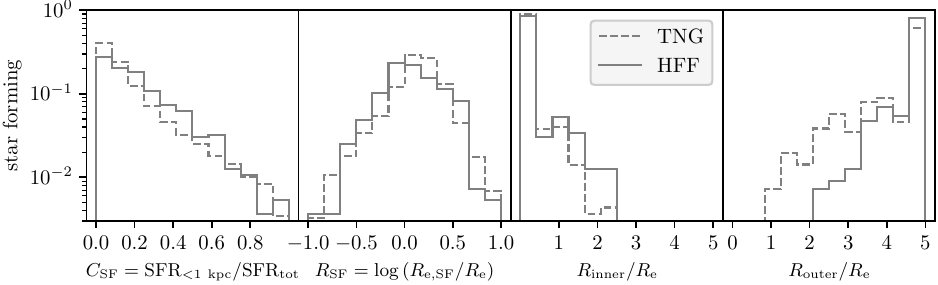}
    \caption{Similar to Figure~\ref{fig:MMs}, the morphological metric distributions (columns) for $k$NN-classified star forming galaxies (solid lines), compared with the corresponding distributions for the simulated galaxies (dashed lines). From left to right, the columns show $C_{\text{SF}}$, $R_{\text{SF}}$, $R_{\text{inner}}$, and $R_{\text{outer}}$, respectively, and each distribution has been normalized so that the $y$-axis can be read as frequency.}
    \label{fig:MMs_SF}
\end{figure*}

During our SFMS determination as described above in Section~\ref{subsec:SFMS}, we additionally investigated the evolution of the SFMS and the resulting fits, by binning the COSMOS-Web galaxies into three redshift bins: $0.25 \leqslant z \leqslant 0.36$, $0.36 \leqslant z \leqslant 0.48$, and $0.48 \leqslant z \leqslant 0.6$. Though we find some evolution across the three considered redshift bins, this evolution is very gentle when we ultimately consider the effect that it has on our sample of Frontier Fields galaxies, where it shifts the number of SFMS and green valley galaxies by a modest amount, on the order of less than tens of galaxies. Given this, we opted to use a single fit for the entire considered redshift range, $0.25 \leqslant z \leqslant 0.6$.

\section{Morphological Metric Distributions for Star Forming Galaxies}\label{app:SF_MM_dists}

Here, we compare the distributions of the morphological metrics for $k$NN-classified star forming galaxies in the Frontier Fields with star forming galaxies drawn from the TNG simulation, as described in \citetalias{lawlor2026a}. These $k$NN-classified star forming galaxies number 556 galaxies from the non-quenched primary sample of 1437 galaxies, and predominantly reside in the field (${\sim}$76\%), though there are a number found in the cluster (131). In Figure~\ref{fig:MMs_SF}, we show the Frontier Fields distributions with solid lines, while the distributions for the simulated galaxies are shown with dashed lines. As in Section~\ref{sec:results}, we find very good agreement between the respective distributions for each morphological metric. As above, applying a two-sample KS test, we find that, for each metric, the distribution for the Frontier Fields galaxies is consistent with being drawn from the same distribution as the simulated galaxies. As stated earlier, though we do not focus on this collection of star forming galaxies, the similarity of these distributions, in combination with the results of the KS test, serve as a broad consistency check. These results are encouraging, and suggest that the $k$NN classification can properly identify star forming galaxies in the Frontier Fields on the basis of their morphological metrics, when trained on unrelated star forming galaxies drawn from the TNG50 simulation.

\section{Overlap with CANUCS and UNCOVER}\label{app:UNCOVER_CANUCS}

Here, we consider the number and prevalence of galaxies that populate the green valley in the SFR--stellar mass plane (e.g., Figure~\ref{fig:SFMS}), given that we might expect a dominant star forming population based on the additional observations of the flanking$/$parallel fields. Correspondingly, this would then produce a minority population of green valley galaxies. In particular, we compare with other large programs targeting the Frontier Fields clusters, and consider the results from CANUCS \citep{willott2022,sarrouh2026} and UNCOVER \citep{bezanson2024}. Both CANUCS and UNCOVER have produced catalogs of stellar population parameters, including stellar masses and integrated star formation rates, for galaxies present in the Frontier Fields \citep{wang2023a,wang2024} and flanking fields \citep{sarrouh2026}. The CANUCS observations include Abell~370, MACS~J0416, and MACS~J1149 \citep{willott2022}, while the UNCOVER observations consider Abell~2744 \citep{bezanson2024}.

Regarding the CANUCS observations, the galaxy properties were determined using the SED fitting code \texttt{Bagpipes} \citep{carnall2018,carnall2019} with a \citet{calzetti2000} dust attenuation law, a \citet{chabrier2003} initial mass function, \citet{bruzual2003} stellar models, and nebular emission from \citet{ferland2017}, as described in \citet{sarrouh2026}. \citet{sarrouh2026} use a parametric double power law star formation history, that allows for a rising and falling component, and use the broadband photometric data from CANUCS, the medium and narrowband follow-up ``JWST in Technicolor'' (GO 3362; PI: A.~Muzzin), and archival HST$/$ACS and HST$/$WFC3 imaging, spanning from the rest-frame ultraviolet to the near-infrared.

Regarding the UNCOVER observations, the galaxy properties are inferred using the \texttt{Prospector} inference code \citep{leja2017,leja2019a,johnson2021} with simple stellar populations drawn from the Flexible Stellar Population Synthesis package \citep{conroy2010}, along with MIST stellar isochrones \citep{choi2016,dotter2016}, using the MILES stellar library \citep{sanchez2006}, as described in \citet{wang2023a,wang2024}. \citet{wang2023a} employ a nonparametric star formation history, using seven logarithmically-spaced time bins \citep{leja2017}, and use the complete set of broadband photometric data from UNCOVER \citep{bezanson2024,weaver2024}, the medium-band follow-up MegaScience \citep{suess2024}, and all available archival HST$/$ACS and HST$/$WFC3 imaging, spanning from the rest-frame ultraviolet to the near-infrared. For both the UNCOVER and CANUCS catalogs, we adopt their median stellar mass estimates, along with their median time-averaged ($\Delta t = 100~\text{Myr}$) star formation rates.

To the best of our knowledge, no team has yet produced large catalogs of galaxy physical properties for galaxies in Abell~S1063 or MACS~J0717, though the Vast Exploration for Nascent Unexplored Sources\footnote{\href{https://jwst-venus.github.io}{https:$//$jwst-venus.github.io}} (VENUS; GO 6882; PIs: S.~Fujimoto and D.~Coe) program has plans to release such catalogs in late 2026. These catalogs will include stellar population parameters for galaxies in the 60 massive lensing clusters that make up the VENUS program, including Abell~S1063 and MACS~J0717. Additionally, though the GLIMPSE program \citep{atek2025} has released photometric catalogs, catalogs tabulating galaxy physical properties based on SED fitting have yet to be published. Nonetheless, by considering the galaxies available from the UNCOVER and CANUCS programs across the four available Frontier Fields clusters, and by adopting similar redshift limits ($0.14 \leqslant z \leqslant 0.67$) and stellar mass limits ($M_{*} \geqslant 10^{8}~M_{\odot}$) as above, we arrive at a supplementary sample of some 4700 galaxies from which we can compare with the distribution of our galaxies in the SFR--stellar mass plane. While we do not have comparison galaxies for Abell~S1063 and MACS~J0717, MACS~J0717 is close in redshift to MACS~J1149 (see Table~\ref{tab:HFF}), that is sampled. As well, Abell~S1063 sits at a redshift intermediate between Abell~2744 and Abell~370, both of which are sampled.

We find substantial overlap in the SFR--stellar mass plane between the locations of our non-quenched primary sample galaxies and galaxies included in UNCOVER and CANUCS. This suggests that though we do not find a dominant star forming population, even when including the parallel fields, this population is similarly absent in the UNCOVER and CANUCS samples. Given this, we are confident that there are no systematic effects in our SED fitting procedure that lead to suppressed SFRs, and that the inferred stellar masses and SFRs are qualitatively robust against different SED fitting procedures. Furthermore, given that the Frontier Fields target massive lensing clusters, the lack of a dominant star forming population is not completely unexpected \citep[e.g.,][]{lewis2002,gomez2003,kauffmann2004}.

\newpage
\section*{ORCID iDs}
\begingroup
\raggedright
Cameron~Lawlor-Forsyth \orcidlink{0000-0002-2958-0593} \\\href{https://orcid.org/0000-0002-2958-0593}{https:$//$orcid.org$/$0000-0002-2958-0593}\par
Michael~L.~Balogh \orcidlink{0000-0003-4849-9536} \href{https://orcid.org/0000-0003-4849-9536}{https:$//$orcid.org$/$0000-0003-4849-9536}\par
Sean~L.~McGee \orcidlink{0000-0003-3255-3139} \href{https://orcid.org/0000-0003-3255-3139}{https:$//$orcid.org$/$0000-0003-3255-3139}\par
Gregory~H.~Rudnick \orcidlink{0000-0001-5851-1856} \\\href{https://orcid.org/0000-0001-5851-1856}{https:$//$orcid.org$/$0000-0001-5851-1856}
\endgroup

\bibliography{references}{}

@ARTICLE{abadi1999,
       author = {{Abadi}, Mario G. and {Moore}, Ben and {Bower}, Richard G.},
        title = "{Ram pressure stripping of spiral galaxies in clusters}",
      journal = {\mnras},
         year = 1999,
        month = oct,
       volume = {308},
       number = {4},
        pages = {947-954},
          doi = {10.1046/j.1365-8711.1999.02715.x},
archivePrefix = {arXiv},
       eprint = {astro-ph/9903436},
 primaryClass = {astro-ph},
       adsurl = {https://ui.adsabs.harvard.edu/abs/1999MNRAS.308..947A}
}

@ARTICLE{abdurrouf2018,
       author = {{Abdurro'uf} and {Akiyama}, Masayuki},
        title = "{Evolution of spatially resolved star formation main sequence and surface density profiles in massive disc galaxies at 0 {\ensuremath{\lesssim}} z {\ensuremath{\lesssim}} 1: inside-out stellar mass buildup and quenching}",
      journal = {\mnras},
         year = 2018,
        month = oct,
       volume = {479},
       number = {4},
        pages = {5083-5100},
          doi = {10.1093/mnras/sty1771},
archivePrefix = {arXiv},
       eprint = {1802.03782},
 primaryClass = {astro-ph.GA},
       adsurl = {https://ui.adsabs.harvard.edu/abs/2018MNRAS.479.5083A}
}

@ARTICLE{abdurrouf2022a,
       author = {{Abdurro'uf} and {Lin}, Yen-Ting and {Hirashita}, Hiroyuki and {Morishita}, Takahiro and {Tacchella}, Sandro and {Akiyama}, Masayuki and {Takeuchi}, Tsutomu T. and {Wu}, Po-Feng},
        title = "{Dissecting Nearby Galaxies with piXedfit. I. Spatially Resolved Properties of Stars, Dust, and Gas as Revealed by Panchromatic SED Fitting}",
      journal = {\apj},
         year = 2022,
        month = feb,
       volume = {926},
       number = {1},
          eid = {81},
        pages = {81},
          doi = {10.3847/1538-4357/ac439a},
archivePrefix = {arXiv},
       eprint = {2110.03158},
 primaryClass = {astro-ph.GA},
       adsurl = {https://ui.adsabs.harvard.edu/abs/2022ApJ...926...81A}
}

@ARTICLE{abdurrouf2023,
       author = {{Abdurro'uf} and {Coe}, Dan and {Jung}, Intae and {Ferguson}, Henry C. and {Brammer}, Gabriel and {Iyer}, Kartheik G. and {Bradley}, Larry D. and {Dayal}, Pratika and {Windhorst}, Rogier A. and {Zitrin}, Adi and {Meena}, Ashish Kumar and {Oguri}, Masamune and {Diego}, Jose M. and {Kokorev}, Vasily and {Dimauro}, Paola and {Adamo}, Angela and {Conselice}, Christopher J. and {Welch}, Brian and {Vanzella}, Eros and {Hsiao}, Tiger Yu-Yang and {Xu}, Xinfeng and {Roy}, Namrata and {Mulcahey}, Celia R.},
        title = "{Spatially Resolved Stellar Populations of 0.3 < z < 6.0 Galaxies in WHL 0137-08 and MACS 0647+70 Clusters as Revealed by JWST: How Do Galaxies Grow and Quench over Cosmic Time?}",
      journal = {\apj},
         year = 2023,
        month = mar,
       volume = {945},
       number = {2},
          eid = {117},
        pages = {117},
          doi = {10.3847/1538-4357/acba06},
archivePrefix = {arXiv},
       eprint = {2301.02209},
 primaryClass = {astro-ph.GA},
       adsurl = {https://ui.adsabs.harvard.edu/abs/2023ApJ...945..117A}
}

@ARTICLE{abell1958,
       author = {{Abell}, George O.},
        title = "{The Distribution of Rich Clusters of Galaxies.}",
      journal = {\apjs},
         year = 1958,
        month = may,
       volume = {3},
        pages = {211},
          doi = {10.1086/190036},
       adsurl = {https://ui.adsabs.harvard.edu/abs/1958ApJS....3..211A}
}

@ARTICLE{abell1989,
       author = {{Abell}, George O. and {Corwin}, Jr., Harold G. and {Olowin}, Ronald P.},
        title = "{A Catalog of Rich Clusters of Galaxies}",
      journal = {\apjs},
         year = 1989,
        month = may,
       volume = {70},
        pages = {1},
          doi = {10.1086/191333},
       adsurl = {https://ui.adsabs.harvard.edu/abs/1989ApJS...70....1A}
}

@ARTICLE{akeson2019,
       author = {{Akeson}, Rachel and {Armus}, Lee and {Bachelet}, Etienne and {Bailey}, Vanessa and {Bartusek}, Lisa and {Bellini}, Andrea and {Benford}, Dominic and {Bennett}, David and {Bhattacharya}, Aparna and {Bohlin}, Ralph and {Boyer}, Martha and {Bozza}, Valerio and {Bryden}, Geoffrey and {Calchi Novati}, Sebastiano and {Carpenter}, Kenneth and {Casertano}, Stefano and {Choi}, Ami and {Content}, David and {Dayal}, Pratika and {Dressler}, Alan and {Dor{\'e}}, Olivier and {Fall}, S. Michael and {Fan}, Xiaohui and {Fang}, Xiao and {Filippenko}, Alexei and {Finkelstein}, Steven and {Foley}, Ryan and {Furlanetto}, Steven and {Kalirai}, Jason and {Gaudi}, B. Scott and {Gilbert}, Karoline and {Girard}, Julien and {Grady}, Kevin and {Greene}, Jenny and {Guhathakurta}, Puragra and {Heinrich}, Chen and {Hemmati}, Shoubaneh and {Hendel}, David and {Henderson}, Calen and {Henning}, Thomas and {Hirata}, Christopher and {Ho}, Shirley and {Huff}, Eric and {Hutter}, Anne and {Jansen}, Rolf and {Jha}, Saurabh and {Johnson}, Samson and {Jones}, David and {Kasdin}, Jeremy and {Kelly}, Patrick and {Kirshner}, Robert and {Koekemoer}, Anton and {Kruk}, Jeffrey and {Lewis}, Nikole and {Macintosh}, Bruce and {Madau}, Piero and {Malhotra}, Sangeeta and {Mandel}, Kaisey and {Massara}, Elena and {Masters}, Daniel and {McEnery}, Julie and {McQuinn}, Kristen and {Melchior}, Peter and {Melton}, Mark and {Mennesson}, Bertrand and {Peeples}, Molly and {Penny}, Matthew and {Perlmutter}, Saul and {Pisani}, Alice and {Plazas}, Andr{\'e}s and {Poleski}, Radek and {Postman}, Marc and {Ranc}, Cl{\'e}ment and {Rauscher}, Bernard and {Rest}, Armin and {Roberge}, Aki and {Robertson}, Brant and {Rodney}, Steven and {Rhoads}, James and {Rhodes}, Jason and {Ryan}, Jr., Russell and {Sahu}, Kailash and {Sand}, David and {Scolnic}, Dan and {Seth}, Anil and {Shvartzvald}, Yossi and {Siellez}, Karelle and {Smith}, Arfon and {Spergel}, David and {Stassun}, Keivan and {Street}, Rachel and {Strolger}, Louis-Gregory and {Szalay}, Alexander and {Trauger}, John and {Troxel}, M.~A. and {Turnbull}, Margaret and {van der Marel}, Roeland and {von der Linden}, Anja and {Wang}, Yun and {Weinberg}, David and {Williams}, Benjamin and {Windhorst}, Rogier and {Wollack}, Edward and {Wu}, Hao-Yi and {Yee}, Jennifer and {Zimmerman}, Neil},
        title = "{The Wide Field Infrared Survey Telescope: 100 Hubbles for the 2020s}",
      journal = {arXiv e-prints},
         year = 2019,
        month = feb,
          eid = {arXiv:1902.05569},
        pages = {arXiv:1902.05569},
          doi = {10.48550/arXiv.1902.05569},
archivePrefix = {arXiv},
       eprint = {1902.05569},
 primaryClass = {astro-ph.IM},
       adsurl = {https://ui.adsabs.harvard.edu/abs/2019arXiv190205569A}
}

@ARTICLE{alongi1993,
       author = {{Alongi}, M. and {Bertelli}, G. and {Bressan}, A. and {Chiosi}, C. and {Fagotto}, F. and {Greggio}, L. and {Nasi}, E.},
        title = "{Evolutionary sequences of stellar models with semiconvection and convective overshoot. I. Z=0.008.}",
      journal = {\aaps},
         year = 1993,
        month = mar,
       volume = {97},
        pages = {851-871},
       adsurl = {https://ui.adsabs.harvard.edu/abs/1993A&AS...97..851A}
}

@ARTICLE{astropy2013,
       author = {{Astropy Collaboration} and {Robitaille}, Thomas P. and {Tollerud}, Erik J. and {Greenfield}, Perry and {Droettboom}, Michael and {Bray}, Erik and {Aldcroft}, Tom and {Davis}, Matt and {Ginsburg}, Adam and {Price-Whelan}, Adrian M. and {Kerzendorf}, Wolfgang E. and {Conley}, Alexander and {Crighton}, Neil and {Barbary}, Kyle and {Muna}, Demitri and {Ferguson}, Henry and {Grollier}, Fr{\'e}d{\'e}ric and {Parikh}, Madhura M. and {Nair}, Prasanth H. and {Unther}, Hans M. and {Deil}, Christoph and {Woillez}, Julien and {Conseil}, Simon and {Kramer}, Roban and {Turner}, James E.~H. and {Singer}, Leo and {Fox}, Ryan and {Weaver}, Benjamin A. and {Zabalza}, Victor and {Edwards}, Zachary I. and {Azalee Bostroem}, K. and {Burke}, D.~J. and {Casey}, Andrew R. and {Crawford}, Steven M. and {Dencheva}, Nadia and {Ely}, Justin and {Jenness}, Tim and {Labrie}, Kathleen and {Lim}, Pey Lian and {Pierfederici}, Francesco and {Pontzen}, Andrew and {Ptak}, Andy and {Refsdal}, Brian and {Servillat}, Mathieu and {Streicher}, Ole},
        title = "{Astropy: A community Python package for astronomy}",
      journal = {\aap},
         year = 2013,
        month = oct,
       volume = {558},
          eid = {A33},
        pages = {A33},
          doi = {10.1051/0004-6361/201322068},
archivePrefix = {arXiv},
       eprint = {1307.6212},
 primaryClass = {astro-ph.IM},
       adsurl = {https://ui.adsabs.harvard.edu/abs/2013A&A...558A..33A}
}

@ARTICLE{astropy2018,
       author = {{Astropy Collaboration} and {Price-Whelan}, A.~M. and {Sip{\H{o}}cz}, B.~M. and {G{\"u}nther}, H.~M. and {Lim}, P.~L. and {Crawford}, S.~M. and {Conseil}, S. and {Shupe}, D.~L. and {Craig}, M.~W. and {Dencheva}, N. and {Ginsburg}, A. and {VanderPlas}, J.~T. and {Bradley}, L.~D. and {P{\'e}rez-Su{\'a}rez}, D. and {de Val-Borro}, M. and {Aldcroft}, T.~L. and {Cruz}, K.~L. and {Robitaille}, T.~P. and {Tollerud}, E.~J. and {Ardelean}, C. and {Babej}, T. and {Bach}, Y.~P. and {Bachetti}, M. and {Bakanov}, A.~V. and {Bamford}, S.~P. and {Barentsen}, G. and {Barmby}, P. and {Baumbach}, A. and {Berry}, K.~L. and {Biscani}, F. and {Boquien}, M. and {Bostroem}, K.~A. and {Bouma}, L.~G. and {Brammer}, G.~B. and {Bray}, E.~M. and {Breytenbach}, H. and {Buddelmeijer}, H. and {Burke}, D.~J. and {Calderone}, G. and {Cano Rodr{\'\i}guez}, J.~L. and {Cara}, M. and {Cardoso}, J.~V.~M. and {Cheedella}, S. and {Copin}, Y. and {Corrales}, L. and {Crichton}, D. and {D'Avella}, D. and {Deil}, C. and {Depagne}, {\'E}. and {Dietrich}, J.~P. and {Donath}, A. and {Droettboom}, M. and {Earl}, N. and {Erben}, T. and {Fabbro}, S. and {Ferreira}, L.~A. and {Finethy}, T. and {Fox}, R.~T. and {Garrison}, L.~H. and {Gibbons}, S.~L.~J. and {Goldstein}, D.~A. and {Gommers}, R. and {Greco}, J.~P. and {Greenfield}, P. and {Groener}, A.~M. and {Grollier}, F. and {Hagen}, A. and {Hirst}, P. and {Homeier}, D. and {Horton}, A.~J. and {Hosseinzadeh}, G. and {Hu}, L. and {Hunkeler}, J.~S. and {Ivezi{\'c}}, {\v{Z}}. and {Jain}, A. and {Jenness}, T. and {Kanarek}, G. and {Kendrew}, S. and {Kern}, N.~S. and {Kerzendorf}, W.~E. and {Khvalko}, A. and {King}, J. and {Kirkby}, D. and {Kulkarni}, A.~M. and {Kumar}, A. and {Lee}, A. and {Lenz}, D. and {Littlefair}, S.~P. and {Ma}, Z. and {Macleod}, D.~M. and {Mastropietro}, M. and {McCully}, C. and {Montagnac}, S. and {Morris}, B.~M. and {Mueller}, M. and {Mumford}, S.~J. and {Muna}, D. and {Murphy}, N.~A. and {Nelson}, S. and {Nguyen}, G.~H. and {Ninan}, J.~P. and {N{\"o}the}, M. and {Ogaz}, S. and {Oh}, S. and {Parejko}, J.~K. and {Parley}, N. and {Pascual}, S. and {Patil}, R. and {Patil}, A.~A. and {Plunkett}, A.~L. and {Prochaska}, J.~X. and {Rastogi}, T. and {Reddy Janga}, V. and {Sabater}, J. and {Sakurikar}, P. and {Seifert}, M. and {Sherbert}, L.~E. and {Sherwood-Taylor}, H. and {Shih}, A.~Y. and {Sick}, J. and {Silbiger}, M.~T. and {Singanamalla}, S. and {Singer}, L.~P. and {Sladen}, P.~H. and {Sooley}, K.~A. and {Sornarajah}, S. and {Streicher}, O. and {Teuben}, P. and {Thomas}, S.~W. and {Tremblay}, G.~R. and {Turner}, J.~E.~H. and {Terr{\'o}n}, V. and {van Kerkwijk}, M.~H. and {de la Vega}, A. and {Watkins}, L.~L. and {Weaver}, B.~A. and {Whitmore}, J.~B. and {Woillez}, J. and {Zabalza}, V. and {Astropy Contributors}},
        title = "{The Astropy Project: Building an Open-science Project and Status of the v2.0 Core Package}",
      journal = {\aj},
         year = 2018,
        month = sep,
       volume = {156},
       number = {3},
          eid = {123},
        pages = {123},
          doi = {10.3847/1538-3881/aabc4f},
archivePrefix = {arXiv},
       eprint = {1801.02634},
 primaryClass = {astro-ph.IM},
       adsurl = {https://ui.adsabs.harvard.edu/abs/2018AJ....156..123A}
}

@ARTICLE{astropy2022,
       author = {{Astropy Collaboration} and {Price-Whelan}, Adrian M. and {Lim}, Pey Lian and {Earl}, Nicholas and {Starkman}, Nathaniel and {Bradley}, Larry and {Shupe}, David L. and {Patil}, Aarya A. and {Corrales}, Lia and {Brasseur}, C.~E. and {N{\"o}the}, Maximilian and {Donath}, Axel and {Tollerud}, Erik and {Morris}, Brett M. and {Ginsburg}, Adam and {Vaher}, Eero and {Weaver}, Benjamin A. and {Tocknell}, James and {Jamieson}, William and {van Kerkwijk}, Marten H. and {Robitaille}, Thomas P. and {Merry}, Bruce and {Bachetti}, Matteo and {G{\"u}nther}, H. Moritz and {Aldcroft}, Thomas L. and {Alvarado-Montes}, Jaime A. and {Archibald}, Anne M. and {B{\'o}di}, Attila and {Bapat}, Shreyas and {Barentsen}, Geert and {Baz{\'a}n}, Juanjo and {Biswas}, Manish and {Boquien}, M{\'e}d{\'e}ric and {Burke}, D.~J. and {Cara}, Daria and {Cara}, Mihai and {Conroy}, Kyle E. and {Conseil}, Simon and {Craig}, Matthew W. and {Cross}, Robert M. and {Cruz}, Kelle L. and {D'Eugenio}, Francesco and {Dencheva}, Nadia and {Devillepoix}, Hadrien A.~R. and {Dietrich}, J{\"o}rg P. and {Eigenbrot}, Arthur Davis and {Erben}, Thomas and {Ferreira}, Leonardo and {Foreman-Mackey}, Daniel and {Fox}, Ryan and {Freij}, Nabil and {Garg}, Suyog and {Geda}, Robel and {Glattly}, Lauren and {Gondhalekar}, Yash and {Gordon}, Karl D. and {Grant}, David and {Greenfield}, Perry and {Groener}, Austen M. and {Guest}, Steve and {Gurovich}, Sebastian and {Handberg}, Rasmus and {Hart}, Akeem and {Hatfield-Dodds}, Zac and {Homeier}, Derek and {Hosseinzadeh}, Griffin and {Jenness}, Tim and {Jones}, Craig K. and {Joseph}, Prajwel and {Kalmbach}, J. Bryce and {Karamehmetoglu}, Emir and {Ka{\l}uszy{\'n}ski}, Miko{\l}aj and {Kelley}, Michael S.~P. and {Kern}, Nicholas and {Kerzendorf}, Wolfgang E. and {Koch}, Eric W. and {Kulumani}, Shankar and {Lee}, Antony and {Ly}, Chun and {Ma}, Zhiyuan and {MacBride}, Conor and {Maljaars}, Jakob M. and {Muna}, Demitri and {Murphy}, N.~A. and {Norman}, Henrik and {O'Steen}, Richard and {Oman}, Kyle A. and {Pacifici}, Camilla and {Pascual}, Sergio and {Pascual-Granado}, J. and {Patil}, Rohit R. and {Perren}, Gabriel I. and {Pickering}, Timothy E. and {Rastogi}, Tanuj and {Roulston}, Benjamin R. and {Ryan}, Daniel F. and {Rykoff}, Eli S. and {Sabater}, Jose and {Sakurikar}, Parikshit and {Salgado}, Jes{\'u}s and {Sanghi}, Aniket and {Saunders}, Nicholas and {Savchenko}, Volodymyr and {Schwardt}, Ludwig and {Seifert-Eckert}, Michael and {Shih}, Albert Y. and {Jain}, Anany Shrey and {Shukla}, Gyanendra and {Sick}, Jonathan and {Simpson}, Chris and {Singanamalla}, Sudheesh and {Singer}, Leo P. and {Singhal}, Jaladh and {Sinha}, Manodeep and {Sip{\H{o}}cz}, Brigitta M. and {Spitler}, Lee R. and {Stansby}, David and {Streicher}, Ole and {{\v{S}}umak}, Jani and {Swinbank}, John D. and {Taranu}, Dan S. and {Tewary}, Nikita and {Tremblay}, Grant R. and {de Val-Borro}, Miguel and {Van Kooten}, Samuel J. and {Vasovi{\'c}}, Zlatan and {Verma}, Shresth and {de Miranda Cardoso}, Jos{\'e} Vin{\'\i}cius and {Williams}, Peter K.~G. and {Wilson}, Tom J. and {Winkel}, Benjamin and {Wood-Vasey}, W.~M. and {Xue}, Rui and {Yoachim}, Peter and {Zhang}, Chen and {Zonca}, Andrea and {Astropy Project Contributors}},
        title = "{The Astropy Project: Sustaining and Growing a Community-oriented Open-source Project and the Latest Major Release (v5.0) of the Core Package}",
      journal = {\apj},
         year = 2022,
        month = aug,
       volume = {935},
       number = {2},
          eid = {167},
        pages = {167},
          doi = {10.3847/1538-4357/ac7c74},
archivePrefix = {arXiv},
       eprint = {2206.14220},
 primaryClass = {astro-ph.IM},
       adsurl = {https://ui.adsabs.harvard.edu/abs/2022ApJ...935..167A}
}

@ARTICLE{balogh2000,
       author = {{Balogh}, Michael L. and {Navarro}, Julio F. and {Morris}, Simon L.},
        title = "{The Origin of Star Formation Gradients in Rich Galaxy Clusters}",
      journal = {\apj},
         year = 2000,
        month = sep,
       volume = {540},
       number = {1},
        pages = {113-121},
          doi = {10.1086/309323},
archivePrefix = {arXiv},
       eprint = {astro-ph/0004078},
 primaryClass = {astro-ph},
       adsurl = {https://ui.adsabs.harvard.edu/abs/2000ApJ...540..113B}
}

@ARTICLE{belfiore2018,
       author = {{Belfiore}, Francesco and {Maiolino}, Roberto and {Bundy}, Kevin and {Masters}, Karen and {Bershady}, Matthew and {Oyarz{\'u}n}, Grecco A. and {Lin}, Lihwai and {Cano-Diaz}, Mariana and {Wake}, David and {Spindler}, Ashley and {Thomas}, Daniel and {Brownstein}, Joel R. and {Drory}, Niv and {Yan}, Renbin},
        title = "{SDSS IV MaNGA - sSFR profiles and the slow quenching of discs in green valley galaxies}",
      journal = {\mnras},
         year = 2018,
        month = jul,
       volume = {477},
       number = {3},
        pages = {3014-3029},
          doi = {10.1093/mnras/sty768},
archivePrefix = {arXiv},
       eprint = {1710.05034},
 primaryClass = {astro-ph.GA},
       adsurl = {https://ui.adsabs.harvard.edu/abs/2018MNRAS.477.3014B}
}

@ARTICLE{bell2004,
       author = {{Bell}, Eric F. and {Wolf}, Christian and {Meisenheimer}, Klaus and {Rix}, Hans-Walter and {Borch}, Andrea and {Dye}, Simon and {Kleinheinrich}, Martina and {Wisotzki}, Lutz and {McIntosh}, Daniel H.},
        title = "{Nearly 5000 Distant Early-Type Galaxies in COMBO-17: A Red Sequence and Its Evolution since z\raisebox{-0.5ex}\textasciitilde1}",
      journal = {\apj},
         year = 2004,
        month = jun,
       volume = {608},
       number = {2},
        pages = {752-767},
          doi = {10.1086/420778},
archivePrefix = {arXiv},
       eprint = {astro-ph/0303394},
 primaryClass = {astro-ph},
       adsurl = {https://ui.adsabs.harvard.edu/abs/2004ApJ...608..752B}
}

@ARTICLE{bertin1996,
       author = {{Bertin}, E. and {Arnouts}, S.},
        title = "{SExtractor: Software for source extraction.}",
      journal = {\aaps},
         year = 1996,
        month = jun,
       volume = {117},
        pages = {393-404},
          doi = {10.1051/aas:1996164},
       adsurl = {https://ui.adsabs.harvard.edu/abs/1996A&AS..117..393B}
}

@ARTICLE{bezanson2024,
       author = {{Bezanson}, Rachel and {Labbe}, Ivo and {Whitaker}, Katherine E. and {Leja}, Joel and {Price}, Sedona H. and {Franx}, Marijn and {Brammer}, Gabriel and {Marchesini}, Danilo and {Zitrin}, Adi and {Wang}, Bingjie and {Weaver}, John R. and {Furtak}, Lukas J. and {Atek}, Hakim and {Coe}, Dan and {Cutler}, Sam E. and {Dayal}, Pratika and {van Dokkum}, Pieter and {Feldmann}, Robert and {F{\"o}rster Schreiber}, Natascha M. and {Fujimoto}, Seiji and {Geha}, Marla and {Glazebrook}, Karl and {de Graaff}, Anna and {Greene}, Jenny E. and {Juneau}, St{\'e}phanie and {Kassin}, Susan and {Kriek}, Mariska and {Khullar}, Gourav and {Maseda}, Michael and {Mowla}, Lamiya A. and {Muzzin}, Adam and {Nanayakkara}, Themiya and {Nelson}, Erica J. and {Oesch}, Pascal A. and {Pacifici}, Camilla and {Pan}, Richard and {Papovich}, Casey and {Setton}, David J. and {Shapley}, Alice E. and {Smit}, Renske and {Stefanon}, Mauro and {Taylor}, Edward N. and {Williams}, Christina C.},
        title = "{The JWST UNCOVER Treasury Survey: Ultradeep NIRSpec and NIRCam Observations before the Epoch of Reionization}",
      journal = {\apj},
         year = 2024,
        month = oct,
       volume = {974},
       number = {1},
          eid = {92},
        pages = {92},
          doi = {10.3847/1538-4357/ad66cf},
archivePrefix = {arXiv},
       eprint = {2212.04026},
 primaryClass = {astro-ph.GA},
       adsurl = {https://ui.adsabs.harvard.edu/abs/2024ApJ...974...92B}
}

@ARTICLE{bluck2020a,
       author = {{Bluck}, Asa F.~L. and {Maiolino}, Roberto and {S{\'a}nchez}, Sebastian F. and {Ellison}, Sara L. and {Thorp}, Mallory D. and {Piotrowska}, Joanna M. and {Teimoorinia}, Hossen and {Bundy}, Kevin A.},
        title = "{Are galactic star formation and quenching governed by local, global, or environmental phenomena?}",
      journal = {\mnras},
         year = 2020,
        month = feb,
       volume = {492},
       number = {1},
        pages = {96-139},
          doi = {10.1093/mnras/stz3264},
archivePrefix = {arXiv},
       eprint = {1911.08857},
 primaryClass = {astro-ph.GA},
       adsurl = {https://ui.adsabs.harvard.edu/abs/2020MNRAS.492...96B}
}

@ARTICLE{bluck2020b,
       author = {{Bluck}, Asa F.~L. and {Maiolino}, Roberto and {Piotrowska}, Joanna M. and {Trussler}, James and {Ellison}, Sara L. and {S{\'a}nchez}, Sebastian F. and {Thorp}, Mallory D. and {Teimoorinia}, Hossen and {Moreno}, Jorge and {Conselice}, Christopher J.},
        title = "{How do central and satellite galaxies quench? - Insights from spatially resolved spectroscopy in the MaNGA survey}",
      journal = {\mnras},
         year = 2020,
        month = nov,
       volume = {499},
       number = {1},
        pages = {230-268},
          doi = {10.1093/mnras/staa2806},
archivePrefix = {arXiv},
       eprint = {2009.05341},
 primaryClass = {astro-ph.GA},
       adsurl = {https://ui.adsabs.harvard.edu/abs/2020MNRAS.499..230B}
}

@ARTICLE{bohringer2004,
       author = {{B{\"o}hringer}, H. and {Schuecker}, P. and {Guzzo}, L. and {Collins}, C.~A. and {Voges}, W. and {Cruddace}, R.~G. and {Ortiz-Gil}, A. and {Chincarini}, G. and {De Grandi}, S. and {Edge}, A.~C. and {MacGillivray}, H.~T. and {Neumann}, D.~M. and {Schindler}, S. and {Shaver}, P.},
        title = "{The ROSAT-ESO Flux Limited X-ray (REFLEX) Galaxy cluster survey. V. The cluster catalogue}",
      journal = {\aap},
         year = 2004,
        month = oct,
       volume = {425},
        pages = {367-383},
          doi = {10.1051/0004-6361:20034484},
archivePrefix = {arXiv},
       eprint = {astro-ph/0405546},
 primaryClass = {astro-ph},
       adsurl = {https://ui.adsabs.harvard.edu/abs/2004A&A...425..367B}
}

@ARTICLE{boselli2022,
       author = {{Boselli}, Alessandro and {Fossati}, Matteo and {Sun}, Ming},
        title = "{Ram pressure stripping in high-density environments}",
      journal = {\aapr},
         year = 2022,
        month = dec,
       volume = {30},
       number = {1},
          eid = {3},
        pages = {3},
          doi = {10.1007/s00159-022-00140-3},
archivePrefix = {arXiv},
       eprint = {2109.13614},
 primaryClass = {astro-ph.GA},
       adsurl = {https://ui.adsabs.harvard.edu/abs/2022A&ARv..30....3B}
}

@ARTICLE{bottrell2024,
       author = {{Bottrell}, Connor and {Yesuf}, Hassen M. and {Popping}, Gerg{\"o} and {Omori}, Kiyoaki Christopher and {Tang}, Shenli and {Ding}, Xuheng and {Pillepich}, Annalisa and {Nelson}, Dylan and {Eisert}, Lukas and {Gao}, Hua and {Goulding}, Andy D. and {Kalita}, Boris S. and {Luo}, Wentao and {Greene}, Jenny E. and {Shi}, Jingjing and {Silverman}, John D.},
        title = "{IllustrisTNG in the HSC-SSP: image data release and the major role of mini mergers as drivers of asymmetry and star formation}",
      journal = {\mnras},
         year = 2024,
        month = jan,
       volume = {527},
       number = {3},
        pages = {6506-6539},
          doi = {10.1093/mnras/stad2971},
archivePrefix = {arXiv},
       eprint = {2308.14793},
 primaryClass = {astro-ph.GA},
       adsurl = {https://ui.adsabs.harvard.edu/abs/2024MNRAS.527.6506B}
}

@ARTICLE{bower2006,
       author = {{Bower}, R.~G. and {Benson}, A.~J. and {Malbon}, R. and {Helly}, J.~C. and {Frenk}, C.~S. and {Baugh}, C.~M. and {Cole}, S. and {Lacey}, C.~G.},
        title = "{Breaking the hierarchy of galaxy formation}",
      journal = {\mnras},
         year = 2006,
        month = aug,
       volume = {370},
       number = {2},
        pages = {645-655},
          doi = {10.1111/j.1365-2966.2006.10519.x},
archivePrefix = {arXiv},
       eprint = {astro-ph/0511338},
 primaryClass = {astro-ph},
       adsurl = {https://ui.adsabs.harvard.edu/abs/2006MNRAS.370..645B}
}

@ARTICLE{bradac2019,
       author = {{Brada{\v{c}}}, Maru{\v{s}}a and {Huang}, Kuang-Han and {Fontana}, Adriano and {Castellano}, Marco and {Merlin}, Emiliano and {Amor{\'\i}n}, Ricardo and {Hoag}, Austin and {Strait}, Victoria and {Santini}, Paola and {Ryan}, Russell E. and {Casertano}, Stefano and {Lemaux}, Brian C. and {Lubin}, Lori M. and {Schmidt}, Kasper B. and {Schrabback}, Tim and {Treu}, Tommaso and {von der Linden}, Anja and {Mason}, Charlotte A. and {Wang}, Xin},
        title = "{Hubble Frontier Field photometric catalogues of Abell 370 and RXC J2248.7-4431: multiwavelength photometry, photometric redshifts, and stellar properties}",
      journal = {\mnras},
         year = 2019,
        month = oct,
       volume = {489},
       number = {1},
        pages = {99-107},
          doi = {10.1093/mnras/stz2119},
archivePrefix = {arXiv},
       eprint = {1906.01725},
 primaryClass = {astro-ph.GA},
       adsurl = {https://ui.adsabs.harvard.edu/abs/2019MNRAS.489...99B}
}

@MISC{bradley2025,
       author = {{Bradley}, Larry and {Sip{\H{o}}cz}, Brigitta and {Robitaille}, Thomas and {Tollerud}, Erik and {Vin{\'\i}cius}, Z{\'e} and {Deil}, Christoph and {Barbary}, Kyle and {Wilson}, Tom J and {Busko}, Ivo and {Donath}, Axel and {G{\"u}nther}, Hans Moritz and {Cara}, Mihai and {Lim}, P.~L. and {Me{\ss}linger}, Sebastian and {Conseil}, Simon and {Droettboom}, Michael and {Bostroem}, Azalee and {Bray}, E.~M. and {Andersen Bratholm}, Lars and {Burnett}, Zach and {Jamieson}, William and {Ginsburg}, Adam and {Taranu}, Dan and {Barentsen}, Geert and {Craig}, Matt and {Morris}, Brett M. and {Perrin}, Marshall and {Rathi}, Shivangee},
        title = "{astropy/photutils: 2.3.0}",
         year = 2025,
        month = sep,
          eid = {10.5281/zenodo.17129028},
          doi = {10.5281/zenodo.17129028},
      version = {2.3.0},
    publisher = {Zenodo},
       adsurl = {https://ui.adsabs.harvard.edu/abs/2025zndo..17129028B}
}

@ARTICLE{brammer2012,
       author = {{Brammer}, Gabriel B. and {van Dokkum}, Pieter G. and {Franx}, Marijn and {Fumagalli}, Mattia and {Patel}, Shannon and {Rix}, Hans-Walter and {Skelton}, Rosalind E. and {Kriek}, Mariska and {Nelson}, Erica and {Schmidt}, Kasper B. and {Bezanson}, Rachel and {da Cunha}, Elisabete and {Erb}, Dawn K. and {Fan}, Xiaohui and {F{\"o}rster Schreiber}, Natascha and {Illingworth}, Garth D. and {Labb{\'e}}, Ivo and {Leja}, Joel and {Lundgren}, Britt and {Magee}, Dan and {Marchesini}, Danilo and {McCarthy}, Patrick and {Momcheva}, Ivelina and {Muzzin}, Adam and {Quadri}, Ryan and {Steidel}, Charles C. and {Tal}, Tomer and {Wake}, David and {Whitaker}, Katherine E. and {Williams}, Anna},
        title = "{3D-HST: A Wide-field Grism Spectroscopic Survey with the Hubble Space Telescope}",
      journal = {\apjs},
         year = 2012,
        month = jun,
       volume = {200},
       number = {2},
          eid = {13},
        pages = {13},
          doi = {10.1088/0067-0049/200/2/13},
archivePrefix = {arXiv},
       eprint = {1204.2829},
 primaryClass = {astro-ph.CO},
       adsurl = {https://ui.adsabs.harvard.edu/abs/2012ApJS..200...13B}
}

@ARTICLE{bremer2018,
       author = {{Bremer}, M.~N. and {Phillipps}, S. and {Kelvin}, L.~S. and {De Propris}, R. and {Kennedy}, Rebecca and {Moffett}, Amanda J. and {Bamford}, S. and {Davies}, L.~J.~M. and {Driver}, S.~P. and {H{\"a}u{\ss}ler}, B. and {Holwerda}, B. and {Hopkins}, A. and {James}, P.~A. and {Liske}, J. and {Percival}, S. and {Taylor}, E.~N.},
        title = "{Galaxy and Mass Assembly (GAMA): Morphological transformation of galaxies across the green valley}",
      journal = {\mnras},
         year = 2018,
        month = may,
       volume = {476},
       number = {1},
        pages = {12-26},
          doi = {10.1093/mnras/sty124},
archivePrefix = {arXiv},
       eprint = {1801.04277},
 primaryClass = {astro-ph.GA},
       adsurl = {https://ui.adsabs.harvard.edu/abs/2018MNRAS.476...12B}
}

@ARTICLE{bressan1993,
       author = {{Bressan}, A. and {Fagotto}, F. and {Bertelli}, G. and {Chiosi}, C.},
        title = "{Evolutionary Sequences of Stellar Models with New Radiative Opacities. II. Z = 0.02}",
      journal = {\aaps},
         year = 1993,
        month = sep,
       volume = {100},
        pages = {647},
       adsurl = {https://ui.adsabs.harvard.edu/abs/1993A&AS..100..647B}
}

@ARTICLE{brinchmann2004,
       author = {{Brinchmann}, J. and {Charlot}, S. and {White}, S.~D.~M. and {Tremonti}, C. and {Kauffmann}, G. and {Heckman}, T. and {Brinkmann}, J.},
        title = "{The physical properties of star-forming galaxies in the low-redshift Universe}",
      journal = {\mnras},
         year = 2004,
        month = jul,
       volume = {351},
       number = {4},
        pages = {1151-1179},
          doi = {10.1111/j.1365-2966.2004.07881.x},
archivePrefix = {arXiv},
       eprint = {astro-ph/0311060},
 primaryClass = {astro-ph},
       adsurl = {https://ui.adsabs.harvard.edu/abs/2004MNRAS.351.1151B}
}

@ARTICLE{bruzual2003,
       author = {{Bruzual}, G. and {Charlot}, S.},
        title = "{Stellar population synthesis at the resolution of 2003}",
      journal = {\mnras},
         year = 2003,
        month = oct,
       volume = {344},
       number = {4},
        pages = {1000-1028},
          doi = {10.1046/j.1365-8711.2003.06897.x},
archivePrefix = {arXiv},
       eprint = {astro-ph/0309134},
 primaryClass = {astro-ph},
       adsurl = {https://ui.adsabs.harvard.edu/abs/2003MNRAS.344.1000B}
}

@ARTICLE{calzetti2000,
       author = {{Calzetti}, Daniela and {Armus}, Lee and {Bohlin}, Ralph C. and {Kinney}, Anne L. and {Koornneef}, Jan and {Storchi-Bergmann}, Thaisa},
        title = "{The Dust Content and Opacity of Actively Star-forming Galaxies}",
      journal = {\apj},
         year = 2000,
        month = apr,
       volume = {533},
       number = {2},
        pages = {682-695},
          doi = {10.1086/308692},
archivePrefix = {arXiv},
       eprint = {astro-ph/9911459},
 primaryClass = {astro-ph},
       adsurl = {https://ui.adsabs.harvard.edu/abs/2000ApJ...533..682C}
}

@INCOLLECTION{calzetti2013,
       author = {{Calzetti}, Daniela},
        title = "{Star Formation Rate Indicators}",
    booktitle = {Secular Evolution of Galaxies},
         year = 2013,
       editor = {{Falc{\'o}n-Barroso}, Jes{\'u}s and {Knapen}, Johan H.},
       volume = {419},
    publisher = {Cambridge Univ. Press},
       adsurl = {https://ui.adsabs.harvard.edu/abs/2013seg..book..419C}
}

@ARTICLE{cappellari2003,
       author = {{Cappellari}, Michele and {Copin}, Yannick},
        title = "{Adaptive spatial binning of integral-field spectroscopic data using Voronoi tessellations}",
      journal = {\mnras},
         year = 2003,
        month = jun,
       volume = {342},
       number = {2},
        pages = {345-354},
          doi = {10.1046/j.1365-8711.2003.06541.x},
archivePrefix = {arXiv},
       eprint = {astro-ph/0302262},
 primaryClass = {astro-ph},
       adsurl = {https://ui.adsabs.harvard.edu/abs/2003MNRAS.342..345C}
}

@ARTICLE{carnall2018,
       author = {{Carnall}, A.~C. and {McLure}, R.~J. and {Dunlop}, J.~S. and {Dav{\'e}}, R.},
        title = "{Inferring the star formation histories of massive quiescent galaxies with BAGPIPES: evidence for multiple quenching mechanisms}",
      journal = {\mnras},
         year = 2018,
        month = nov,
       volume = {480},
       number = {4},
        pages = {4379-4401},
          doi = {10.1093/mnras/sty2169},
archivePrefix = {arXiv},
       eprint = {1712.04452},
 primaryClass = {astro-ph.GA},
       adsurl = {https://ui.adsabs.harvard.edu/abs/2018MNRAS.480.4379C}
}

@ARTICLE{carnall2019,
       author = {{Carnall}, Adam C. and {Leja}, Joel and {Johnson}, Benjamin D. and {McLure}, Ross J. and {Dunlop}, James S. and {Conroy}, Charlie},
        title = "{How to Measure Galaxy Star Formation Histories. I. Parametric Models}",
      journal = {\apj},
         year = 2019,
        month = mar,
       volume = {873},
       number = {1},
          eid = {44},
        pages = {44},
          doi = {10.3847/1538-4357/ab04a2},
archivePrefix = {arXiv},
       eprint = {1811.03635},
 primaryClass = {astro-ph.GA},
       adsurl = {https://ui.adsabs.harvard.edu/abs/2019ApJ...873...44C}
}

@ARTICLE{carnall2023a,
       author = {{Carnall}, A.~C. and {McLeod}, D.~J. and {McLure}, R.~J. and {Dunlop}, J.~S. and {Begley}, R. and {Cullen}, F. and {Donnan}, C.~T. and {Hamadouche}, M.~L. and {Jewell}, S.~M. and {Jones}, E.~W. and {Pollock}, C.~L. and {Wild}, V.},
        title = "{A surprising abundance of massive quiescent galaxies at 3 < z < 5 in the first data from JWST CEERS}",
      journal = {\mnras},
         year = 2023,
        month = apr,
       volume = {520},
       number = {3},
        pages = {3974-3985},
          doi = {10.1093/mnras/stad369},
archivePrefix = {arXiv},
       eprint = {2208.00986},
 primaryClass = {astro-ph.GA},
       adsurl = {https://ui.adsabs.harvard.edu/abs/2023MNRAS.520.3974C}
}

@ARTICLE{carnall2024,
       author = {{Carnall}, A.~C. and {Cullen}, F. and {McLure}, R.~J. and {McLeod}, D.~J. and {Begley}, R. and {Donnan}, C.~T. and {Dunlop}, J.~S. and {Shapley}, A.~E. and {Rowlands}, K. and {Almaini}, O. and {Arellano-C{\'o}rdova}, K.~Z. and {Barrufet}, L. and {Cimatti}, A. and {Ellis}, R.~S. and {Grogin}, N.~A. and {Hamadouche}, M.~L. and {Illingworth}, G.~D. and {Koekemoer}, A.~M. and {Leung}, H.-H. and {Lovell}, C.~C. and {P{\'e}rez-Gonz{\'a}lez}, P.~G. and {Santini}, P. and {Stanton}, T.~M. and {Wild}, V.},
        title = "{The JWST EXCELS survey: too much, too young, too fast? Ultra-massive quiescent galaxies at 3 < z < 5}",
      journal = {\mnras},
         year = 2024,
        month = oct,
       volume = {534},
       number = {1},
        pages = {325-348},
          doi = {10.1093/mnras/stae2092},
archivePrefix = {arXiv},
       eprint = {2405.02242},
 primaryClass = {astro-ph.GA},
       adsurl = {https://ui.adsabs.harvard.edu/abs/2024MNRAS.534..325C}
}

@ARTICLE{castellano2016a,
       author = {{Castellano}, M. and {Amor{\'\i}n}, R. and {Merlin}, E. and {Fontana}, A. and {McLure}, R.~J. and {M{\'a}rmol-Queralt{\'o}}, E. and {Mortlock}, A. and {Parsa}, S. and {Dunlop}, J.~S. and {Elbaz}, D. and {Balestra}, I. and {Boucaud}, A. and {Bourne}, N. and {Boutsia}, K. and {Brammer}, G. and {Bruce}, V.~A. and {Buitrago}, F. and {Capak}, P. and {Cappelluti}, N. and {Ciesla}, L. and {Comastri}, A. and {Cullen}, F. and {Derriere}, S. and {Faber}, S.~M. and {Giallongo}, E. and {Grazian}, A. and {Grillo}, C. and {Mercurio}, A. and {Micha{\l}owski}, M.~J. and {Nonino}, M. and {Paris}, D. and {Pentericci}, L. and {Pilo}, S. and {Rosati}, P. and {Santini}, P. and {Schreiber}, C. and {Shu}, X. and {Wang}, T.},
        title = "{The ASTRODEEP Frontier Fields catalogues. II. Photometric redshifts and rest frame properties in Abell-2744 and MACS-J0416}",
      journal = {\aap},
         year = 2016,
        month = may,
       volume = {590},
          eid = {A31},
        pages = {A31},
          doi = {10.1051/0004-6361/201527514},
archivePrefix = {arXiv},
       eprint = {1603.02461},
 primaryClass = {astro-ph.GA},
       adsurl = {https://ui.adsabs.harvard.edu/abs/2016A&A...590A..31C}
}

@ARTICLE{chabrier2003,
       author = {{Chabrier}, Gilles},
        title = "{Galactic Stellar and Substellar Initial Mass Function}",
      journal = {\pasp},
         year = 2003,
        month = jul,
       volume = {115},
       number = {809},
        pages = {763-795},
          doi = {10.1086/376392},
archivePrefix = {arXiv},
       eprint = {astro-ph/0304382},
 primaryClass = {astro-ph},
       adsurl = {https://ui.adsabs.harvard.edu/abs/2003PASP..115..763C}
}

@ARTICLE{chen2024,
       author = {{Chen}, Hongxing and {Xie}, Lizhi and {Wang}, Jie and {Hu}, Wenkai and {De Lucia}, Gabriella and {Fontanot}, Fabio and {Hirschamnn}, Michaela},
        title = "{Environmental effects on satellite galaxies from the perspective of cold gas}",
      journal = {\mnras},
         year = 2024,
        month = feb,
       volume = {528},
       number = {2},
        pages = {2451-2463},
          doi = {10.1093/mnras/stae162},
archivePrefix = {arXiv},
       eprint = {2401.07158},
 primaryClass = {astro-ph.GA},
       adsurl = {https://ui.adsabs.harvard.edu/abs/2024MNRAS.528.2451C}
}

@ARTICLE{cheng2024a,
       author = {{Cheng}, Isaac and {Woods}, Tyrone E. and {C{\^o}t{\'e}}, Patrick and {Glover}, Jennifer and {Bansal}, Dhananjhay and {Amenouche}, Melissa and {Marshall}, Madeline A. and {Amen}, Laurie and {Hutchings}, John and {Ferrarese}, Laura and {Venn}, Kim A. and {Balogh}, Michael and {Blouin}, Simon and {Cloutier}, Ryan and {Dickson}, Nolan and {Gallagher}, Sarah and {Hellmich}, Martin and {H{\'e}nault-Brunet}, Vincent and {Khatu}, Viraja and {Lawlor-Forsyth}, Cameron and {Morgan}, Cameron and {Richer}, Harvey and {Sawicki}, Marcin and {Sorba}, Robert},
        title = "{FORECASTOR. I. Finding Optics Requirements and Exposure Times for the Cosmological Advanced Survey Telescope for Optical and UV Research Mission}",
      journal = {\aj},
         year = 2024,
        month = apr,
       volume = {167},
       number = {4},
          eid = {178},
        pages = {178},
          doi = {10.3847/1538-3881/ad2987},
archivePrefix = {arXiv},
       eprint = {2402.08137},
 primaryClass = {astro-ph.IM},
       adsurl = {https://ui.adsabs.harvard.edu/abs/2024AJ....167..178C}
}

@ARTICLE{choi2016,
       author = {{Choi}, Jieun and {Dotter}, Aaron and {Conroy}, Charlie and {Cantiello}, Matteo and {Paxton}, Bill and {Johnson}, Benjamin D.},
        title = "{Mesa Isochrones and Stellar Tracks (MIST). I. Solar-scaled Models}",
      journal = {\apj},
         year = 2016,
        month = jun,
       volume = {823},
       number = {2},
          eid = {102},
        pages = {102},
          doi = {10.3847/0004-637X/823/2/102},
archivePrefix = {arXiv},
       eprint = {1604.08592},
 primaryClass = {astro-ph.SR},
       adsurl = {https://ui.adsabs.harvard.edu/abs/2016ApJ...823..102C}
}

@ARTICLE{ciesla2016,
       author = {{Ciesla}, L. and {Boselli}, A. and {Elbaz}, D. and {Boissier}, S. and {Buat}, V. and {Charmandaris}, V. and {Schreiber}, C. and {B{\'e}thermin}, M. and {Baes}, M. and {Boquien}, M. and {De Looze}, I. and {Fern{\'a}ndez-Ontiveros}, J.~A. and {Pappalardo}, C. and {Spinoglio}, L. and {Viaene}, S.},
        title = "{The imprint of rapid star formation quenching on the spectral energy distributions of galaxies}",
      journal = {\aap},
         year = 2016,
        month = jan,
       volume = {585},
          eid = {A43},
        pages = {A43},
          doi = {10.1051/0004-6361/201527107},
archivePrefix = {arXiv},
       eprint = {1510.07657},
 primaryClass = {astro-ph.GA},
       adsurl = {https://ui.adsabs.harvard.edu/abs/2016A&A...585A..43C}
}

@ARTICLE{ciesla2017,
       author = {{Ciesla}, L. and {Elbaz}, D. and {Fensch}, J.},
        title = "{The SFR-M$_{{\ensuremath{*}}}$ main sequence archetypal star-formation history and analytical models}",
      journal = {\aap},
         year = 2017,
        month = dec,
       volume = {608},
          eid = {A41},
        pages = {A41},
          doi = {10.1051/0004-6361/201731036},
archivePrefix = {arXiv},
       eprint = {1706.08531},
 primaryClass = {astro-ph.GA},
       adsurl = {https://ui.adsabs.harvard.edu/abs/2017A&A...608A..41C}
}

@ARTICLE{ciesla2018,
       author = {{Ciesla}, L. and {Elbaz}, D. and {Schreiber}, C. and {Daddi}, E. and {Wang}, T.},
        title = "{Identification of galaxies that experienced a recent major drop of star formation}",
      journal = {\aap},
         year = 2018,
        month = jul,
       volume = {615},
          eid = {A61},
        pages = {A61},
          doi = {10.1051/0004-6361/201832715},
archivePrefix = {arXiv},
       eprint = {1803.10239},
 primaryClass = {astro-ph.GA},
       adsurl = {https://ui.adsabs.harvard.edu/abs/2018A&A...615A..61C}
}

@ARTICLE{ciesla2021,
       author = {{Ciesla}, L. and {Buat}, V. and {Boquien}, M. and {Boselli}, A. and {Elbaz}, D. and {Aufort}, G.},
        title = "{Investigating the delay between dust radiation and star-formation in local and distant quenching galaxies}",
      journal = {\aap},
         year = 2021,
        month = sep,
       volume = {653},
          eid = {A6},
        pages = {A6},
          doi = {10.1051/0004-6361/202140762},
archivePrefix = {arXiv},
       eprint = {2106.00017},
 primaryClass = {astro-ph.GA},
       adsurl = {https://ui.adsabs.harvard.edu/abs/2021A&A...653A...6C}
}

@ARTICLE{coe2019,
       author = {{Coe}, Dan and {Salmon}, Brett and {Brada{\v{c}}}, Maru{\v{s}}a and {Bradley}, Larry D. and {Sharon}, Keren and {Zitrin}, Adi and {Acebron}, Ana and {Cerny}, Catherine and {Cibirka}, Nath{\'a}lia and {Strait}, Victoria and {Paterno-Mahler}, Rachel and {Mahler}, Guillaume and {Avila}, Roberto J. and {Ogaz}, Sara and {Huang}, Kuang-Han and {Pelliccia}, Debora and {Stark}, Daniel P. and {Mainali}, Ramesh and {Oesch}, Pascal A. and {Trenti}, Michele and {Carrasco}, Daniela and {Dawson}, William A. and {Rodney}, Steven A. and {Strolger}, Louis-Gregory and {Riess}, Adam G. and {Jones}, Christine and {Frye}, Brenda L. and {Czakon}, Nicole G. and {Umetsu}, Keiichi and {Vulcani}, Benedetta and {Graur}, Or and {Jha}, Saurabh W. and {Graham}, Melissa L. and {Molino}, Alberto and {Nonino}, Mario and {Hjorth}, Jens and {Selsing}, Jonatan and {Christensen}, Lise and {Kikuchihara}, Shotaro and {Ouchi}, Masami and {Oguri}, Masamune and {Welch}, Brian and {Lemaux}, Brian C. and {Andrade-Santos}, Felipe and {Hoag}, Austin T. and {Johnson}, Traci L. and {Peterson}, Avery and {Past}, Matthew and {Fox}, Carter and {Agulli}, Irene and {Livermore}, Rachael and {Ryan}, Russell E. and {Lam}, Daniel and {Sendra-Server}, Irene and {Toft}, Sune and {Lovisari}, Lorenzo and {Su}, Yuanyuan},
        title = "{RELICS: Reionization Lensing Cluster Survey}",
      journal = {\apj},
         year = 2019,
        month = oct,
       volume = {884},
       number = {1},
          eid = {85},
        pages = {85},
          doi = {10.3847/1538-4357/ab412b},
archivePrefix = {arXiv},
       eprint = {1903.02002},
 primaryClass = {astro-ph.GA},
       adsurl = {https://ui.adsabs.harvard.edu/abs/2019ApJ...884...85C}
}

@ARTICLE{cole2000,
       author = {{Cole}, Shaun and {Lacey}, Cedric G. and {Baugh}, Carlton M. and {Frenk}, Carlos S.},
        title = "{Hierarchical galaxy formation}",
      journal = {\mnras},
         year = 2000,
        month = nov,
       volume = {319},
       number = {1},
        pages = {168-204},
          doi = {10.1046/j.1365-8711.2000.03879.x},
archivePrefix = {arXiv},
       eprint = {astro-ph/0007281},
 primaryClass = {astro-ph},
       adsurl = {https://ui.adsabs.harvard.edu/abs/2000MNRAS.319..168C}
}

@MISC{collette2023,
       author = {{Collette}, Andrew and {Kluyver}, Thomas and {Caswell}, Thomas A and {Tocknell}, James and {Kieffer}, Jerome and {Jelenak}, Aleksandar and {Scopatz}, Anthony and {Dale}, Darren and {Chen} and {VINCENT}, Thomas and {Einhorn}, Matt and {Payno} and {Juliagarriga} and {Sciarelli}, Pierlauro and {Valls}, Valentin and {Ghosh}, Satrajit and {Kofoed Pedersen}, Ulrik and {Kittisopikul}, Mark and {Jakirkham} and {Raspaud}, Martin and {Danilevski}, Cyril and {Abbasi}, Hameer and {Readey}, John and {M{\"u}hlbauer}, Kai and {Paramonov}, Andrey and {Chan}, Lawrence and {De Schepper}, Robin and {Sol{\'e}}, V. Armando and {Jialin} and {Hay Guest}, Daniel},
        title = "{h5py/h5py: 3.8.0}",
         year = 2023,
        month = jan,
          eid = {10.5281/zenodo.7560547},
          doi = {10.5281/zenodo.7560547},
      version = {v3.8.0},
    publisher = {Zenodo},
       adsurl = {https://ui.adsabs.harvard.edu/abs/2023zndo...7560547C}
}

@ARTICLE{conroy2010,
       author = {{Conroy}, Charlie and {Gunn}, James E.},
        title = "{The Propagation of Uncertainties in Stellar Population Synthesis Modeling. III. Model Calibration, Comparison, and Evaluation}",
      journal = {\apj},
         year = 2010,
        month = apr,
       volume = {712},
       number = {2},
        pages = {833-857},
          doi = {10.1088/0004-637X/712/2/833},
archivePrefix = {arXiv},
       eprint = {0911.3151},
 primaryClass = {astro-ph.CO},
       adsurl = {https://ui.adsabs.harvard.edu/abs/2010ApJ...712..833C}
}

@ARTICLE{cote2025,
       author = {{C{\^o}t{\'e}}, Patrick and {Woods}, Tyrone E. and {Hutchings}, John B. and {Rhodes}, Jason D. and {S{\'a}nchez-Janssen}, Rub{\'e}n. and {Scott}, Alan D. and {Pazder}, John and {Amenouche}, Melissa and {Balogh}, Michael and {Blouin}, Simon and {Cournoyer}, Alain and {Drout}, Maria R. and {Kuzmin}, Nick and {Mack}, Katherine J. and {Ferrarese}, Laura and {Fraser}, Wesley C. and {Gallagher}, Sarah C. and {Grandmont}, Fr{\'e}d{\'e}ric and {Haggard}, Daryl and {Harrison}, Paul and {H{\'e}nault-Brunet}, Vincent and {Kavelaars}, J.~J. and {Khatu}, Viraja and {Roediger}, Joel C. and {Rowe}, Jason and {Sawicki}, Marcin and {Skottfelt}, Jesper and {Taylor}, Matt and {van Waerbeke}, Ludo and {Amen}, Laurie and {Bansal}, Dhananjhay and {Bergeron}, Martin and {Brown}, Toby and {Burley}, Greg and {Chand}, Hum and {Cheng}, Isaac and {Cloutier}, Ryan and {Dickson}, Nolan and {Djazovski}, Oleg and {Damjanov}, Ivana and {Doherty}, James and {Finner}, Kyle and {Del Valle Espinosa}, Macarena Garc{\'\i}a. and {Glover}, Jennifer and {G{\'o}mez de Castro}, Ana I. and {Graur}, Or and {Hardy}, Tim and {Kao}, Michelle and {Leahy}, Denis and {Lokhorst}, Deborah and {Malz}, Alex I. and {Man}, Allison and {Marshall}, Madeline A. and {McGee}, Sean and {McKenzie}, Ryan and {Michaud}, Kai and {More}, Surhud S. and {Morris}, David and {Morris}, Patrick W. and {Moutard}, Thibaud and {Naqvi}, Wasi and {Nicholl}, Matt and {Noirot}, Ga{\"e}l. and {Oey}, M.~S. and {Opitom}, Cyrielle and {Salim}, Samir and {Scott}, Bryan R. and {Shapiro}, Charles A. and {Stern}, Daniel and {Subramaniam}, Annapurni and {Thilke}, David and {Wevers}, Ivan and {Vorobiev}, Dmitri and {Aaron Yung}, L.~Y. and {Zamkotsian}, Fr{\'e}d{\'e}ric and {Aigrain}, Suzanne and {Alavi}, Anahita and {Barstow}, Martin and {Bartosik}, Peter and {Bluhm}, Hadleigh and {Bovy}, Jo and {Cameron}, Peter and {Carlberg}, Raymond G. and {Christiansen}, Jessie L. and {Chen}, Yuyang and {Crowther}, Paul and {Dage}, Kristen and {Dotter}, Aaron L. and {Dufour}, Patrick and {Dupuis}, Jean and {Dryer}, Ben and {Duara}, Angaraj and {Eadie}, Gwendolyn M. and {Eduardo}, Marielle R. and {Estrada-Carpenter}, Vincente and {Fabbro}, S{\'e}bastien and {Faisst}, Andreas and {Ford}, Nicole M. and {Fraser}, Morgan and {Gaensicke}, Boris T. and {Ganesh}, Shashkiran and {Gandhi}, Poshak and {Graham}, Melissa L. and {Hamel}, Rebecca and {Hellmich}, Martin and {Hennessy}, John and {Hessel}, Kaitlyn and {Heyl}, Jeremy and {Heymans}, Catherine and {Hezaveh}, Yashar and {Hlozek}, Renee and {Hoenk}, Michael E. and {Holland}, Andrew and {Huff}, Eric and {Hutchinson}, Ian and {Iwata}, Ikuru and {Jewell}, April D. and {Johnstone}, Doug and {Jones}, Maia and {Jones}, Todd and {Lang}, Dustin and {Lapington}, Jon and {Larivi{\`e}re}, Justin and {Lawlor-Forsyth}, Cameron and {Laurin}, Denis and {Lee}, Charles and {Legin}, Ronan and {Li}, Ting S. and {Lim}, Sungsoon and {Ludwig}, Bethany and {Kozun}, Matt and {Vivek}, M. and {Mann}, Robert and {McConnachie}, Alan W. and {McDonough}, Evan and {Metchev}, Stanimir and {Miller}, David R. and {Moriya}, Takashi and {Morgan}, Cameron and {Navarro}, Julio and {Naz{\'e}}, Ya{\"e}l. and {Nikzad}, Shouleh and {Oad}, Vivek and {Ouellette}, Nathalie and {Pass}, Emily K. and {Percival}, Will J. and {Levasseur}, Laurence Perreault and {Postma}, Joe and {Raza}, Nayyer and {Richards}, Gordon T. and {Richer}, Harvey and {Robert}, Carmelle and {Rosolowsky}, Erik and {Ruan}, John J. and {Rugheimer}, Sarah and {Safi-Harb}, Samar and {Saha}, Kanak and {Scowcroft}, Vicky and {Sestito}, Federico and {Sharma}, Himanshu and {Sikora}, James and {Sivakoff}, Gregory R. and {Sivarani}, Thirupathi and {Smith}, Patrick and {Soh}, Warren and {Sorba}, Robert and {Subramanian}, Smitha and {Teimoorinia}, Hossen and {Teplitz}, Harry I. and {Thadani}, Shaylin and {Thadani}, Shavon and {Tohuvavohu}, Aaron and {Venn}, Kim A. and {Vieira}, Nicholas and {Webb}, Jeremy J. and {Wiegert}, Paul and {Wierckx}, Ryan and {Wu}, Yanqin and {Yeung}, Jade and {Yi}, Sukyoung K.},
        title = "{The CASTOR mission}",
      journal = {JATIS},
         year = 2025,
        month = oct,
       volume = {11},
          eid = {042202},
        pages = {042202},
          doi = {10.1117/1.JATIS.11.4.042202},
       adsurl = {https://ui.adsabs.harvard.edu/abs/2025JATIS..11d2202C}
}

@ARTICLE{cover1967,
       author = {Cover, T. and Hart, P.},
      journal = {ITIT}, 
        title = {Nearest neighbor pattern classification}, 
         year = {1967},
       volume = {13},
       number = {1},
        pages = {21-27},
          doi = {10.1109/TIT.1967.1053964}
}

@ARTICLE{couch1984,
       author = {{Couch}, W.~J. and {Newell}, E.~B.},
        title = "{Distant clusters of galaxies. I. Uniform photometry of 14 rich clusters.}",
      journal = {\apjs},
         year = 1984,
        month = sep,
       volume = {56},
        pages = {143-192},
          doi = {10.1086/190979},
       adsurl = {https://ui.adsabs.harvard.edu/abs/1984ApJS...56..143C}
}

@ARTICLE{croton2006,
       author = {{Croton}, Darren J. and {Springel}, Volker and {White}, Simon D.~M. and {De Lucia}, G. and {Frenk}, C.~S. and {Gao}, L. and {Jenkins}, A. and {Kauffmann}, G. and {Navarro}, J.~F. and {Yoshida}, N.},
        title = "{The many lives of active galactic nuclei: cooling flows, black holes and the luminosities and colours of galaxies}",
      journal = {\mnras},
         year = 2006,
        month = jan,
       volume = {365},
       number = {1},
        pages = {11-28},
          doi = {10.1111/j.1365-2966.2005.09675.x},
archivePrefix = {arXiv},
       eprint = {astro-ph/0508046},
 primaryClass = {astro-ph},
       adsurl = {https://ui.adsabs.harvard.edu/abs/2006MNRAS.365...11C}
}

@ARTICLE{davies2019,
       author = {{Davies}, L.~J.~M. and {Robotham}, A.~S.~G. and {Lagos}, C. del P. and {Driver}, S.~P. and {Stevens}, A.~R.~H. and {Bah{\'e}}, Y.~M. and {Alpaslan}, M. and {Bremer}, M.~N. and {Brown}, M.~J.~I. and {Brough}, S. and {Bland-Hawthorn}, J. and {Cortese}, L. and {Elahi}, P. and {Grootes}, M.~W. and {Holwerda}, B.~W. and {Ludlow}, A.~D. and {McGee}, S. and {Owers}, M. and {Phillipps}, S.},
        title = "{Galaxy and Mass Assembly (GAMA): environmental quenching of centrals and satellites in groups}",
      journal = {\mnras},
         year = 2019,
        month = mar,
       volume = {483},
       number = {4},
        pages = {5444-5458},
          doi = {10.1093/mnras/sty3393},
archivePrefix = {arXiv},
       eprint = {1901.01640},
 primaryClass = {astro-ph.GA},
       adsurl = {https://ui.adsabs.harvard.edu/abs/2019MNRAS.483.5444D}
}

@ARTICLE{dicriscienzo2017,
       author = {{Di Criscienzo}, M. and {Merlin}, E. and {Castellano}, M. and {Santini}, P. and {Fontana}, A. and {Amorin}, R. and {Boutsia}, K. and {Derriere}, S. and {Dunlop}, J.~S. and {Elbaz}, D. and {Grazian}, A. and {McLure}, R.~J. and {M{\'a}rmol-Queralt{\'o}}, E. and {Michalowski}, M.~J. and {Mortlock}, S. and {Parsa}, S. and {Pentericci}, L.},
        title = "{The ASTRODEEP Frontier Fields catalogues. III. Multiwavelength photometry and rest-frame properties of MACS-J0717 and MACS-J1149}",
      journal = {\aap},
         year = 2017,
        month = oct,
       volume = {607},
          eid = {A30},
        pages = {A30},
          doi = {10.1051/0004-6361/201731172},
archivePrefix = {arXiv},
       eprint = {1706.03790},
 primaryClass = {astro-ph.GA},
       adsurl = {https://ui.adsabs.harvard.edu/abs/2017A&A...607A..30D}
}

@ARTICLE{dimatteo2005,
       author = {{Di Matteo}, Tiziana and {Springel}, Volker and {Hernquist}, Lars},
        title = "{Energy input from quasars regulates the growth and activity of black holes and their host galaxies}",
      journal = {\nat},
         year = 2005,
        month = feb,
       volume = {433},
       number = {7026},
        pages = {604-607},
          doi = {10.1038/nature03335},
archivePrefix = {arXiv},
       eprint = {astro-ph/0502199},
 primaryClass = {astro-ph},
       adsurl = {https://ui.adsabs.harvard.edu/abs/2005Natur.433..604D}
}

@ARTICLE{diemer2019,
       author = {{Diemer}, Benedikt and {Stevens}, Adam R.~H. and {Lagos}, Claudia del P. and {Calette}, A.~R. and {Tacchella}, Sandro and {Hernquist}, Lars and {Marinacci}, Federico and {Nelson}, Dylan and {Pillepich}, Annalisa and {Rodriguez-Gomez}, Vicente and {Villaescusa-Navarro}, Francisco and {Vogelsberger}, Mark},
        title = "{Atomic and molecular gas in IllustrisTNG galaxies at low redshift}",
      journal = {\mnras},
         year = 2019,
        month = aug,
       volume = {487},
       number = {2},
        pages = {1529-1550},
          doi = {10.1093/mnras/stz1323},
archivePrefix = {arXiv},
       eprint = {1902.10714},
 primaryClass = {astro-ph.GA},
       adsurl = {https://ui.adsabs.harvard.edu/abs/2019MNRAS.487.1529D}
}

@ARTICLE{donnari2019,
       author = {{Donnari}, Martina and {Pillepich}, Annalisa and {Nelson}, Dylan and {Vogelsberger}, Mark and {Genel}, Shy and {Weinberger}, Rainer and {Marinacci}, Federico and {Springel}, Volker and {Hernquist}, Lars},
        title = "{The star formation activity of IllustrisTNG galaxies: main sequence, UVJ diagram, quenched fractions, and systematics}",
      journal = {\mnras},
         year = 2019,
        month = jun,
       volume = {485},
       number = {4},
        pages = {4817-4840},
          doi = {10.1093/mnras/stz712},
archivePrefix = {arXiv},
       eprint = {1812.07584},
 primaryClass = {astro-ph.GA},
       adsurl = {https://ui.adsabs.harvard.edu/abs/2019MNRAS.485.4817D}
}

@ARTICLE{donnari2021a,
       author = {{Donnari}, Martina and {Pillepich}, Annalisa and {Joshi}, Gandhali D. and {Nelson}, Dylan and {Genel}, Shy and {Marinacci}, Federico and {Rodriguez-Gomez}, Vicente and {Pakmor}, R{\"u}diger and {Torrey}, Paul and {Vogelsberger}, Mark and {Hernquist}, Lars},
        title = "{Quenched fractions in the IllustrisTNG simulations: the roles of AGN feedback, environment, and pre-processing}",
      journal = {\mnras},
         year = 2021,
        month = jan,
       volume = {500},
       number = {3},
        pages = {4004-4024},
          doi = {10.1093/mnras/staa3006},
archivePrefix = {arXiv},
       eprint = {2008.00005},
 primaryClass = {astro-ph.GA},
       adsurl = {https://ui.adsabs.harvard.edu/abs/2021MNRAS.500.4004D}
}

@ARTICLE{donnari2021b,
       author = {{Donnari}, Martina and {Pillepich}, Annalisa and {Nelson}, Dylan and {Marinacci}, Federico and {Vogelsberger}, Mark and {Hernquist}, Lars},
        title = "{Quenched fractions in the IllustrisTNG simulations: comparison with observations and other theoretical models}",
      journal = {\mnras},
         year = 2021,
        month = oct,
       volume = {506},
       number = {4},
        pages = {4760-4780},
          doi = {10.1093/mnras/stab1950},
archivePrefix = {arXiv},
       eprint = {2008.00004},
 primaryClass = {astro-ph.GA},
       adsurl = {https://ui.adsabs.harvard.edu/abs/2021MNRAS.506.4760D}
}

@ARTICLE{dore2018,
       author = {{Dor{\'e}}, Olivier and {Hirata}, Christopher and {Wang}, Yun and {Weinberg}, David and {Baronchelli}, Ivano and {Benson}, Andrew and {Capak}, Peter and {Choi}, Ami and {Eifler}, Tim and {Hemmati}, Shoubaneh and {Ho}, Shirley and {Izard}, Albert and {Jain}, Bhuvnesh and {Jarvis}, Mike and {Kiessling}, Alina and {Krause}, Elisabeth and {Massara}, Elena and {Masters}, Dan and {Merson}, Alex and {Miyatake}, Hironao and {Plazas Malagon}, Andres and {Mandelbaum}, Rachel and {Samushia}, Lado and {Shapiro}, Chaz and {Simet}, Melanie and {Spergel}, David and {Teplitz}, Harry and {Troxel}, Michael and {Bean}, Rachel and {Colbert}, James and {Heinrich}, Chen He and {Heitmann}, Katrin and {Helou}, George and {Hudson}, Michael and {Huff}, Eric and {Leauthaud}, Alexie and {MacCrann}, Niall and {Padmanabhan}, Nikhil and {Pisani}, Alice and {Rhodes}, Jason and {Rozo}, Eduardo and {Seiffert}, Mike and {Smith}, Kendrick and {Takada}, Masahiro and {von der Linden}, Anja and {Lupton}, Robert and {Yoshida}, Naoki and {Wu}, Hao-Yi and {Zu}, Ying},
        title = "{WFIRST Science Investigation Team ``Cosmology with the High Latitude Survey'' Annual Report 2017}",
      journal = {arXiv e-prints},
         year = 2018,
        month = apr,
          eid = {arXiv:1804.03628},
        pages = {arXiv:1804.03628},
          doi = {10.48550/arXiv.1804.03628},
archivePrefix = {arXiv},
       eprint = {1804.03628},
 primaryClass = {astro-ph.CO},
       adsurl = {https://ui.adsabs.harvard.edu/abs/2018arXiv180403628D}
}

@ARTICLE{dotter2016,
       author = {{Dotter}, Aaron},
        title = "{MESA Isochrones and Stellar Tracks (MIST) 0: Methods for the Construction of Stellar Isochrones}",
      journal = {\apjs},
         year = 2016,
        month = jan,
       volume = {222},
       number = {1},
          eid = {8},
        pages = {8},
          doi = {10.3847/0067-0049/222/1/8},
archivePrefix = {arXiv},
       eprint = {1601.05144},
 primaryClass = {astro-ph.SR},
       adsurl = {https://ui.adsabs.harvard.edu/abs/2016ApJS..222....8D}
}

@ARTICLE{dressler1999,
       author = {{Dressler}, Alan and {Smail}, Ian and {Poggianti}, Bianca M. and {Butcher}, Harvey and {Couch}, Warrick J. and {Ellis}, Richard S. and {Oemler}, Jr., Augustus},
        title = "{A Spectroscopic Catalog of 10 Distant Rich Clusters of Galaxies}",
      journal = {\apjs},
         year = 1999,
        month = may,
       volume = {122},
       number = {1},
        pages = {51-80},
          doi = {10.1086/313213},
archivePrefix = {arXiv},
       eprint = {astro-ph/9901263},
 primaryClass = {astro-ph},
       adsurl = {https://ui.adsabs.harvard.edu/abs/1999ApJS..122...51D}
}

@ARTICLE{ebeling2001,
       author = {{Ebeling}, H. and {Edge}, A.~C. and {Henry}, J.~P.},
        title = "{MACS: A Quest for the Most Massive Galaxy Clusters in the Universe}",
      journal = {\apj},
         year = 2001,
        month = jun,
       volume = {553},
       number = {2},
        pages = {668-676},
          doi = {10.1086/320958},
archivePrefix = {arXiv},
       eprint = {astro-ph/0009101},
 primaryClass = {astro-ph},
       adsurl = {https://ui.adsabs.harvard.edu/abs/2001ApJ...553..668E}
}

@ARTICLE{ebeling2007,
       author = {{Ebeling}, H. and {Barrett}, E. and {Donovan}, D. and {Ma}, C.-J. and {Edge}, A.~C. and {van Speybroeck}, L.},
        title = "{A Complete Sample of 12 Very X-Ray Luminous Galaxy Clusters at z > 0.5}",
      journal = {\apjl},
         year = 2007,
        month = may,
       volume = {661},
       number = {1},
        pages = {L33-L36},
          doi = {10.1086/518603},
archivePrefix = {arXiv},
       eprint = {astro-ph/0703394},
 primaryClass = {astro-ph},
       adsurl = {https://ui.adsabs.harvard.edu/abs/2007ApJ...661L..33E}
}

@ARTICLE{euclid2022,
       author = {{Euclid Collaboration} and {Scaramella}, R. and {Amiaux}, J. and {Mellier}, Y. and {Burigana}, C. and {Carvalho}, C.~S. and {Cuillandre}, J.-C. and {Da Silva}, A. and {Derosa}, A. and {Dinis}, J. and {Maiorano}, E. and {Maris}, M. and {Tereno}, I. and {Laureijs}, R. and {Boenke}, T. and {Buenadicha}, G. and {Dupac}, X. and {Gaspar Venancio}, L.~M. and {G{\'o}mez-{\'A}lvarez}, P. and {Hoar}, J. and {Lorenzo Alvarez}, J. and {Racca}, G.~D. and {Saavedra-Criado}, G. and {Schwartz}, J. and {Vavrek}, R. and {Schirmer}, M. and {Aussel}, H. and {Azzollini}, R. and {Cardone}, V.~F. and {Cropper}, M. and {Ealet}, A. and {Garilli}, B. and {Gillard}, W. and {Granett}, B.~R. and {Guzzo}, L. and {Hoekstra}, H. and {Jahnke}, K. and {Kitching}, T. and {Maciaszek}, T. and {Meneghetti}, M. and {Miller}, L. and {Nakajima}, R. and {Niemi}, S.~M. and {Pasian}, F. and {Percival}, W.~J. and {Pottinger}, S. and {Sauvage}, M. and {Scodeggio}, M. and {Wachter}, S. and {Zacchei}, A. and {Aghanim}, N. and {Amara}, A. and {Auphan}, T. and {Auricchio}, N. and {Awan}, S. and {Balestra}, A. and {Bender}, R. and {Bodendorf}, C. and {Bonino}, D. and {Branchini}, E. and {Brau-Nogue}, S. and {Brescia}, M. and {Candini}, G.~P. and {Capobianco}, V. and {Carbone}, C. and {Carlberg}, R.~G. and {Carretero}, J. and {Casas}, R. and {Castander}, F.~J. and {Castellano}, M. and {Cavuoti}, S. and {Cimatti}, A. and {Cledassou}, R. and {Congedo}, G. and {Conselice}, C.~J. and {Conversi}, L. and {Copin}, Y. and {Corcione}, L. and {Costille}, A. and {Courbin}, F. and {Degaudenzi}, H. and {Douspis}, M. and {Dubath}, F. and {Duncan}, C.~A.~J. and {Dusini}, S. and {Farrens}, S. and {Ferriol}, S. and {Fosalba}, P. and {Fourmanoit}, N. and {Frailis}, M. and {Franceschi}, E. and {Franzetti}, P. and {Fumana}, M. and {Gillis}, B. and {Giocoli}, C. and {Grazian}, A. and {Grupp}, F. and {Haugan}, S.~V.~H. and {Holmes}, W. and {Hormuth}, F. and {Hudelot}, P. and {Kermiche}, S. and {Kiessling}, A. and {Kilbinger}, M. and {Kohley}, R. and {Kubik}, B. and {K{\"u}mmel}, M. and {Kunz}, M. and {Kurki-Suonio}, H. and {Lahav}, O. and {Ligori}, S. and {Lilje}, P.~B. and {Lloro}, I. and {Mansutti}, O. and {Marggraf}, O. and {Markovic}, K. and {Marulli}, F. and {Massey}, R. and {Maurogordato}, S. and {Melchior}, M. and {Merlin}, E. and {Meylan}, G. and {Mohr}, J.~J. and {Moresco}, M. and {Morin}, B. and {Moscardini}, L. and {Munari}, E. and {Nichol}, R.~C. and {Padilla}, C. and {Paltani}, S. and {Peacock}, J. and {Pedersen}, K. and {Pettorino}, V. and {Pires}, S. and {Poncet}, M. and {Popa}, L. and {Pozzetti}, L. and {Raison}, F. and {Rebolo}, R. and {Rhodes}, J. and {Rix}, H.-W. and {Roncarelli}, M. and {Rossetti}, E. and {Saglia}, R. and {Schneider}, P. and {Schrabback}, T. and {Secroun}, A. and {Seidel}, G. and {Serrano}, S. and {Sirignano}, C. and {Sirri}, G. and {Skottfelt}, J. and {Stanco}, L. and {Starck}, J.~L. and {Tallada-Cresp{\'\i}}, P. and {Tavagnacco}, D. and {Taylor}, A.~N. and {Teplitz}, H.~I. and {Toledo-Moreo}, R. and {Torradeflot}, F. and {Trifoglio}, M. and {Valentijn}, E.~A. and {Valenziano}, L. and {Verdoes Kleijn}, G.~A. and {Wang}, Y. and {Welikala}, N. and {Weller}, J. and {Wetzstein}, M. and {Zamorani}, G. and {Zoubian}, J. and {Andreon}, S. and {Baldi}, M. and {Bardelli}, S. and {Boucaud}, A. and {Camera}, S. and {Di Ferdinando}, D. and {Fabbian}, G. and {Farinelli}, R. and {Galeotta}, S. and {Graci{\'a}-Carpio}, J. and {Maino}, D. and {Medinaceli}, E. and {Mei}, S. and {Neissner}, C. and {Polenta}, G. and {Renzi}, A. and {Romelli}, E. and {Rosset}, C. and {Sureau}, F. and {Tenti}, M. and {Vassallo}, T. and {Zucca}, E. and {Baccigalupi}, C. and {Balaguera-Antol{\'\i}nez}, A. and {Battaglia}, P. and {Biviano}, A. and {Borgani}, S. and {Bozzo}, E. and {Cabanac}, R. and {Cappi}, A.},
        title = "{Euclid preparation. I. The Euclid Wide Survey}",
      journal = {\aap},
         year = 2022,
        month = jun,
       volume = {662},
          eid = {A112},
        pages = {A112},
          doi = {10.1051/0004-6361/202141938},
archivePrefix = {arXiv},
       eprint = {2108.01201},
 primaryClass = {astro-ph.CO},
       adsurl = {https://ui.adsabs.harvard.edu/abs/2022A&A...662A.112E}
}

@ARTICLE{euclid2025a,
       author = {{Euclid Collaboration} and {Mellier}, Y. and {Abdurro'uf} and {Acevedo Barroso}, J.~A. and {Ach{\'u}carro}, A. and {Adamek}, J. and {Adam}, R. and {Addison}, G.~E. and {Aghanim}, N. and {Aguena}, M. and {Ajani}, V. and {Akrami}, Y. and {Al-Bahlawan}, A. and {Alavi}, A. and {Albuquerque}, I.~S. and {Alestas}, G. and {Alguero}, G. and {Allaoui}, A. and {Allen}, S.~W. and {Allevato}, V. and {Alonso-Tetilla}, A.~V. and {Altieri}, B. and {Alvarez-Candal}, A. and {Alvi}, S. and {Amara}, A. and {Amendola}, L. and {Amiaux}, J. and {Andika}, I.~T. and {Andreon}, S. and {Andrews}, A. and {Angora}, G. and {Angulo}, R.~E. and {Annibali}, F. and {Anselmi}, A. and {Anselmi}, S. and {Arcari}, S. and {Archidiacono}, M. and {Aric{\`o}}, G. and {Arnaud}, M. and {Arnouts}, S. and {Asgari}, M. and {Asorey}, J. and {Atayde}, L. and {Atek}, H. and {Atrio-Barandela}, F. and {Aubert}, M. and {Aubourg}, E. and {Auphan}, T. and {Auricchio}, N. and {Aussel}, B. and {Aussel}, H. and {Avelino}, P.~P. and {Avgoustidis}, A. and {Avila}, S. and {Awan}, S. and {Azzollini}, R. and {Baccigalupi}, C. and {Bachelet}, E. and {Bacon}, D. and {Baes}, M. and {Bagley}, M.~B. and {Bahr-Kalus}, B. and {Balaguera-Antolinez}, A. and {Balbinot}, E. and {Balcells}, M. and {Baldi}, M. and {Baldry}, I. and {Balestra}, A. and {Ballardini}, M. and {Ballester}, O. and {Balogh}, M. and {Ba{\~n}ados}, E. and {Barbier}, R. and {Bardelli}, S. and {Baron}, M. and {Barreiro}, T. and {Barrena}, R. and {Barriere}, J.-C. and {Barros}, B.~J. and {Barthelemy}, A. and {Bartolo}, N. and {Basset}, A. and {Battaglia}, P. and {Battisti}, A.~J. and {Baugh}, C.~M. and {Baumont}, L. and {Bazzanini}, L. and {Beaulieu}, J.-P. and {Beckmann}, V. and {Belikov}, A.~N. and {Bel}, J. and {Bellagamba}, F. and {Bella}, M. and {Bellini}, E. and {Benabed}, K. and {Bender}, R. and {Benevento}, G. and {Bennett}, C.~L. and {Benson}, K. and {Bergamini}, P. and {Bermejo-Climent}, J.~R. and {Bernardeau}, F. and {Bertacca}, D. and {Berthe}, M. and {Berthier}, J. and {Bethermin}, M. and {Beutler}, F. and {Bevillon}, C. and {Bhargava}, S. and {Bhatawdekar}, R. and {Bianchi}, D. and {Bisigello}, L. and {Biviano}, A. and {Blake}, R.~P. and {Blanchard}, A. and {Blazek}, J. and {Blot}, L. and {Bosco}, A. and {Bodendorf}, C. and {Boenke}, T. and {B{\"o}hringer}, H. and {Boldrini}, P. and {Bolzonella}, M. and {Bonchi}, A. and {Bonici}, M. and {Bonino}, D. and {Bonino}, L. and {Bonvin}, C. and {Bon}, W. and {Booth}, J.~T. and {Borgani}, S. and {Borlaff}, A.~S. and {Borsato}, E. and {Bose}, B. and {Botticella}, M.~T. and {Boucaud}, A. and {Bouche}, F. and {Boucher}, J.~S. and {Boutigny}, D. and {Bouvard}, T. and {Bouwens}, R. and {Bouy}, H. and {Bowler}, R.~A.~A. and {Bozza}, V. and {Bozzo}, E. and {Branchini}, E. and {Brando}, G. and {Brau-Nogue}, S. and {Brekke}, P. and {Bremer}, M.~N. and {Brescia}, M. and {Breton}, M.-A. and {Brinchmann}, J. and {Brinckmann}, T. and {Brockley-Blatt}, C. and {Brodwin}, M. and {Brouard}, L. and {Brown}, M.~L. and {Bruton}, S. and {Bucko}, J. and {Buddelmeijer}, H. and {Buenadicha}, G. and {Buitrago}, F. and {Burger}, P. and {Burigana}, C. and {Busillo}, V. and {Busonero}, D. and {Cabanac}, R. and {Cabayol-Garcia}, L. and {Cagliari}, M.~S. and {Caillat}, A. and {Caillat}, L. and {Calabrese}, M. and {Calabro}, A. and {Calderone}, G. and {Calura}, F. and {Camacho Quevedo}, B. and {Camera}, S. and {Campos}, L. and {Ca{\~n}as-Herrera}, G. and {Candini}, G.~P. and {Cantiello}, M. and {Capobianco}, V. and {Cappellaro}, E. and {Cappelluti}, N. and {Cappi}, A. and {Caputi}, K.~I. and {Cara}, C. and {Carbone}, C. and {Cardone}, V.~F. and {Carella}, E. and {Carlberg}, R.~G. and {Carle}, M. and {Carminati}, L. and {Caro}, F. and {Carrasco}, J.~M. and {Carretero}, J. and {Carrilho}, P. and {Carron Duque}, J. and {Carry}, B.},
        title = "{Euclid: I. Overview of the Euclid mission}",
      journal = {\aap},
         year = 2025,
        month = may,
       volume = {697},
          eid = {A1},
        pages = {A1},
          doi = {10.1051/0004-6361/202450810},
archivePrefix = {arXiv},
       eprint = {2405.13491},
 primaryClass = {astro-ph.CO},
       adsurl = {https://ui.adsabs.harvard.edu/abs/2025A&A...697A...1E}
}

@ARTICLE{faber2007,
       author = {{Faber}, S.~M. and {Willmer}, C.~N.~A. and {Wolf}, C. and {Koo}, D.~C. and {Weiner}, B.~J. and {Newman}, J.~A. and {Im}, M. and {Coil}, A.~L. and {Conroy}, C. and {Cooper}, M.~C. and {Davis}, M. and {Finkbeiner}, D.~P. and {Gerke}, B.~F. and {Gebhardt}, K. and {Groth}, E.~J. and {Guhathakurta}, P. and {Harker}, J. and {Kaiser}, N. and {Kassin}, S. and {Kleinheinrich}, M. and {Konidaris}, N.~P. and {Kron}, R.~G. and {Lin}, L. and {Luppino}, G. and {Madgwick}, D.~S. and {Meisenheimer}, K. and {Noeske}, K.~G. and {Phillips}, A.~C. and {Sarajedini}, V.~L. and {Schiavon}, R.~P. and {Simard}, L. and {Szalay}, A.~S. and {Vogt}, N.~P. and {Yan}, R.},
        title = "{Galaxy Luminosity Functions to z\raisebox{-0.5ex}\textasciitilde1 from DEEP2 and COMBO-17: Implications for Red Galaxy Formation}",
      journal = {\apj},
         year = 2007,
        month = aug,
       volume = {665},
       number = {1},
        pages = {265-294},
          doi = {10.1086/519294},
archivePrefix = {arXiv},
       eprint = {astro-ph/0506044},
 primaryClass = {astro-ph},
       adsurl = {https://ui.adsabs.harvard.edu/abs/2007ApJ...665..265F}
}

@ARTICLE{fabian2012,
       author = {{Fabian}, A.~C.},
        title = "{Observational Evidence of Active Galactic Nuclei Feedback}",
      journal = {\araa},
         year = 2012,
        month = sep,
       volume = {50},
        pages = {455-489},
          doi = {10.1146/annurev-astro-081811-125521},
archivePrefix = {arXiv},
       eprint = {1204.4114},
 primaryClass = {astro-ph.CO},
       adsurl = {https://ui.adsabs.harvard.edu/abs/2012ARA&A..50..455F}
}

@ARTICLE{fagotto1994a,
       author = {{Fagotto}, F. and {Bressan}, A. and {Bertelli}, G. and {Chiosi}, C.},
        title = "{Evolutionary sequences of stellar models with new radiative opacities. III. Z=0.0004 and Z=0.05}",
      journal = {\aaps},
         year = 1994,
        month = apr,
       volume = {104},
        pages = {365-376},
       adsurl = {https://ui.adsabs.harvard.edu/abs/1994A&AS..104..365F}
}

@ARTICLE{fagotto1994b,
       author = {{Fagotto}, F. and {Bressan}, A. and {Bertelli}, G. and {Chiosi}, C.},
        title = "{Evolutionary sequences of stellar models with new radiative opacities. IV. Z=0.004 and Z=0.008}",
      journal = {\aaps},
         year = 1994,
        month = may,
       volume = {105},
        pages = {29-38},
       adsurl = {https://ui.adsabs.harvard.edu/abs/1994A&AS..105...29F}
}

@ARTICLE{fagotto1994c,
       author = {{Fagotto}, F. and {Bressan}, A. and {Bertelli}, G. and {Chiosi}, C.},
        title = "{Evolutionary sequences of stellar models with very high metallicity. V. Z=0.1}",
      journal = {\aaps},
         year = 1994,
        month = may,
       volume = {105},
        pages = {39-45},
       adsurl = {https://ui.adsabs.harvard.edu/abs/1994A&AS..105...39F}
}

@TECHREPORT{fix1951,
     author = {Evelyn Fix and J. L. Hodges},
     title  = "{Discriminatory Analysis---Nonparametric Discrimination: Consistency Properties}",
     institution = {USAF School of Aviation Medicine},
     address = {Randolph Field, TX},
     year = {1951},
     month = feb,
     number = {4: 21-49-004}
}

@ARTICLE{fix1989,
     author = {Evelyn Fix and J. L. Hodges},
     title = "{Discriminatory Analysis - Nonparametric Discrimination: Consistency Properties}",
     journal = {International Statistical Review},
     year = {1989},
     month = dec,
     volume = {57},
     number = {3},
     pages = {238--247},
     doi = {10.2307/1403797},
     ISSN = {03067734, 17515823},
     URL = {http://www.jstor.org/stable/1403797},
     publisher = {[Wiley, International Statistical Institute (ISI)]}
}

@ARTICLE{fumagalli2011,
       author = {{Fumagalli}, Mattia and {Gavazzi}, G. and {Scaramella}, R. and {Franzetti}, P.},
        title = "{Constraining the ages of the fireballs in the wake of the dIrr galaxy VCC 1217/IC 3418}",
      journal = {\aap},
         year = 2011,
        month = apr,
       volume = {528},
          eid = {A46},
        pages = {A46},
          doi = {10.1051/0004-6361/201015463},
archivePrefix = {arXiv},
       eprint = {1011.1665},
 primaryClass = {astro-ph.CO},
       adsurl = {https://ui.adsabs.harvard.edu/abs/2011A&A...528A..46F}
}

@ARTICLE{gardner2006,
       author = {{Gardner}, Jonathan P. and {Mather}, John C. and {Clampin}, Mark and {Doyon}, Rene and {Greenhouse}, Matthew A. and {Hammel}, Heidi B. and {Hutchings}, John B. and {Jakobsen}, Peter and {Lilly}, Simon J. and {Long}, Knox S. and {Lunine}, Jonathan I. and {McCaughrean}, Mark J. and {Mountain}, Matt and {Nella}, John and {Rieke}, George H. and {Rieke}, Marcia J. and {Rix}, Hans-Walter and {Smith}, Eric P. and {Sonneborn}, George and {Stiavelli}, Massimo and {Stockman}, H.~S. and {Windhorst}, Rogier A. and {Wright}, Gillian S.},
        title = "{The James Webb Space Telescope}",
      journal = {\ssr},
         year = 2006,
        month = apr,
       volume = {123},
       number = {4},
        pages = {485-606},
          doi = {10.1007/s11214-006-8315-7},
archivePrefix = {arXiv},
       eprint = {astro-ph/0606175},
 primaryClass = {astro-ph},
       adsurl = {https://ui.adsabs.harvard.edu/abs/2006SSRv..123..485G}
}

@ARTICLE{gardner2023,
       author = {{Gardner}, Jonathan P. and {Mather}, John C. and {Abbott}, Randy and {Abell}, James S. and {Abernathy}, Mark and {Abney}, Faith E. and {Abraham}, John G. and {Abraham}, Roberto and {Abul-Huda}, Yasin M. and {Acton}, Scott and {Adams}, Cynthia K. and {Adams}, Evan and {Adler}, David S. and {Adriaensen}, Maarten and {Aguilar}, Jonathan Albert and {Ahmed}, Mansoor and {Ahmed}, Nasif S. and {Ahmed}, Tanjira and {Albat}, R{\"u}deger and {Albert}, Lo{\"\i}c and {Alberts}, Stacey and {Aldridge}, David and {Allen}, Mary Marsha and {Allen}, Shaune S. and {Altenburg}, Martin and {Altunc}, Serhat and {Alvarez}, Jose Lorenzo and {{\'A}lvarez-M{\'a}rquez}, Javier and {Alves de Oliveira}, Catarina and {Ambrose}, Leslie L. and {Anandakrishnan}, Satya M. and {Andersen}, Gregory C. and {Anderson}, Harry James and {Anderson}, Jay and {Anderson}, Kristen and {Anderson}, Sara M. and {Aprea}, Julio and {Archer}, Benita J. and {Arenberg}, Jonathan W. and {Argyriou}, Ioannis and {Arribas}, Santiago and {Artigau}, {\'E}tienne and {Arvai}, Amanda Rose and {Atcheson}, Paul and {Atkinson}, Charles B. and {Averbukh}, Jesse and {Aymergen}, Cagatay and {Bacinski}, John J. and {Baggett}, Wayne E. and {Bagnasco}, Giorgio and {Baker}, Lynn L. and {Balzano}, Vicki Ann and {Banks}, Kimberly A. and {Baran}, David A. and {Barker}, Elizabeth A. and {Barrett}, Larry K. and {Barringer}, Bruce O. and {Barto}, Allison and {Bast}, William and {Baudoz}, Pierre and {Baum}, Stefi and {Beatty}, Thomas G. and {Beaulieu}, Mathilde and {Bechtold}, Kathryn and {Beck}, Tracy and {Beddard}, Megan M. and {Beichman}, Charles and {Bellagama}, Larry and {Bely}, Pierre and {Berger}, Timothy W. and {Bergeron}, Louis E. and {Bernier}, Antoine-Darveau and {Bertch}, Maria D. and {Beskow}, Charlotte and {Betz}, Laura E. and {Biagetti}, Carl P. and {Birkmann}, Stephan and {Bjorklund}, Kurt F. and {Blackwood}, James D. and {Blazek}, Ronald Paul and {Blossfeld}, Stephen and {Bluth}, Marcel and {Boccaletti}, Anthony and {Boegner}, Jr., Martin E. and {Bohlin}, Ralph C. and {Boia}, John Joseph and {B{\"o}ker}, Torsten and {Bonaventura}, N. and {Bond}, Nicholas A. and {Bosley}, Kari Ann and {Boucarut}, Rene A. and {Bouchet}, Patrice and {Bouwman}, Jeroen and {Bower}, Gary and {Bowers}, Ariel S. and {Bowers}, Charles W. and {Boyce}, Leslye A. and {Boyer}, Christine T. and {Boyer}, Martha L. and {Boyer}, Michael and {Boyer}, Robert and {Bradley}, Larry D. and {Brady}, Gregory R. and {Brandl}, Bernhard R. and {Brannen}, Judith L. and {Breda}, David and {Bremmer}, Harold G. and {Brennan}, David and {Bresnahan}, Pamela A. and {Bright}, Stacey N. and {Broiles}, Brian J. and {Bromenschenkel}, Asa and {Brooks}, Brian H. and {Brooks}, Keira J. and {Brown}, Bob and {Brown}, Bruce and {Brown}, Thomas M. and {Bruce}, Barry W. and {Bryson}, Jonathan G. and {Bujanda}, Edwin D. and {Bullock}, Blake M. and {Bunker}, A.~J. and {Bureo}, Rafael and {Burt}, Irving J. and {Bush}, James Aaron and {Bushouse}, Howard A. and {Bussman}, Marie C. and {Cabaud}, Olivier and {Cale}, Steven and {Calhoon}, Charles D. and {Calvani}, Humberto and {Canipe}, Alicia M. and {Caputo}, Francis M. and {Cara}, Mihai and {Carey}, Larkin and {Case}, Michael Eli and {Cesari}, Thaddeus and {Cetorelli}, Lee D. and {Chance}, Don R. and {Chandler}, Lynn and {Chaney}, Dave and {Chapman}, George N. and {Charlot}, S. and {Chayer}, Pierre and {Cheezum}, Jeffrey I. and {Chen}, Bin and {Chen}, Christine H. and {Cherinka}, Brian and {Chichester}, Sarah C. and {Chilton}, Zachary S. and {Chittiraibalan}, Dharini and {Clampin}, Mark and {Clark}, Charles R. and {Clark}, Kerry W. and {Clark}, Stephanie M. and {Claybrooks}, Edward E. and {Cleveland}, Keith A. and {Cohen}, Andrew L. and {Cohen}, Lester M. and {Col{\'o}n}, Knicole D. and {Coleman}, Benee L. and {Colina}, Luis and {Comber}, Brian J. and {Comeau}, Thomas M. and {Comer}, Thomas and {Conde Reis}, Alain and {Connolly}, Dennis C. and {Conroy}, Kyle E. and {Contos}, Adam R. and {Contreras}, James and {Cook}, Neil J. and {Cooper}, James L. and {Cooper}, Rachel Aviva and {Correia}, Michael F. and {Correnti}, Matteo and {Cossou}, Christophe and {Costanza}, Brian F. and {Coulais}, Alain and {Cox}, Colin R. and {Coyle}, Ray T. and {Cracraft}, Misty M. and {Crew}, Keith A. and {Curtis}, Gary J. and {Cusveller}, Bianca and {Da Costa Maciel}, Cleyciane and {Dailey}, Christopher T. and {Daugeron}, Fr{\'e}d{\'e}ric and {Davidson}, Greg S. and {Davies}, James E. and {Davis}, Katherine Anne and {Davis}, Michael S. and {Day}, Ratna and {de Chambure}, Daniel and {de Jong}, Pauline and {De Marchi}, Guido and {Dean}, Bruce H. and {Decker}, John E. and {Delisa}, Amy S. and {Dell}, Lawrence C. and {Dellagatta}, Gail},
        title = "{The James Webb Space Telescope Mission}",
      journal = {\pasp},
         year = 2023,
        month = jun,
       volume = {135},
       number = {1048},
          eid = {068001},
        pages = {068001},
          doi = {10.1088/1538-3873/acd1b5},
archivePrefix = {arXiv},
       eprint = {2304.04869},
 primaryClass = {astro-ph.IM},
       adsurl = {https://ui.adsabs.harvard.edu/abs/2023PASP..135f8001G}
}

@ARTICLE{gavazzi2013,
       author = {{Gavazzi}, G. and {Fumagalli}, M. and {Fossati}, M. and {Galardo}, V. and {Grossetti}, F. and {Boselli}, A. and {Giovanelli}, R. and {Haynes}, M.~P.},
        title = "{H{\ensuremath{\alpha}}3: an H{\ensuremath{\alpha}} imaging survey of HI selected galaxies from ALFALFA. II. Star formation properties of galaxies in the Virgo cluster and surroundings}",
      journal = {\aap},
         year = 2013,
        month = may,
       volume = {553},
          eid = {A89},
        pages = {A89},
          doi = {10.1051/0004-6361/201218789},
archivePrefix = {arXiv},
       eprint = {1303.2846},
 primaryClass = {astro-ph.CO},
       adsurl = {https://ui.adsabs.harvard.edu/abs/2013A&A...553A..89G}
}

@ARTICLE{gensior2020,
       author = {{Gensior}, Jindra and {Kruijssen}, J.~M. Diederik and {Keller}, Benjamin W.},
        title = "{Heart of darkness: the influence of galactic dynamics on quenching star formation in galaxy spheroids}",
      journal = {\mnras},
         year = 2020,
        month = jun,
       volume = {495},
       number = {1},
        pages = {199-223},
          doi = {10.1093/mnras/staa1184},
archivePrefix = {arXiv},
       eprint = {2002.01484},
 primaryClass = {astro-ph.GA},
       adsurl = {https://ui.adsabs.harvard.edu/abs/2020MNRAS.495..199G}
}

@ARTICLE{girardi1996,
       author = {{Girardi}, L. and {Bressan}, A. and {Chiosi}, C. and {Bertelli}, G. and {Nasi}, E.},
        title = "{Evolutionary sequences of stellar models with new radiative opacities. VI. Z=0.0001.}",
      journal = {\aaps},
         year = 1996,
        month = may,
       volume = {117},
        pages = {113-125},
       adsurl = {https://ui.adsabs.harvard.edu/abs/1996A&AS..117..113G}
}

@ARTICLE{goddard2017,
       author = {{Goddard}, D. and {Thomas}, D. and {Maraston}, C. and {Westfall}, K. and {Etherington}, J. and {Riffel}, R. and {Mallmann}, N.~D. and {Zheng}, Z. and {Argudo-Fern{\'a}ndez}, M. and {Lian}, J. and {Bershady}, M. and {Bundy}, K. and {Drory}, N. and {Law}, D. and {Yan}, R. and {Wake}, D. and {Weijmans}, A. and {Bizyaev}, D. and {Brownstein}, J. and {Lane}, R.~R. and {Maiolino}, R. and {Masters}, K. and {Merrifield}, M. and {Nitschelm}, C. and {Pan}, K. and {Roman-Lopes}, A. and {Storchi-Bergmann}, T. and {Schneider}, D.~P.},
        title = "{SDSS-IV MaNGA: Spatially resolved star formation histories in galaxies as a function of galaxy mass and type}",
      journal = {\mnras},
         year = 2017,
        month = apr,
       volume = {466},
       number = {4},
        pages = {4731-4758},
          doi = {10.1093/mnras/stw3371},
archivePrefix = {arXiv},
       eprint = {1612.01546},
 primaryClass = {astro-ph.GA},
       adsurl = {https://ui.adsabs.harvard.edu/abs/2017MNRAS.466.4731G}
}

@ARTICLE{gomez2012,
       author = {{G{\'o}mez}, P.~L. and {Valkonen}, L.~E. and {Romer}, A.~K. and {Lloyd-Davies}, E. and {Verdugo}, T. and {Cantalupo}, C.~M. and {Daub}, M.~D. and {Goldstein}, J.~H. and {Kuo}, C.~L. and {Lange}, A.~E. and {Lueker}, M. and {Holzapfel}, W.~L. and {Peterson}, J.~B. and {Ruhl}, J. and {Runyan}, M.~C. and {Reichardt}, C.~L. and {Sabirli}, K.},
        title = "{Optical and X-Ray Observations of the Merging Cluster AS1063}",
      journal = {\aj},
         year = 2012,
        month = sep,
       volume = {144},
       number = {3},
          eid = {79},
        pages = {79},
          doi = {10.1088/0004-6256/144/3/79},
       adsurl = {https://ui.adsabs.harvard.edu/abs/2012AJ....144...79G}
}

@ARTICLE{gonzalez2015,
       author = {{Gonz{\'a}lez Delgado}, R.~M. and {Garc{\'\i}a-Benito}, R. and {P{\'e}rez}, E. and {Cid Fernandes}, R. and {de Amorim}, A.~L. and {Cortijo-Ferrero}, C. and {Lacerda}, E.~A.~D. and {L{\'o}pez Fern{\'a}ndez}, R. and {Vale-Asari}, N. and {S{\'a}nchez}, S.~F. and {Moll{\'a}}, M. and {Ruiz-Lara}, T. and {S{\'a}nchez-Bl{\'a}zquez}, P. and {Walcher}, C.~J. and {Alves}, J. and {Aguerri}, J.~A.~L. and {Bekerait{\'e}}, S. and {Bland-Hawthorn}, J. and {Galbany}, L. and {Gallazzi}, A. and {Husemann}, B. and {Iglesias-P{\'a}ramo}, J. and {Kalinova}, V. and {L{\'o}pez-S{\'a}nchez}, A.~R. and {Marino}, R.~A. and {M{\'a}rquez}, I. and {Masegosa}, J. and {Mast}, D. and {M{\'e}ndez-Abreu}, J. and {Mendoza}, A. and {del Olmo}, A. and {P{\'e}rez}, I. and {Quirrenbach}, A. and {Zibetti}, S.},
        title = "{The CALIFA survey across the Hubble sequence. Spatially resolved stellar population properties in galaxies}",
      journal = {\aap},
         year = 2015,
        month = sep,
       volume = {581},
          eid = {A103},
        pages = {A103},
          doi = {10.1051/0004-6361/201525938},
archivePrefix = {arXiv},
       eprint = {1506.04157},
 primaryClass = {astro-ph.GA},
       adsurl = {https://ui.adsabs.harvard.edu/abs/2015A&A...581A.103G}
}

@ARTICLE{gonzalez2016,
       author = {{Gonz{\'a}lez Delgado}, R.~M. and {Cid Fernandes}, R. and {P{\'e}rez}, E. and {Garc{\'\i}a-Benito}, R. and {L{\'o}pez Fern{\'a}ndez}, R. and {Lacerda}, E.~A.~D. and {Cortijo-Ferrero}, C. and {de Amorim}, A.~L. and {Vale Asari}, N. and {S{\'a}nchez}, S.~F. and {Walcher}, C.~J. and {Wisotzki}, L. and {Mast}, D. and {Alves}, J. and {Ascasibar}, Y. and {Bland-Hawthorn}, J. and {Galbany}, L. and {Kennicutt}, R.~C. and {M{\'a}rquez}, I. and {Masegosa}, J. and {Moll{\'a}}, M. and {S{\'a}nchez-Bl{\'a}zquez}, P. and {V{\'\i}lchez}, J.~M.},
        title = "{Star formation along the Hubble sequence. Radial structure of the star formation of CALIFA galaxies}",
      journal = {\aap},
         year = 2016,
        month = may,
       volume = {590},
          eid = {A44},
        pages = {A44},
          doi = {10.1051/0004-6361/201628174},
archivePrefix = {arXiv},
       eprint = {1603.00874},
 primaryClass = {astro-ph.GA},
       adsurl = {https://ui.adsabs.harvard.edu/abs/2016A&A...590A..44G}
}

@ARTICLE{grogin2011,
       author = {{Grogin}, Norman A. and {Kocevski}, Dale D. and {Faber}, S.~M. and {Ferguson}, Henry C. and {Koekemoer}, Anton M. and {Riess}, Adam G. and {Acquaviva}, Viviana and {Alexander}, David M. and {Almaini}, Omar and {Ashby}, Matthew L.~N. and {Barden}, Marco and {Bell}, Eric F. and {Bournaud}, Fr{\'e}d{\'e}ric and {Brown}, Thomas M. and {Caputi}, Karina I. and {Casertano}, Stefano and {Cassata}, Paolo and {Castellano}, Marco and {Challis}, Peter and {Chary}, Ranga-Ram and {Cheung}, Edmond and {Cirasuolo}, Michele and {Conselice}, Christopher J. and {Roshan Cooray}, Asantha and {Croton}, Darren J. and {Daddi}, Emanuele and {Dahlen}, Tomas and {Dav{\'e}}, Romeel and {de Mello}, Du{\'\i}lia F. and {Dekel}, Avishai and {Dickinson}, Mark and {Dolch}, Timothy and {Donley}, Jennifer L. and {Dunlop}, James S. and {Dutton}, Aaron A. and {Elbaz}, David and {Fazio}, Giovanni G. and {Filippenko}, Alexei V. and {Finkelstein}, Steven L. and {Fontana}, Adriano and {Gardner}, Jonathan P. and {Garnavich}, Peter M. and {Gawiser}, Eric and {Giavalisco}, Mauro and {Grazian}, Andrea and {Guo}, Yicheng and {Hathi}, Nimish P. and {H{\"a}ussler}, Boris and {Hopkins}, Philip F. and {Huang}, Jia-Sheng and {Huang}, Kuang-Han and {Jha}, Saurabh W. and {Kartaltepe}, Jeyhan S. and {Kirshner}, Robert P. and {Koo}, David C. and {Lai}, Kamson and {Lee}, Kyoung-Soo and {Li}, Weidong and {Lotz}, Jennifer M. and {Lucas}, Ray A. and {Madau}, Piero and {McCarthy}, Patrick J. and {McGrath}, Elizabeth J. and {McIntosh}, Daniel H. and {McLure}, Ross J. and {Mobasher}, Bahram and {Moustakas}, Leonidas A. and {Mozena}, Mark and {Nandra}, Kirpal and {Newman}, Jeffrey A. and {Niemi}, Sami-Matias and {Noeske}, Kai G. and {Papovich}, Casey J. and {Pentericci}, Laura and {Pope}, Alexandra and {Primack}, Joel R. and {Rajan}, Abhijith and {Ravindranath}, Swara and {Reddy}, Naveen A. and {Renzini}, Alvio and {Rix}, Hans-Walter and {Robaina}, Aday R. and {Rodney}, Steven A. and {Rosario}, David J. and {Rosati}, Piero and {Salimbeni}, Sara and {Scarlata}, Claudia and {Siana}, Brian and {Simard}, Luc and {Smidt}, Joseph and {Somerville}, Rachel S. and {Spinrad}, Hyron and {Straughn}, Amber N. and {Strolger}, Louis-Gregory and {Telford}, Olivia and {Teplitz}, Harry I. and {Trump}, Jonathan R. and {van der Wel}, Arjen and {Villforth}, Carolin and {Wechsler}, Risa H. and {Weiner}, Benjamin J. and {Wiklind}, Tommy and {Wild}, Vivienne and {Wilson}, Grant and {Wuyts}, Stijn and {Yan}, Hao-Jing and {Yun}, Min S.},
        title = "{CANDELS: The Cosmic Assembly Near-infrared Deep Extragalactic Legacy Survey}",
      journal = {\apjs},
         year = 2011,
        month = dec,
       volume = {197},
       number = {2},
          eid = {35},
        pages = {35},
          doi = {10.1088/0067-0049/197/2/35},
archivePrefix = {arXiv},
       eprint = {1105.3753},
 primaryClass = {astro-ph.CO},
       adsurl = {https://ui.adsabs.harvard.edu/abs/2011ApJS..197...35G}
}

@ARTICLE{gunn1972,
       author = {{Gunn}, James E. and {Gott}, III, J. Richard},
        title = "{On the Infall of Matter Into Clusters of Galaxies and Some Effects on Their Evolution}",
      journal = {\apj},
         year = 1972,
        month = aug,
       volume = {176},
        pages = {1},
          doi = {10.1086/151605},
       adsurl = {https://ui.adsabs.harvard.edu/abs/1972ApJ...176....1G}
}

@ARTICLE{guo2019,
       author = {{Guo}, Kexin and {Peng}, Yingjie and {Shao}, Li and {Fu}, Hai and {Catinella}, Barbara and {Cortese}, Luca and {Yuan}, Feng and {Yan}, Renbin and {Zhang}, Chengpeng and {Dou}, Jing},
        title = "{SDSS-IV MaNGA: The Roles of AGNs and Dynamical Processes in Star Formation Quenching in Nearby Disk Galaxies}",
      journal = {\apj},
         year = 2019,
        month = jan,
       volume = {870},
       number = {1},
          eid = {19},
        pages = {19},
          doi = {10.3847/1538-4357/aaee88},
archivePrefix = {arXiv},
       eprint = {1811.01957},
 primaryClass = {astro-ph.GA},
       adsurl = {https://ui.adsabs.harvard.edu/abs/2019ApJ...870...19G}
}

@ARTICLE{habouzit2022,
       author = {{Habouzit}, M{\'e}lanie and {Somerville}, Rachel S. and {Li}, Yuan and {Genel}, Shy and {Aird}, James and {Angl{\'e}s-Alc{\'a}zar}, Daniel and {Dav{\'e}}, Romeel and {Georgiev}, Iskren Y. and {McAlpine}, Stuart and {Rosas-Guevara}, Yetli and {Dubois}, Yohan and {Nelson}, Dylan and {Banados}, Eduardo and {Hernquist}, Lars and {Peirani}, S{\'e}bastien and {Vogelsberger}, Mark},
        title = "{Supermassive black holes in cosmological simulations - II: the AGN population and predictions for upcoming X-ray missions}",
      journal = {\mnras},
         year = 2022,
        month = jan,
       volume = {509},
       number = {2},
        pages = {3015-3042},
          doi = {10.1093/mnras/stab3147},
archivePrefix = {arXiv},
       eprint = {2111.01802},
 primaryClass = {astro-ph.GA},
       adsurl = {https://ui.adsabs.harvard.edu/abs/2022MNRAS.509.3015H}
}

@ARTICLE{harris2020,
       author = {{Harris}, Charles R. and {Millman}, K. Jarrod and {van der Walt}, St{\'e}fan J. and {Gommers}, Ralf and {Virtanen}, Pauli and {Cournapeau}, David and {Wieser}, Eric and {Taylor}, Julian and {Berg}, Sebastian and {Smith}, Nathaniel J. and {Kern}, Robert and {Picus}, Matti and {Hoyer}, Stephan and {van Kerkwijk}, Marten H. and {Brett}, Matthew and {Haldane}, Allan and {del R{\'\i}o}, Jaime Fern{\'a}ndez and {Wiebe}, Mark and {Peterson}, Pearu and {G{\'e}rard-Marchant}, Pierre and {Sheppard}, Kevin and {Reddy}, Tyler and {Weckesser}, Warren and {Abbasi}, Hameer and {Gohlke}, Christoph and {Oliphant}, Travis E.},
        title = "{Array programming with NumPy}",
      journal = {\nat},
         year = 2020,
        month = sep,
       volume = {585},
       number = {7825},
        pages = {357-362},
          doi = {10.1038/s41586-020-2649-2},
archivePrefix = {arXiv},
       eprint = {2006.10256},
 primaryClass = {cs.MS},
       adsurl = {https://ui.adsabs.harvard.edu/abs/2020Natur.585..357H}
}

@ARTICLE{hunter2007,
       author = {{Hunter}, John D.},
        title = "{Matplotlib: A 2D Graphics Environment}",
      journal = {CSE},
         year = 2007,
        month = jan,
       volume = {9},
       number = {3},
        pages = {90-95},
          doi = {10.1109/MCSE.2007.55},
       adsurl = {https://ui.adsabs.harvard.edu/abs/2007CSE.....9...90H}
}

@ARTICLE{jauzac2014,
       author = {{Jauzac}, M. and {Cl{\'e}ment}, B. and {Limousin}, M. and {Richard}, J. and {Jullo}, E. and {Ebeling}, H. and {Atek}, H. and {Kneib}, J.-P. and {Knowles}, K. and {Natarajan}, P. and {Eckert}, D. and {Egami}, E. and {Massey}, R. and {Rexroth}, M.},
        title = "{Hubble Frontier Fields: a high-precision strong-lensing analysis of galaxy cluster MACSJ0416.1-2403 using {\ensuremath{\sim}}200 multiple images}",
      journal = {\mnras},
         year = 2014,
        month = sep,
       volume = {443},
       number = {2},
        pages = {1549-1554},
          doi = {10.1093/mnras/stu1355},
archivePrefix = {arXiv},
       eprint = {1405.3582},
 primaryClass = {astro-ph.CO},
       adsurl = {https://ui.adsabs.harvard.edu/abs/2014MNRAS.443.1549J}
}

@ARTICLE{johnson2021,
       author = {{Johnson}, Benjamin D. and {Leja}, Joel and {Conroy}, Charlie and {Speagle}, Joshua S.},
        title = "{Stellar Population Inference with Prospector}",
      journal = {\apjs},
         year = 2021,
        month = jun,
       volume = {254},
       number = {2},
          eid = {22},
        pages = {22},
          doi = {10.3847/1538-4365/abef67},
archivePrefix = {arXiv},
       eprint = {2012.01426},
 primaryClass = {astro-ph.GA},
       adsurl = {https://ui.adsabs.harvard.edu/abs/2021ApJS..254...22J}
}

@ARTICLE{kawata2008,
       author = {{Kawata}, Daisuke and {Mulchaey}, John S.},
        title = "{Strangulation in Galaxy Groups}",
      journal = {\apjl},
         year = 2008,
        month = jan,
       volume = {672},
       number = {2},
        pages = {L103},
          doi = {10.1086/526544},
archivePrefix = {arXiv},
       eprint = {0707.3814},
 primaryClass = {astro-ph},
       adsurl = {https://ui.adsabs.harvard.edu/abs/2008ApJ...672L.103K}
}

@ARTICLE{kennicutt1998a,
       author = {{Kennicutt}, Jr., Robert C.},
        title = "{The Global Schmidt Law in Star-forming Galaxies}",
      journal = {\apj},
         year = 1998,
        month = may,
       volume = {498},
       number = {2},
        pages = {541-552},
          doi = {10.1086/305588},
archivePrefix = {arXiv},
       eprint = {astro-ph/9712213},
 primaryClass = {astro-ph},
       adsurl = {https://ui.adsabs.harvard.edu/abs/1998ApJ...498..541K}
}

@ARTICLE{kennicutt1998b,
       author = {{Kennicutt}, Jr., Robert C.},
        title = "{Star Formation in Galaxies Along the Hubble Sequence}",
      journal = {\araa},
         year = 1998,
        month = jan,
       volume = {36},
        pages = {189-232},
          doi = {10.1146/annurev.astro.36.1.189},
archivePrefix = {arXiv},
       eprint = {astro-ph/9807187},
 primaryClass = {astro-ph},
       adsurl = {https://ui.adsabs.harvard.edu/abs/1998ARA&A..36..189K}
}

@ARTICLE{kennicutt2012,
       author = {{Kennicutt}, Robert C. and {Evans}, Neal J.},
        title = "{Star Formation in the Milky Way and Nearby Galaxies}",
      journal = {\araa},
         year = 2012,
        month = sep,
       volume = {50},
        pages = {531-608},
          doi = {10.1146/annurev-astro-081811-125610},
archivePrefix = {arXiv},
       eprint = {1204.3552},
 primaryClass = {astro-ph.GA},
       adsurl = {https://ui.adsabs.harvard.edu/abs/2012ARA&A..50..531K}
}

@ARTICLE{koekemoer2011,
       author = {{Koekemoer}, Anton M. and {Faber}, S.~M. and {Ferguson}, Henry C. and {Grogin}, Norman A. and {Kocevski}, Dale D. and {Koo}, David C. and {Lai}, Kamson and {Lotz}, Jennifer M. and {Lucas}, Ray A. and {McGrath}, Elizabeth J. and {Ogaz}, Sara and {Rajan}, Abhijith and {Riess}, Adam G. and {Rodney}, Steve A. and {Strolger}, Louis and {Casertano}, Stefano and {Castellano}, Marco and {Dahlen}, Tomas and {Dickinson}, Mark and {Dolch}, Timothy and {Fontana}, Adriano and {Giavalisco}, Mauro and {Grazian}, Andrea and {Guo}, Yicheng and {Hathi}, Nimish P. and {Huang}, Kuang-Han and {van der Wel}, Arjen and {Yan}, Hao-Jing and {Acquaviva}, Viviana and {Alexander}, David M. and {Almaini}, Omar and {Ashby}, Matthew L.~N. and {Barden}, Marco and {Bell}, Eric F. and {Bournaud}, Fr{\'e}d{\'e}ric and {Brown}, Thomas M. and {Caputi}, Karina I. and {Cassata}, Paolo and {Challis}, Peter J. and {Chary}, Ranga-Ram and {Cheung}, Edmond and {Cirasuolo}, Michele and {Conselice}, Christopher J. and {Roshan Cooray}, Asantha and {Croton}, Darren J. and {Daddi}, Emanuele and {Dav{\'e}}, Romeel and {de Mello}, Duilia F. and {de Ravel}, Loic and {Dekel}, Avishai and {Donley}, Jennifer L. and {Dunlop}, James S. and {Dutton}, Aaron A. and {Elbaz}, David and {Fazio}, Giovanni G. and {Filippenko}, Alexei V. and {Finkelstein}, Steven L. and {Frazer}, Chris and {Gardner}, Jonathan P. and {Garnavich}, Peter M. and {Gawiser}, Eric and {Gruetzbauch}, Ruth and {Hartley}, Will G. and {H{\"a}ussler}, Boris and {Herrington}, Jessica and {Hopkins}, Philip F. and {Huang}, Jia-Sheng and {Jha}, Saurabh W. and {Johnson}, Andrew and {Kartaltepe}, Jeyhan S. and {Khostovan}, Ali A. and {Kirshner}, Robert P. and {Lani}, Caterina and {Lee}, Kyoung-Soo and {Li}, Weidong and {Madau}, Piero and {McCarthy}, Patrick J. and {McIntosh}, Daniel H. and {McLure}, Ross J. and {McPartland}, Conor and {Mobasher}, Bahram and {Moreira}, Heidi and {Mortlock}, Alice and {Moustakas}, Leonidas A. and {Mozena}, Mark and {Nandra}, Kirpal and {Newman}, Jeffrey A. and {Nielsen}, Jennifer L. and {Niemi}, Sami and {Noeske}, Kai G. and {Papovich}, Casey J. and {Pentericci}, Laura and {Pope}, Alexandra and {Primack}, Joel R. and {Ravindranath}, Swara and {Reddy}, Naveen A. and {Renzini}, Alvio and {Rix}, Hans-Walter and {Robaina}, Aday R. and {Rosario}, David J. and {Rosati}, Piero and {Salimbeni}, Sara and {Scarlata}, Claudia and {Siana}, Brian and {Simard}, Luc and {Smidt}, Joseph and {Snyder}, Diana and {Somerville}, Rachel S. and {Spinrad}, Hyron and {Straughn}, Amber N. and {Telford}, Olivia and {Teplitz}, Harry I. and {Trump}, Jonathan R. and {Vargas}, Carlos and {Villforth}, Carolin and {Wagner}, Cory R. and {Wandro}, Pat and {Wechsler}, Risa H. and {Weiner}, Benjamin J. and {Wiklind}, Tommy and {Wild}, Vivienne and {Wilson}, Grant and {Wuyts}, Stijn and {Yun}, Min S.},
        title = "{CANDELS: The Cosmic Assembly Near-infrared Deep Extragalactic Legacy Survey{\textemdash}The Hubble Space Telescope Observations, Imaging Data Products, and Mosaics}",
      journal = {\apjs},
         year = 2011,
        month = dec,
       volume = {197},
       number = {2},
          eid = {36},
        pages = {36},
          doi = {10.1088/0067-0049/197/2/36},
archivePrefix = {arXiv},
       eprint = {1105.3754},
 primaryClass = {astro-ph.CO},
       adsurl = {https://ui.adsabs.harvard.edu/abs/2011ApJS..197...36K}
}

@ARTICLE{koopmann2004a,
       author = {{Koopmann}, Rebecca A. and {Kenney}, Jeffrey D.~P.},
        title = "{Massive Star Formation Rates and Radial Distributions from H{\ensuremath{\alpha}} Imaging of 84 Virgo Cluster and Isolated Spiral Galaxies}",
      journal = {\apj},
         year = 2004,
        month = oct,
       volume = {613},
       number = {2},
        pages = {851-865},
          doi = {10.1086/423190},
archivePrefix = {arXiv},
       eprint = {astro-ph/0209547},
 primaryClass = {astro-ph},
       adsurl = {https://ui.adsabs.harvard.edu/abs/2004ApJ...613..851K}
}

@ARTICLE{koopmann2004b,
       author = {{Koopmann}, Rebecca A. and {Kenney}, Jeffrey D.~P.},
        title = "{H{\ensuremath{\alpha}} Morphologies and Environmental Effects in Virgo Cluster Spiral Galaxies}",
      journal = {\apj},
         year = 2004,
        month = oct,
       volume = {613},
       number = {2},
        pages = {866-885},
          doi = {10.1086/423191},
archivePrefix = {arXiv},
       eprint = {astro-ph/0406243},
 primaryClass = {astro-ph},
       adsurl = {https://ui.adsabs.harvard.edu/abs/2004ApJ...613..866K}
}

@ARTICLE{kriek2009,
       author = {{Kriek}, Mariska and {van Dokkum}, Pieter G. and {Labb{\'e}}, Ivo and {Franx}, Marijn and {Illingworth}, Garth D. and {Marchesini}, Danilo and {Quadri}, Ryan F.},
        title = "{An Ultra-Deep Near-Infrared Spectrum of a Compact Quiescent Galaxy at z = 2.2}",
      journal = {\apj},
         year = 2009,
        month = jul,
       volume = {700},
       number = {1},
        pages = {221-231},
          doi = {10.1088/0004-637X/700/1/221},
archivePrefix = {arXiv},
       eprint = {0905.1692},
 primaryClass = {astro-ph.CO},
       adsurl = {https://ui.adsabs.harvard.edu/abs/2009ApJ...700..221K}
}

@ARTICLE{kristian1978,
       author = {{Kristian}, J. and {Sandage}, A. and {Westphal}, J.~A.},
        title = "{The extension of the Hubble diagram. II. New redshifts and photometry of very distant galaxy clusters: first indication of a deviation of the Hubble diagram from a straight line.}",
      journal = {\apj},
         year = 1978,
        month = apr,
       volume = {221},
        pages = {383-394},
          doi = {10.1086/156038},
       adsurl = {https://ui.adsabs.harvard.edu/abs/1978ApJ...221..383K}
}

@ARTICLE{kulkarni2021,
       author = {{Kulkarni}, S.~R. and {Harrison}, Fiona A. and {Grefenstette}, Brian W. and {Earnshaw}, Hannah P. and {Andreoni}, Igor and {Berg}, Danielle A. and {Bloom}, Joshua S. and {Cenko}, S. Bradley and {Chornock}, Ryan and {Christiansen}, Jessie L. and {Coughlin}, Michael W. and {Wuollet Criswell}, Alexander and {Darvish}, Behnam and {Das}, Kaustav K. and {De}, Kishalay and {Dessart}, Luc and {Dixon}, Don and {Dorsman}, Bas and {El-Badry}, Kareem and {Evans}, Christopher and {Ford}, K.~E. Saavik and {Fremling}, Christoffer and {Gansicke}, Boris T. and {Gezari}, Suvi and {Goetberg}, Y. and {Green}, Gregory M. and {Graham}, Matthew J. and {Heida}, Marianne and {Ho}, Anna Y.~Q. and {Jaodand}, Amruta D. and {Johns-Krull}, Christopher M. and {Kasliwal}, Mansi M. and {Lazzarini}, Margaret and {Lu}, Wenbin and {Margutti}, Raffaella and {Martin}, D. Christopher and {Masters}, Daniel Charles and {McKernan}, Barry and {Naze}, Yael and {Nissanke}, Samaya M. and {Parazin}, B. and {Perley}, Daniel A. and {Phinney}, E. Sterl and {Piro}, Anthony L. and {Raaijmakers}, G. and {Rauw}, Gregor and {Rodriguez}, Antonio C. and {Sana}, Hugues and {Senchyna}, Peter and {Singer}, Leo P. and {Spake}, Jessica J. and {Stassun}, Keivan G. and {Stern}, Daniel and {Teplitz}, Harry I. and {Weisz}, Daniel R. and {Yao}, Yuhan},
        title = "{Science with the Ultraviolet Explorer (UVEX)}",
      journal = {arXiv e-prints},
         year = 2021,
        month = nov,
          eid = {arXiv:2111.15608},
        pages = {arXiv:2111.15608},
          doi = {10.48550/arXiv.2111.15608},
archivePrefix = {arXiv},
       eprint = {2111.15608},
 primaryClass = {astro-ph.GA},
       adsurl = {https://ui.adsabs.harvard.edu/abs/2021arXiv211115608K}
}

@ARTICLE{larson1980,
       author = {{Larson}, R.~B. and {Tinsley}, B.~M. and {Caldwell}, C.~N.},
        title = "{The evolution of disk galaxies and the origin of S0 galaxies}",
      journal = {\apj},
         year = 1980,
        month = may,
       volume = {237},
        pages = {692-707},
          doi = {10.1086/157917},
       adsurl = {https://ui.adsabs.harvard.edu/abs/1980ApJ...237..692L}
}

@ARTICLE{lawlor2026a,
       author = {{Lawlor-Forsyth}, Cameron and {Balogh}, Michael L. and {Sazonova}, Elizaveta and {Morgan}, Cameron R. and {McGee}, Sean L. and {Rudnick}, Gregory H.},
        title = "{Identifying and Distinguishing Quenching Galaxies with Spatially Resolved Star Formation in TNG50}",
      journal = {\apj},
         year = 2026,
        month = mar,
       volume = {1000},
       number = {1},
          eid = {61},
        pages = {61},
          doi = {10.3847/1538-4357/ae4502},
archivePrefix = {arXiv},
       eprint = {2603.00444},
 primaryClass = {astro-ph.GA},
       adsurl = {https://ui.adsabs.harvard.edu/abs/2026ApJ..1000...61L}
}

@ARTICLE{lawlor2026b,
       author = {{Lawlor-Forsyth}, Cameron and {Balogh}, Michael L. and {McGee}, Sean L. and {Rudnick}, Gregory H.},
        title = "{Identifying and Distinguishing Quenching Galaxies with Spatially Resolved Star Formation in Mock CASTOR and NGRST Observations}",
      journal = {arXiv e-prints},
         year = 2026,
        month = jul,
          eid = {arXiv:2607.15638},
        pages = {arXiv:2607.15638},
          doi = {10.48550/arXiv.2607.15638},
archivePrefix = {arXiv},
       eprint = {2607.15638},
 primaryClass = {astro-ph.GA},
       adsurl = {https://ui.adsabs.harvard.edu/abs/2026arXiv260715638L}
}

@ARTICLE{lawlor2026c,
       author = {{Lawlor-Forsyth}, Cameron},
        title = "{Estimating Cluster Galaxy Infall Time from Phase Space in TNG}",
      journal = {arXiv e-prints},
         year = 2026,
        month = jul,
          eid = {arXiv:2607.20406},
        pages = {arXiv:2607.20406},
archivePrefix = {arXiv},
       eprint = {2607.20406},
 primaryClass = {astro-ph.GA},
       adsurl = {https://ui.adsabs.harvard.edu/abs/2026arXiv260720406L}
}

@ARTICLE{leja2017,
       author = {{Leja}, Joel and {Johnson}, Benjamin D. and {Conroy}, Charlie and {van Dokkum}, Pieter G. and {Byler}, Nell},
        title = "{Deriving Physical Properties from Broadband Photometry with Prospector: Description of the Model and a Demonstration of its Accuracy Using 129 Galaxies in the Local Universe}",
      journal = {\apj},
         year = 2017,
        month = mar,
       volume = {837},
       number = {2},
          eid = {170},
        pages = {170},
          doi = {10.3847/1538-4357/aa5ffe},
archivePrefix = {arXiv},
       eprint = {1609.09073},
 primaryClass = {astro-ph.GA},
       adsurl = {https://ui.adsabs.harvard.edu/abs/2017ApJ...837..170L}
}

@ARTICLE{leja2019a,
       author = {{Leja}, Joel and {Carnall}, Adam C. and {Johnson}, Benjamin D. and {Conroy}, Charlie and {Speagle}, Joshua S.},
        title = "{How to Measure Galaxy Star Formation Histories. II. Nonparametric Models}",
      journal = {\apj},
         year = 2019,
        month = may,
       volume = {876},
       number = {1},
          eid = {3},
        pages = {3},
          doi = {10.3847/1538-4357/ab133c},
archivePrefix = {arXiv},
       eprint = {1811.03637},
 primaryClass = {astro-ph.GA},
       adsurl = {https://ui.adsabs.harvard.edu/abs/2019ApJ...876....3L}
}

@ARTICLE{lin2019,
       author = {{Lin}, Lihwai and {Hsieh}, Bau-Ching and {Pan}, Hsi-An and {Rembold}, Sandro B. and {S{\'a}nchez}, Sebasti{\'a}n F. and {Argudo-Fern{\'a}ndez}, Maria and {Rowlands}, Kate and {Belfiore}, Francesco and {Bizyaev}, Dmitry and {Lacerna}, Ivan and {Riffel}, Rogr{\'e}io and {Rong}, Yu and {Yuan}, Fangting and {Drory}, Niv and {Maiolino}, Roberto and {Wilcots}, Eric},
        title = "{SDSS-IV MaNGA: Inside-out versus Outside-in Quenching of Galaxies in Different Local Environments}",
      journal = {\apj},
         year = 2019,
        month = feb,
       volume = {872},
       number = {1},
          eid = {50},
        pages = {50},
          doi = {10.3847/1538-4357/aafa84},
archivePrefix = {arXiv},
       eprint = {1901.05126},
 primaryClass = {astro-ph.GA},
       adsurl = {https://ui.adsabs.harvard.edu/abs/2019ApJ...872...50L}
}

@ARTICLE{lintott2008,
       author = {{Lintott}, Chris J. and {Schawinski}, Kevin and {Slosar}, An{\v{z}}e and {Land}, Kate and {Bamford}, Steven and {Thomas}, Daniel and {Raddick}, M. Jordan and {Nichol}, Robert C. and {Szalay}, Alex and {Andreescu}, Dan and {Murray}, Phil and {Vandenberg}, Jan},
        title = "{Galaxy Zoo: morphologies derived from visual inspection of galaxies from the Sloan Digital Sky Survey}",
      journal = {\mnras},
         year = 2008,
        month = sep,
       volume = {389},
       number = {3},
        pages = {1179-1189},
          doi = {10.1111/j.1365-2966.2008.13689.x},
archivePrefix = {arXiv},
       eprint = {0804.4483},
 primaryClass = {astro-ph},
       adsurl = {https://ui.adsabs.harvard.edu/abs/2008MNRAS.389.1179L}
}

@ARTICLE{lotz2017,
       author = {{Lotz}, J.~M. and {Koekemoer}, A. and {Coe}, D. and {Grogin}, N. and {Capak}, P. and {Mack}, J. and {Anderson}, J. and {Avila}, R. and {Barker}, E.~A. and {Borncamp}, D. and {Brammer}, G. and {Durbin}, M. and {Gunning}, H. and {Hilbert}, B. and {Jenkner}, H. and {Khandrika}, H. and {Levay}, Z. and {Lucas}, R.~A. and {MacKenty}, J. and {Ogaz}, S. and {Porterfield}, B. and {Reid}, N. and {Robberto}, M. and {Royle}, P. and {Smith}, L.~J. and {Storrie-Lombardi}, L.~J. and {Sunnquist}, B. and {Surace}, J. and {Taylor}, D.~C. and {Williams}, R. and {Bullock}, J. and {Dickinson}, M. and {Finkelstein}, S. and {Natarajan}, P. and {Richard}, J. and {Robertson}, B. and {Tumlinson}, J. and {Zitrin}, A. and {Flanagan}, K. and {Sembach}, K. and {Soifer}, B.~T. and {Mountain}, M.},
        title = "{The Frontier Fields: Survey Design and Initial Results}",
      journal = {\apj},
         year = 2017,
        month = mar,
       volume = {837},
       number = {1},
          eid = {97},
        pages = {97},
          doi = {10.3847/1538-4357/837/1/97},
archivePrefix = {arXiv},
       eprint = {1605.06567},
 primaryClass = {astro-ph.GA},
       adsurl = {https://ui.adsabs.harvard.edu/abs/2017ApJ...837...97L}
}

@ARTICLE{lsst2009,
       author = {{LSST Science Collaboration} and {Abell}, Paul A. and {Allison}, Julius and {Anderson}, Scott F. and {Andrew}, John R. and {Angel}, J. Roger P. and {Armus}, Lee and {Arnett}, David and {Asztalos}, S.~J. and {Axelrod}, Tim S. and {Bailey}, Stephen and {Ballantyne}, D.~R. and {Bankert}, Justin R. and {Barkhouse}, Wayne A. and {Barr}, Jeffrey D. and {Barrientos}, L. Felipe and {Barth}, Aaron J. and {Bartlett}, James G. and {Becker}, Andrew C. and {Becla}, Jacek and {Beers}, Timothy C. and {Bernstein}, Joseph P. and {Biswas}, Rahul and {Blanton}, Michael R. and {Bloom}, Joshua S. and {Bochanski}, John J. and {Boeshaar}, Pat and {Borne}, Kirk D. and {Bradac}, Marusa and {Brandt}, W.~N. and {Bridge}, Carrie R. and {Brown}, Michael E. and {Brunner}, Robert J. and {Bullock}, James S. and {Burgasser}, Adam J. and {Burge}, James H. and {Burke}, David L. and {Cargile}, Phillip A. and {Chandrasekharan}, Srinivasan and {Chartas}, George and {Chesley}, Steven R. and {Chu}, You-Hua and {Cinabro}, David and {Claire}, Mark W. and {Claver}, Charles F. and {Clowe}, Douglas and {Connolly}, A.~J. and {Cook}, Kem H. and {Cooke}, Jeff and {Cooray}, Asantha and {Covey}, Kevin R. and {Culliton}, Christopher S. and {de Jong}, Roelof and {de Vries}, Willem H. and {Debattista}, Victor P. and {Delgado}, Francisco and {Dell'Antonio}, Ian P. and {Dhital}, Saurav and {Di Stefano}, Rosanne and {Dickinson}, Mark and {Dilday}, Benjamin and {Djorgovski}, S.~G. and {Dobler}, Gregory and {Donalek}, Ciro and {Dubois-Felsmann}, Gregory and {Durech}, Josef and {Eliasdottir}, Ardis and {Eracleous}, Michael and {Eyer}, Laurent and {Falco}, Emilio E. and {Fan}, Xiaohui and {Fassnacht}, Christopher D. and {Ferguson}, Harry C. and {Fernandez}, Yanga R. and {Fields}, Brian D. and {Finkbeiner}, Douglas and {Figueroa}, Eduardo E. and {Fox}, Derek B. and {Francke}, Harold and {Frank}, James S. and {Frieman}, Josh and {Fromenteau}, Sebastien and {Furqan}, Muhammad and {Galaz}, Gaspar and {Gal-Yam}, A. and {Garnavich}, Peter and {Gawiser}, Eric and {Geary}, John and {Gee}, Perry and {Gibson}, Robert R. and {Gilmore}, Kirk and {Grace}, Emily A. and {Green}, Richard F. and {Gressler}, William J. and {Grillmair}, Carl J. and {Habib}, Salman and {Haggerty}, J.~S. and {Hamuy}, Mario and {Harris}, Alan W. and {Hawley}, Suzanne L. and {Heavens}, Alan F. and {Hebb}, Leslie and {Henry}, Todd J. and {Hileman}, Edward and {Hilton}, Eric J. and {Hoadley}, Keri and {Holberg}, J.~B. and {Holman}, Matt J. and {Howell}, Steve B. and {Infante}, Leopoldo and {Ivezic}, Zeljko and {Jacoby}, Suzanne H. and {Jain}, Bhuvnesh and {R} and {Jedicke} and {Jee}, M. James and {Garrett Jernigan}, J. and {Jha}, Saurabh W. and {Johnston}, Kathryn V. and {Jones}, R. Lynne and {Juric}, Mario and {Kaasalainen}, Mikko and {Styliani} and {Kafka} and {Kahn}, Steven M. and {Kaib}, Nathan A. and {Kalirai}, Jason and {Kantor}, Jeff and {Kasliwal}, Mansi M. and {Keeton}, Charles R. and {Kessler}, Richard and {Knezevic}, Zoran and {Kowalski}, Adam and {Krabbendam}, Victor L. and {Krughoff}, K. Simon and {Kulkarni}, Shrinivas and {Kuhlman}, Stephen and {Lacy}, Mark and {Lepine}, Sebastien and {Liang}, Ming and {Lien}, Amy and {Lira}, Paulina and {Long}, Knox S. and {Lorenz}, Suzanne and {Lotz}, Jennifer M. and {Lupton}, R.~H. and {Lutz}, Julie and {Macri}, Lucas M. and {Mahabal}, Ashish A. and {Mandelbaum}, Rachel and {Marshall}, Phil and {May}, Morgan and {McGehee}, Peregrine M. and {Meadows}, Brian T. and {Meert}, Alan and {Milani}, Andrea and {Miller}, Christopher J. and {Miller}, Michelle and {Mills}, David and {Minniti}, Dante and {Monet}, David and {Mukadam}, Anjum S. and {Nakar}, Ehud and {Neill}, Douglas R. and {Newman}, Jeffrey A. and {Nikolaev}, Sergei and {Nordby}, Martin and {O'Connor}, Paul and {Oguri}, Masamune and {Oliver}, John and {Olivier}, Scot S. and {Olsen}, Julia K. and {Olsen}, Knut and {Olszewski}, Edward W. and {Oluseyi}, Hakeem and {Padilla}, Nelson D. and {Parker}, Alex and {Pepper}, Joshua and {Peterson}, John R. and {Petry}, Catherine and {Pinto}, Philip A. and {Pizagno}, James L. and {Popescu}, Bogdan and {Prsa}, Andrej and {Radcka}, Veljko and {Raddick}, M. Jordan and {Rasmussen}, Andrew and {Rau}, Arne and {Rho}, Jeonghee and {Rhoads}, James E. and {Richards}, Gordon T. and {Ridgway}, Stephen T. and {Robertson}, Brant E. and {Roskar}, Rok and {Saha}, Abhijit and {Sarajedini}, Ata and {Scannapieco}, Evan and {Schalk}, Terry and {Schindler}, Rafe and {Schmidt}, Samuel},
        title = "{LSST Science Book, Version 2.0}",
      journal = {arXiv e-prints},
         year = 2009,
        month = dec,
          eid = {arXiv:0912.0201},
        pages = {arXiv:0912.0201},
          doi = {10.48550/arXiv.0912.0201},
archivePrefix = {arXiv},
       eprint = {0912.0201},
 primaryClass = {astro-ph.IM},
       adsurl = {https://ui.adsabs.harvard.edu/abs/2009arXiv0912.0201L}
}

@ARTICLE{ma2022,
       author = {{Ma}, Wenlin and {Liu}, Kexin and {Guo}, Hong and {Cui}, Weiguang and {Jones}, Michael G. and {Wang}, Jing and {Zhang}, Le and {Dav{\'e}}, Romeel},
        title = "{Effects of Active Galactic Nucleus Feedback on Cold Gas Depletion and Quenching of Central Galaxies}",
      journal = {\apj},
         year = 2022,
        month = dec,
       volume = {941},
       number = {2},
          eid = {205},
        pages = {205},
          doi = {10.3847/1538-4357/aca326},
archivePrefix = {arXiv},
       eprint = {2211.09969},
 primaryClass = {astro-ph.GA},
       adsurl = {https://ui.adsabs.harvard.edu/abs/2022ApJ...941..205M}
}

@ARTICLE{madau2014,
       author = {{Madau}, Piero and {Dickinson}, Mark},
        title = "{Cosmic Star-Formation History}",
      journal = {\araa},
         year = 2014,
        month = aug,
       volume = {52},
        pages = {415-486},
          doi = {10.1146/annurev-astro-081811-125615},
archivePrefix = {arXiv},
       eprint = {1403.0007},
 primaryClass = {astro-ph.CO},
       adsurl = {https://ui.adsabs.harvard.edu/abs/2014ARA&A..52..415M}
}

@ARTICLE{mann2012,
       author = {{Mann}, Andrew W. and {Ebeling}, Harald},
        title = "{X-ray-optical classification of cluster mergers and the evolution of the cluster merger fraction}",
      journal = {\mnras},
         year = 2012,
        month = mar,
       volume = {420},
       number = {3},
        pages = {2120-2138},
          doi = {10.1111/j.1365-2966.2011.20170.x},
archivePrefix = {arXiv},
       eprint = {1111.2396},
 primaryClass = {astro-ph.CO},
       adsurl = {https://ui.adsabs.harvard.edu/abs/2012MNRAS.420.2120M}
}

@INCOLLECTION{marinelli2025,
       author = {{Marinelli}, Mariarosa and {Green}, Joel and others},
        title = "{Wide Field Camera 3 Instrument Handbook, Version 18.0}",
    booktitle = {Wide Field Camera 3 Instrument Handbook, Version 18.0},
         year = 2025,
    publisher = {Baltimore: STScI},
       adsurl = {https://ui.adsabs.harvard.edu/abs/2025wfci.book...18M}
}

@ARTICLE{marshall2025,
       author = {{Marshall}, Madeline A. and {Amen}, Laurie and {Woods}, Tyrone E. and {C{\^o}t{\'e}}, Patrick and {Yung}, L.~Y. Aaron and {Amenouche}, Melissa and {Pass}, Emily K. and {Balogh}, Michael L. and {Salim}, Samir and {Moutard}, Thibaud},
        title = "{FORECASTOR - II. Simulating galaxy surveys with the Cosmological Advanced Survey Telescope for Optical and UV Research}",
      journal = {\mnras},
         year = 2025,
        month = feb,
       volume = {537},
       number = {2},
        pages = {1703-1719},
          doi = {10.1093/mnras/staf065},
archivePrefix = {arXiv},
       eprint = {2402.17163},
 primaryClass = {astro-ph.GA},
       adsurl = {https://ui.adsabs.harvard.edu/abs/2025MNRAS.537.1703M}
}

@ARTICLE{martig2009,
       author = {{Martig}, Marie and {Bournaud}, Fr{\'e}d{\'e}ric and {Teyssier}, Romain and {Dekel}, Avishai},
        title = "{Morphological Quenching of Star Formation: Making Early-Type Galaxies Red}",
      journal = {\apj},
         year = 2009,
        month = dec,
       volume = {707},
       number = {1},
        pages = {250-267},
          doi = {10.1088/0004-637X/707/1/250},
archivePrefix = {arXiv},
       eprint = {0905.4669},
 primaryClass = {astro-ph.CO},
       adsurl = {https://ui.adsabs.harvard.edu/abs/2009ApJ...707..250M}
}

@ARTICLE{martin2005,
       author = {{Martin}, D. Christopher and {Fanson}, James and {Schiminovich}, David and {Morrissey}, Patrick and {Friedman}, Peter G. and {Barlow}, Tom A. and {Conrow}, Tim and {Grange}, Robert and {Jelinsky}, Patrick N. and {Milliard}, Bruno and {Siegmund}, Oswald H.~W. and {Bianchi}, Luciana and {Byun}, Yong-Ik and {Donas}, Jose and {Forster}, Karl and {Heckman}, Timothy M. and {Lee}, Young-Wook and {Madore}, Barry F. and {Malina}, Roger F. and {Neff}, Susan G. and {Rich}, R. Michael and {Small}, Todd and {Surber}, Frank and {Szalay}, Alex S. and {Welsh}, Barry and {Wyder}, Ted K.},
        title = "{The Galaxy Evolution Explorer: A Space Ultraviolet Survey Mission}",
      journal = {\apjl},
         year = 2005,
        month = jan,
       volume = {619},
       number = {1},
        pages = {L1-L6},
          doi = {10.1086/426387},
archivePrefix = {arXiv},
       eprint = {astro-ph/0411302},
 primaryClass = {astro-ph},
       adsurl = {https://ui.adsabs.harvard.edu/abs/2005ApJ...619L...1M}
}

@ARTICLE{martin2007,
       author = {{Martin}, D. Christopher and {Wyder}, Ted K. and {Schiminovich}, David and {Barlow}, Tom A. and {Forster}, Karl and {Friedman}, Peter G. and {Morrissey}, Patrick and {Neff}, Susan G. and {Seibert}, Mark and {Small}, Todd and {Welsh}, Barry Y. and {Bianchi}, Luciana and {Donas}, Jos{\'e} and {Heckman}, Timothy M. and {Lee}, Young-Wook and {Madore}, Barry F. and {Milliard}, Bruno and {Rich}, R. Michael and {Szalay}, Alex S. and {Yi}, Sukyoung K.},
        title = "{The UV-Optical Galaxy Color-Magnitude Diagram. III. Constraints on Evolution from the Blue to the Red Sequence}",
      journal = {\apjs},
         year = 2007,
        month = dec,
       volume = {173},
       number = {2},
        pages = {342-356},
          doi = {10.1086/516639},
archivePrefix = {arXiv},
       eprint = {astro-ph/0703281},
 primaryClass = {astro-ph},
       adsurl = {https://ui.adsabs.harvard.edu/abs/2007ApJS..173..342M}
}

@ARTICLE{matharu2021,
       author = {{Matharu}, Jasleen and {Muzzin}, Adam and {Brammer}, Gabriel B. and {Nelson}, Erica J. and {Auger}, Matthew W. and {Hewett}, Paul C. and {van der Burg}, Remco and {Balogh}, Michael and {Demarco}, Ricardo and {Marchesini}, Danilo and {Noble}, Allison G. and {Rudnick}, Gregory and {van der Wel}, Arjen and {Wilson}, Gillian and {Yee}, Howard K.~C.},
        title = "{HST/WFC3 Grism Observations of z   1 Clusters: Evidence for Rapid Outside-in Environmental Quenching from Spatially Resolved H{\ensuremath{\alpha}} Maps}",
      journal = {\apj},
         year = 2021,
        month = dec,
       volume = {923},
       number = {2},
          eid = {222},
        pages = {222},
          doi = {10.3847/1538-4357/ac26c3},
archivePrefix = {arXiv},
       eprint = {2109.06186},
 primaryClass = {astro-ph.GA},
       adsurl = {https://ui.adsabs.harvard.edu/abs/2021ApJ...923..222M}
}

@ARTICLE{matharu2022,
       author = {{Matharu}, Jasleen and {Papovich}, Casey and {Simons}, Raymond C. and {Momcheva}, Ivelina and {Brammer}, Gabriel and {Ji}, Zhiyuan and {Backhaus}, Bren E. and {Cleri}, Nikko J. and {Estrada-Carpenter}, Vicente and {Finkelstein}, Steven L. and {Finlator}, Kristian and {Giavalisco}, Mauro and {Jung}, Intae and {Muzzin}, Adam and {Nelson}, Erica J. and {Pillepich}, Annalisa and {Trump}, Jonathan R. and {Weiner}, Benjamin},
        title = "{CLEAR: The Evolution of Spatially Resolved Star Formation in Galaxies between 0.5 {\ensuremath{\lesssim}} z {\ensuremath{\lesssim}} 1.7 Using H{\ensuremath{\alpha}} Emission Line Maps}",
      journal = {\apj},
         year = 2022,
        month = sep,
       volume = {937},
       number = {1},
          eid = {16},
        pages = {16},
          doi = {10.3847/1538-4357/ac8471},
archivePrefix = {arXiv},
       eprint = {2205.08543},
 primaryClass = {astro-ph.GA},
       adsurl = {https://ui.adsabs.harvard.edu/abs/2022ApJ...937...16M}
}

@ARTICLE{mcdonough2025,
       author = {{McDonough}, Bryanne and {Curtis}, Olivia and {Brainerd}, Tereasa G.},
        title = "{Intrinsic and Environmental Effects on the Distribution of Star Formation in TNG100 Galaxies}",
      journal = {\apj},
         year = 2025,
        month = jan,
       volume = {978},
       number = {1},
          eid = {67},
        pages = {67},
          doi = {10.3847/1538-4357/ad9747},
archivePrefix = {arXiv},
       eprint = {2411.13666},
 primaryClass = {astro-ph.GA},
       adsurl = {https://ui.adsabs.harvard.edu/abs/2025ApJ...978...67M}
}

@ARTICLE{mcnamara2007,
       author = {{McNamara}, B.~R. and {Nulsen}, P.~E.~J.},
        title = "{Heating Hot Atmospheres with Active Galactic Nuclei}",
      journal = {\araa},
         year = 2007,
        month = sep,
       volume = {45},
       number = {1},
        pages = {117-175},
          doi = {10.1146/annurev.astro.45.051806.110625},
archivePrefix = {arXiv},
       eprint = {0709.2152},
 primaryClass = {astro-ph},
       adsurl = {https://ui.adsabs.harvard.edu/abs/2007ARA&A..45..117M}
}

@ARTICLE{mcnamara2012,
       author = {{McNamara}, B.~R. and {Nulsen}, P.~E.~J.},
        title = "{Mechanical feedback from active galactic nuclei in galaxies, groups and clusters}",
      journal = {New Journal of Physics},
         year = 2012,
        month = may,
       volume = {14},
       number = {5},
          eid = {055023},
        pages = {055023},
          doi = {10.1088/1367-2630/14/5/055023},
archivePrefix = {arXiv},
       eprint = {1204.0006},
 primaryClass = {astro-ph.CO},
       adsurl = {https://ui.adsabs.harvard.edu/abs/2012NJPh...14e5023M}
}

@ARTICLE{medezinski2016,
       author = {{Medezinski}, Elinor and {Umetsu}, Keiichi and {Okabe}, Nobuhiro and {Nonino}, Mario and {Molnar}, Sandor and {Massey}, Richard and {Dupke}, Renato and {Merten}, Julian},
        title = "{Frontier Fields: Subaru Weak-Lensing Analysis of the Merging Galaxy Cluster A2744}",
      journal = {\apj},
         year = 2016,
        month = jan,
       volume = {817},
       number = {1},
          eid = {24},
        pages = {24},
          doi = {10.3847/0004-637X/817/1/24},
archivePrefix = {arXiv},
       eprint = {1507.03992},
 primaryClass = {astro-ph.CO},
       adsurl = {https://ui.adsabs.harvard.edu/abs/2016ApJ...817...24M}
}

@ARTICLE{merlin2016,
       author = {{Merlin}, E. and {Amor{\'\i}n}, R. and {Castellano}, M. and {Fontana}, A. and {Buitrago}, F. and {Dunlop}, J.~S. and {Elbaz}, D. and {Boucaud}, A. and {Bourne}, N. and {Boutsia}, K. and {Brammer}, G. and {Bruce}, V.~A. and {Capak}, P. and {Cappelluti}, N. and {Ciesla}, L. and {Comastri}, A. and {Cullen}, F. and {Derriere}, S. and {Faber}, S.~M. and {Ferguson}, H.~C. and {Giallongo}, E. and {Grazian}, A. and {Lotz}, J. and {Micha{\l}owski}, M.~J. and {Paris}, D. and {Pentericci}, L. and {Pilo}, S. and {Santini}, P. and {Schreiber}, C. and {Shu}, X. and {Wang}, T.},
        title = "{The ASTRODEEP Frontier Fields catalogues. I. Multiwavelength photometry of Abell-2744 and MACS-J0416}",
      journal = {\aap},
         year = 2016,
        month = may,
       volume = {590},
          eid = {A30},
        pages = {A30},
          doi = {10.1051/0004-6361/201527513},
archivePrefix = {arXiv},
       eprint = {1603.02460},
 primaryClass = {astro-ph.GA},
       adsurl = {https://ui.adsabs.harvard.edu/abs/2016A&A...590A..30M}
}

@ARTICLE{mitchell2020,
       author = {{Mitchell}, Peter D. and {Schaye}, Joop and {Bower}, Richard G. and {Crain}, Robert A.},
        title = "{Galactic outflow rates in the EAGLE simulations}",
      journal = {\mnras},
         year = 2020,
        month = may,
       volume = {494},
       number = {3},
        pages = {3971-3997},
          doi = {10.1093/mnras/staa938},
archivePrefix = {arXiv},
       eprint = {1910.09566},
 primaryClass = {astro-ph.GA},
       adsurl = {https://ui.adsabs.harvard.edu/abs/2020MNRAS.494.3971M}
}

@ARTICLE{moore1996,
       author = {{Moore}, Ben and {Katz}, Neal and {Lake}, George and {Dressler}, Alan and {Oemler}, Augustus},
        title = "{Galaxy harassment and the evolution of clusters of galaxies}",
      journal = {\nat},
         year = 1996,
        month = feb,
       volume = {379},
       number = {6566},
        pages = {613-616},
          doi = {10.1038/379613a0},
archivePrefix = {arXiv},
       eprint = {astro-ph/9510034},
 primaryClass = {astro-ph},
       adsurl = {https://ui.adsabs.harvard.edu/abs/1996Natur.379..613M}
}

@ARTICLE{moore1998,
       author = {{Moore}, Ben and {Lake}, George and {Katz}, Neal},
        title = "{Morphological Transformation from Galaxy Harassment}",
      journal = {\apj},
         year = 1998,
        month = mar,
       volume = {495},
       number = {1},
        pages = {139-151},
          doi = {10.1086/305264},
archivePrefix = {arXiv},
       eprint = {astro-ph/9701211},
 primaryClass = {astro-ph},
       adsurl = {https://ui.adsabs.harvard.edu/abs/1998ApJ...495..139M}
}

@ARTICLE{moore1999,
       author = {{Moore}, Ben and {Lake}, George and {Quinn}, Thomas and {Stadel}, Joachim},
        title = "{On the survival and destruction of spiral galaxies in clusters}",
      journal = {\mnras},
         year = 1999,
        month = apr,
       volume = {304},
       number = {3},
        pages = {465-474},
          doi = {10.1046/j.1365-8711.1999.02345.x},
archivePrefix = {arXiv},
       eprint = {astro-ph/9811127},
 primaryClass = {astro-ph},
       adsurl = {https://ui.adsabs.harvard.edu/abs/1999MNRAS.304..465M}
}

@ARTICLE{morselli2019,
       author = {{Morselli}, L. and {Popesso}, P. and {Cibinel}, A. and {Oesch}, P.~A. and {Montes}, M. and {Atek}, H. and {Illingworth}, G.~D. and {Holden}, B.},
        title = "{Spatial distribution of stellar mass and star formation activity at 0.2 < z < 1.2 across and along the main sequence}",
      journal = {\aap},
         year = 2019,
        month = jun,
       volume = {626},
          eid = {A61},
        pages = {A61},
          doi = {10.1051/0004-6361/201834559},
archivePrefix = {arXiv},
       eprint = {1812.08561},
 primaryClass = {astro-ph.GA},
       adsurl = {https://ui.adsabs.harvard.edu/abs/2019A&A...626A..61M}
}

@ARTICLE{moustakas2013,
       author = {{Moustakas}, John and {Coil}, Alison L. and {Aird}, James and {Blanton}, Michael R. and {Cool}, Richard J. and {Eisenstein}, Daniel J. and {Mendez}, Alexander J. and {Wong}, Kenneth C. and {Zhu}, Guangtun and {Arnouts}, St{\'e}phane},
        title = "{PRIMUS: Constraints on Star Formation Quenching and Galaxy Merging, and the Evolution of the Stellar Mass Function from z = 0-1}",
      journal = {\apj},
         year = 2013,
        month = apr,
       volume = {767},
       number = {1},
          eid = {50},
        pages = {50},
          doi = {10.1088/0004-637X/767/1/50},
archivePrefix = {arXiv},
       eprint = {1301.1688},
 primaryClass = {astro-ph.CO},
       adsurl = {https://ui.adsabs.harvard.edu/abs/2013ApJ...767...50M}
}

@ARTICLE{nelson2018,
       author = {{Nelson}, Dylan and {Pillepich}, Annalisa and {Springel}, Volker and {Weinberger}, Rainer and {Hernquist}, Lars and {Pakmor}, R{\"u}diger and {Genel}, Shy and {Torrey}, Paul and {Vogelsberger}, Mark and {Kauffmann}, Guinevere and {Marinacci}, Federico and {Naiman}, Jill},
        title = "{First results from the IllustrisTNG simulations: the galaxy colour bimodality}",
      journal = {\mnras},
         year = 2018,
        month = mar,
       volume = {475},
       number = {1},
        pages = {624-647},
          doi = {10.1093/mnras/stx3040},
archivePrefix = {arXiv},
       eprint = {1707.03395},
 primaryClass = {astro-ph.GA},
       adsurl = {https://ui.adsabs.harvard.edu/abs/2018MNRAS.475..624N}
}

@ARTICLE{nelson2019a,
       author = {{Nelson}, Dylan and {Springel}, Volker and {Pillepich}, Annalisa and {Rodriguez-Gomez}, Vicente and {Torrey}, Paul and {Genel}, Shy and {Vogelsberger}, Mark and {Pakmor}, Ruediger and {Marinacci}, Federico and {Weinberger}, Rainer and {Kelley}, Luke and {Lovell}, Mark and {Diemer}, Benedikt and {Hernquist}, Lars},
        title = "{The IllustrisTNG simulations: public data release}",
      journal = {ComAC},
         year = 2019,
        month = may,
       volume = {6},
       number = {1},
          eid = {2},
        pages = {2},
          doi = {10.1186/s40668-019-0028-x},
archivePrefix = {arXiv},
       eprint = {1812.05609},
 primaryClass = {astro-ph.GA},
       adsurl = {https://ui.adsabs.harvard.edu/abs/2019ComAC...6....2N}
}

@ARTICLE{nelson2019b,
       author = {{Nelson}, Dylan and {Pillepich}, Annalisa and {Springel}, Volker and {Pakmor}, R{\"u}diger and {Weinberger}, Rainer and {Genel}, Shy and {Torrey}, Paul and {Vogelsberger}, Mark and {Marinacci}, Federico and {Hernquist}, Lars},
        title = "{First results from the TNG50 simulation: galactic outflows driven by supernovae and black hole feedback}",
      journal = {\mnras},
         year = 2019,
        month = dec,
       volume = {490},
       number = {3},
        pages = {3234-3261},
          doi = {10.1093/mnras/stz2306},
archivePrefix = {arXiv},
       eprint = {1902.05554},
 primaryClass = {astro-ph.GA},
       adsurl = {https://ui.adsabs.harvard.edu/abs/2019MNRAS.490.3234N}
}

@ARTICLE{nelson2021,
       author = {{Nelson}, Erica J. and {Tacchella}, Sandro and {Diemer}, Benedikt and {Leja}, Joel and {Hernquist}, Lars and {Whitaker}, Katherine E. and {Weinberger}, Rainer and {Pillepich}, Annalisa and {Nelson}, Dylan and {Terrazas}, Bryan A. and {Nevin}, Rebecca and {Brammer}, Gabriel B. and {Burkhart}, Blakesley and {Cochrane}, Rachel K. and {van Dokkum}, Pieter and {Johnson}, Benjamin D. and {Marinacci}, Federico and {Mowla}, Lamiya and {Pakmor}, R{\"u}diger and {Skelton}, Rosalind E. and {Speagle}, Joshua and {Springel}, Volker and {Torrey}, Paul and {Vogelsberger}, Mark and {Wuyts}, Stijn},
        title = "{Spatially resolved star formation and inside-out quenching in the TNG50 simulation and 3D-HST observations}",
      journal = {\mnras},
         year = 2021,
        month = nov,
       volume = {508},
       number = {1},
        pages = {219-235},
          doi = {10.1093/mnras/stab2131},
archivePrefix = {arXiv},
       eprint = {2101.12212},
 primaryClass = {astro-ph.GA},
       adsurl = {https://ui.adsabs.harvard.edu/abs/2021MNRAS.508..219N}
}

@ARTICLE{oman2013,
       author = {{Oman}, Kyle A. and {Hudson}, Michael J. and {Behroozi}, Peter S.},
        title = "{Disentangling satellite galaxy populations using orbit tracking in simulations}",
      journal = {\mnras},
         year = 2013,
        month = may,
       volume = {431},
       number = {3},
        pages = {2307-2316},
          doi = {10.1093/mnras/stt328},
archivePrefix = {arXiv},
       eprint = {1301.6757},
 primaryClass = {astro-ph.CO},
       adsurl = {https://ui.adsabs.harvard.edu/abs/2013MNRAS.431.2307O}
}

@ARTICLE{oman2016,
       author = {{Oman}, Kyle A. and {Hudson}, Michael J.},
        title = "{Satellite quenching time-scales in clusters from projected phase space measurements matched to simulated orbits}",
      journal = {\mnras},
         year = 2016,
        month = dec,
       volume = {463},
       number = {3},
        pages = {3083-3095},
          doi = {10.1093/mnras/stw2195},
archivePrefix = {arXiv},
       eprint = {1607.07934},
 primaryClass = {astro-ph.GA},
       adsurl = {https://ui.adsabs.harvard.edu/abs/2016MNRAS.463.3083O}
}

@ARTICLE{owers2011,
       author = {{Owers}, Matt S. and {Randall}, Scott W. and {Nulsen}, Paul E.~J. and {Couch}, Warrick J. and {David}, Laurence P. and {Kempner}, Joshua C.},
        title = "{The Dissection of Abell 2744: A Rich Cluster Growing Through Major and Minor Mergers}",
      journal = {\apj},
         year = 2011,
        month = feb,
       volume = {728},
       number = {1},
          eid = {27},
        pages = {27},
          doi = {10.1088/0004-637X/728/1/27},
archivePrefix = {arXiv},
       eprint = {1012.1315},
 primaryClass = {astro-ph.CO},
       adsurl = {https://ui.adsabs.harvard.edu/abs/2011ApJ...728...27O}
}

@ARTICLE{pasquali2019,
       author = {{Pasquali}, A. and {Smith}, R. and {Gallazzi}, A. and {De Lucia}, G. and {Zibetti}, S. and {Hirschmann}, M. and {Yi}, S.~K.},
        title = "{Physical properties of SDSS satellite galaxies in projected phase space}",
      journal = {\mnras},
         year = 2019,
        month = apr,
       volume = {484},
       number = {2},
        pages = {1702-1723},
          doi = {10.1093/mnras/sty3530},
archivePrefix = {arXiv},
       eprint = {1901.04238},
 primaryClass = {astro-ph.GA},
       adsurl = {https://ui.adsabs.harvard.edu/abs/2019MNRAS.484.1702P}
}

@ARTICLE{pedregosa2011,
       author = {{Pedregosa}, Fabian and {Varoquaux}, Ga{\"e}l and {Gramfort}, Alexandre and {Michel}, Vincent and {Thirion}, Bertrand and {Grisel}, Olivier and {Blondel}, Mathieu and {M{\"u}ller}, Andreas and {Nothman}, Joel and {Louppe}, Gilles and {Prettenhofer}, Peter and {Weiss}, Ron and {Dubourg}, Vincent and {Vanderplas}, Jake and {Passos}, Alexandre and {Cournapeau}, David and {Brucher}, Matthieu and {Perrot}, Matthieu and {Duchesnay}, {\'E}douard},
        title = "{Scikit-learn: Machine Learning in Python}",
      journal = {JMLR},
         year = 2011,
        month = oct,
       volume = {12},
        pages = {2825-2830},
          doi = {10.48550/arXiv.1201.0490},
archivePrefix = {arXiv},
       eprint = {1201.0490},
 primaryClass = {cs.LG},
       adsurl = {https://ui.adsabs.harvard.edu/abs/2011JMLR...12.2825P}
}

@INCOLLECTION{stark2025,
       author = {{Stark}, D.~V. and others},
        title = "{Advanced Camera for Surveys Instrument Handbook, Version 25.0}",
    booktitle = {Advanced Camera for Surveys Instrument Handbook, Version 25.0},
         year = 2025,
    publisher = {Baltimore: STScI},
       adsurl = {https://ui.adsabs.harvard.edu/abs/2025acsi.book...25S}
}
\bibliographystyle{aasjournal}

\end{document}